\documentclass[prd,nofootinbib,preprint,superscriptaddress]{revtex4}
\usepackage[a4paper,top=2cm,bottom=2cm,left=2cm,right=2cm,marginparwidth=2cm]{geometry}
\usepackage{amssymb}
\usepackage{amsmath}
\usepackage{bbold}
\usepackage{epsfig}
\usepackage{graphicx}
\usepackage{tikz}
\usetikzlibrary{snakes}
\usepackage{subfigure}
\usepackage{soul}
\usepackage{pifont}
\newcommand{\xmark}{\ding{55}}

\usepackage[colorlinks=true, allcolors=blue]{hyperref}

\newcommand{\pg}{\textcolor{red}}

\begin{document}

\title{A singlet scalar assisted $N_{2}$ Leptogenesis and Pseudo-Scalar Dark Matter}

\author{Dilip Kumar Ghosh}
\email{dilipghoshjal@gmail.com}
\affiliation{School of Physical Sciences, Indian Association for the Cultivation of Science, Kolkata-700032, India}

\author{Purusottam Ghosh}
\email{pghoshiitg@gmail.com}
\affiliation{Institute of Mathematical Sciences,
Taramani, Chennai 600 113, India}
\affiliation{Homi Bhabha National Institute,
Anushakti Nagar, Mumbai 400094, India}

\author{Koustav Mukherjee}
\email{koustav.physics1995@gmail.com}
\affiliation{School of Physical Sciences, Indian Association for the Cultivation of Science, Kolkata-700032, India}

\author{Nimmala Narendra}
\email{nimmalanarendra@gmail.com}
\affiliation{School of Physical Sciences, Indian Association for the Cultivation of Science, Kolkata-700032, India}

\begin{abstract}
We study the Leptogenesis and Dark Matter in the presence of an extra singlet complex scalar field in an extended discrete $\mathcal{Z_{\rm 3}}$ symmetry. The vacuum expectation value of the new scalar spontaneously breaks the $\mathcal{Z_{\rm 3}}$ symmetry. A remnant CP-like $\mathcal{Z_{\rm 2}}$ symmetry stabilizes the imaginary part of the complex scalar field, which can act as a pseudo-Goldstone Dark Matter. The real part of the complex scalar couples to RHN opens up new decay channels, which can lead to a larger CP-violation in generating the lepton asymmetry. Thus, the singlet complex scalar plays a crucial role in understanding the Leptogenesis and Dark Matter parameter space.  This singlet complex scalar is also responsible for the First-Order Phase Transition (FOPT), which may provide observable stochastic Gravitational wave signatures. We discuss the possible correlations among these three phenomena.
\end{abstract}
\maketitle
\section{Introduction}\label{sec:Intro}

Cosmological observations suggest that the number of baryons in the observable universe is not equal to the number of anti-baryons. In other words, according to our current understanding of particle physics and cosmology, all large-scale structures in the visible universe are composed of matter, consisting of protons and electrons, with no significant presence of antimatter, such as anti-protons or positrons. This asymmetry between matter and antimatter of the Universe can be expressed in terms of the baryon-to-photon ratio $\eta_B=\frac{n_B- {n}_{\overline{B}}}{n_\gamma}=(6.09 \pm 0.06) \times 10^{-10}$ \cite{Planck:2018vyg}. 
Leptogenesis is an attractive mechanism to generate such cosmological baryon asymmetry of the Universe \cite{Fukugita:1986hr}. Once the lepton asymmetry is generated, a portion of it converts to Baryon Asymmetry of the Universe (BAU) via Electroweak sphaleron processes \cite{Kuzmin:1985mm}. Leptogenesis has received special attention ever since the evidence of non-zero neutrino masses \cite{SNO:2002tuh, Bahcall:2004mz, Super-Kamiokande:2001bfk, KamLAND:2002uet}. 
In the Type-I seesaw mechanism\cite{Minkowski:1977sc, Yanagida:1979as, Mohapatra:1979ia,PhysRevD.22.2227,PhysRevD.25.774,Paul:2024iie}, the heavy right-handed neutrinos that couple to SM particles through the Dirac Yukawa interaction can decay and generate the lepton asymmetry by satisfying Sakharov conditions\cite{Sakharov:1967dj}.

Another long-standing puzzle that appears in cosmological evolution is Dark Matter (DM). The existence of DM is supported by several astrophysical and cosmological observations based on its Gravitational interaction, including the anisotropies of the Cosmic Microwave Background (CMB) \cite{Planck:2018vyg}, Gravitational lensing, {galaxy rotation curves in spiral galaxies \cite{Rubin1983}, and the motion of galaxy clusters \cite{Zwicky:1933gu}}. Analysis of anisotropies in CMB data reveals that approximately one-fourth of the Universe consists of DM, which is non-baryonic and non-luminous \cite{Hu:2001bc}.
Based on CMB observation, the PLANCK collaboration reported the observed relic density of DM to be $\Omega_\text{DM} h^2=0.120\pm 0.001$\cite{Planck:2018vyg}. However, the nature of DM, its non-gravitational interactions, and its production mechanism remain unknown. Over the years, different types of production mechanisms of DM  in the early Universe have been proposed based on its interaction strength with the visible sector. The WIMP (Weakly Interacting Massive Particle) like DM scenarios are widely studied in the literature \cite{Kolb:1990vq}. The WIMP is assumed to be in thermal equilibrium with the visible sector particles in the early Universe at a temperature above its mass scale. The WIMP freezes out from the thermal bath as the Universe expands and the temperature falls below its mass scale. The sizeable interaction with the visible sector enables WIMPs to be detected through direct (XENON1T\cite{XENON:2018voc}, PANDAX 4T\cite{PandaX-4T:2021bab},  LUX-ZEPLIN (LZ) \cite{LZ:2022ufs,LZ:2024zvo}, etc.), indirect (FERMI LAT, MAGIC \cite{Fermi-LAT:2015att,MAGIC:2016xys}), and collider (LHC, ILC, etc. \cite{Kahlhoefer:2017dnp}) search experiments. The non-observation of DM in these experiments imposes constraints on WIMP-like scenarios.

In the seesaw mechanism, the explanation behind the light neutrino masses requires a high-energy scale for the right-handed neutrinos that are presently beyond the reach of current or near-future collider experiments. Several attempts have been made in the literature to bring down this scale, which are Akhmedov-Rubakov-Smirnov (ARS) mechanism\cite{Akhmedov:1998qx}, Resonant Leptogenesis \cite{Pilaftsis:2003gt}, Leptogenesis in the scotogenic model of radiative neutrino masses \cite{Hugle:2018qbw, Ma:2006km}{\it, etc}. In recent work, \cite{Alanne:2018brf}, authors considered a simple real scalar extension claiming the right-handed neutrino mass below the TeV scale so that they can be searched at present and future colliders. For the first time, the general mechanism has been defined in \cite{LeDall:2014too}. A real singlet scalar can couple to a pair of right-handed neutrinos. This vertex allows new decay channels leading to a larger CP violation. Thus, the lowest right-handed neutrino mass can be brought to a TeV scale. The Leptogenesis parameter space can be expressed in low energy parameters using the Casas-Ibarra parameterization \cite{Casas:2001sr}. The advantage of this mechanism is that it can evade the lower bound (Davidson-Ibarra bound \cite{Davidson:2002qv}) on the lowest right-handed neutrino mass. Here, we consider a complex singlet scalar instead of a real scalar; once it gets a vacuum expectation value (vev), the real part of the scalar can play a role in achieving Leptogenesis as discussed above. The imaginary part of the complex scalar can serve as a pseudo-scalar dark matter candidate, with its stability ensured by an appropriate discrete symmetry. 
This allows us to explore the 
study of Leptogenesis and DM simultaneously within a unified framework while also facilitating the realization of Leptogenesis at the TeV scale.

In the present work, we consider a simple scenario that extends the SM symmetry with a discrete $\mathcal{Z_{\rm 3}}$ symmetry and the minimal particle content with two right-handed neutrinos $N_{1,2}$ and a singlet complex scalar. {The cubic terms in the scalar potential allow a strong first-order phase transition and lead to spontaneous breaking of the \(\mathcal{Z}_3\) symmetry down to \(\mathcal{Z}_2\), which stabilizes the DM  \cite{Kannike:2019mzk,Ghosh:2024ing}.}
 With this minimal setup, we study the $N_2$-Leptogenesis \cite{Engelhard:2006yg, Alanne:2018brf, LeDall:2014too} in the presence of a viable pseudo-scalar DM. To keep our analysis simple, we ignore neutrino flavor effects. The singlet complex scalar of this scenario plays a key role in understanding both the phenomenological concepts: Baryon asymmetry of the Universe and relic density of DM. The singlet scalar that couples with the right-handed neutrinos allows an additional contribution to the total CP asymmetry after it gets a vev. The CP-odd state of the singlet scalar acts as a pseudo-scalar DM candidate of the model due to a discrete $\mathcal{Z_{\rm 2}}$ like symmetry of the potential. The CP even state mixes with the Standard Model (SM) Higgs and allows the annihilation of DM to SM particles in addressing the observed relic density of DM. The quartic coupling between the singlet scalar and the SM Higgs, and the vev of the singlet scalar play a crucial role together in both the Leptogenesis and DM sectors. Thus, this model establishes a common parameter space that can be explored through Higgs searches, DM direct detection (DD) experiments, and upcoming collider experiments.

The groundbreaking discovery of Gravitational Waves (GW) by the LIGO \cite{Abbott_2009} has ushered in a new era of cosmological exploration. Stochastic Gravitational waves, generated during the early Universe, can arise from the strong first-order phase transitions (SFOPT). In our current theoretical framework, the introduction of an additional scalar field coupled to the Standard Model Higgs boson enhances the possibility of such a phase transition. We investigate the parameter space of this extended model that is consistent with both Leptogenesis and DM phenomena while allowing for a strong first-order phase transition \footnote{{Note that the first order phase transition can have an additional contribution to electroweak baryogenesis. However, it is well known
that addressing the observed baryon asymmetry requires sufficient CP violation, which is absent in the SM with a BSM scalar \cite{Vaskonen:2016yiu}.}}.
Inclusion of the $\mu_3$ term in the scalar potential introduces a barrier in the tree-level potential, facilitating the occurrence of the phase transition. These transitions could produce Gravitational waves 
that may be detectable by future Gravitational wave observatories like
LISA\cite{amaroseoane2017laser}, LIGO\cite{Abbott_2009}, BBO\cite{Crowder_2005}, DECIGO\cite{PhysRevD.73.064006}{\it, etc.}. Studies related to the FOPT of the ${\mathcal {Z_{\rm 3}}}$ model have already been discussed in the literature  \cite{Kang_2018,Ghosh:2022fzp}. The FOPT and GW spectrum studies have been conducted with $\mathcal{Z_{\rm 3}}$ symmetry in a pNGB DM model \cite{Kannike:2019mzk}. 
Motivated by the above studies, our primary goal in this analysis is to identify a unified parameter space that can simultaneously support low-scale Leptogenesis and a strong first-order phase transition while meeting all current DM constraints. To illustrate the SFOPT phenomenon and its potential observable signatures of stochastic Gravitational waves, we select a few representative benchmark points that comply with theoretical and existing experimental constraints. Our analysis demonstrates that  SFOPT, primarily driven by the SM Higgs boson, is highly improbable in this parameter space, even in scenarios with under-abundant DM relic densities, while SFOPT driven by BSM Higgs boson remains viable as 
discussed in Sec.\ref{sec:-results}.

The paper is organized as follows. In Sec.\ref{sec:-model}, we introduce the model and discuss the scalar sector of the model. In Sec.\ref{sec:-Leptogenesis}, we discuss the $N_2$ Leptogenesis and its parameter space in the presence of a singlet scalar. The DM relic density and its detection aspects are discussed in Sec.\ref{sec:-DM}.
In Sec.\ref{sec:-PT}, we examine the strong first-order phase transition (FOPT), which results in the generation of gravitational waves, as discussed in detail in Sec. \ref{sec:-GW}. In Sec.\ref{sec:-results}, we discuss and 
analyse our results. Finally, we summarize and conclude in Sec.\ref{sec:-summary}.
\section{The Model} \label{sec:-model}
\begin{table}[ht]
\begin{center}
\begin{tabular}{|c|c|c|c|}
\hline
Fields & $SU(2)_L$ & $U(1)_Y$ & $\mathcal{Z_{\rm 3}}$  \\
\hline
$N^{'}$ & 1 & 0 &  $N^{'} \to N^{'}$      \\
$\Phi$     & 1 & 0 & $\Phi \to e^{i 2\pi/3}\, \Phi$  \\
\hline
\end{tabular}
\end{center}
\caption{\it Charge assignment of the content of the additional fields under the gauge group $G_{\rm SM} \otimes \mathcal{Z}_3$. }
\label{table}
\end{table}

We consider the SM with an extra complex singlet scalar $\Phi$ and two right-handed neutrinos (RHN) $N_{i}^{'},\,(i=1,2)$ under an extended discrete symmetry $\mathcal{Z_{\rm 3}}$ \cite{Kannike:2019mzk, Gross:2017dan}. Note the $\mathcal{Z}_3$ transformation leaves the SM fields unchanged. The complex scalar couples with right-handed neutrinos via a dimension-five operator, given in Eq.\ref{Eq:-IniLagrangian}. The assigned charges of the new particle content under the $\mathcal{Z_{\rm 3}}$ symmetry are described in Table \ref{table}. 

The relevant interaction Lagrangian  under the extended gauge group can be written as
\begin{eqnarray}
\mathcal{L} &\supset &  -\left[ \lambda_{\alpha i}'\, \overline{L_\alpha} \tilde{H} N_i' +\frac{1}{2} ( M_{ij}' +\frac{y_{ij}'}{\Lambda} \Phi^{\dagger}\Phi) \overline{N_{i}^{'c}} N_{j}' + h.c.\right]-V(H,\Phi) ,
\label{Eq:-IniLagrangian}
\end{eqnarray}

where $\lambda_{\alpha i}'$ denote the RHN-lepton-Higgs Yukawa matrix with $\{\alpha=e,\mu,\tau\}$ and $\{i=1,2\}$. 
${L_\alpha}=
\begin{pmatrix}
\nu_\alpha &  \alpha_L
\end{pmatrix}^T $ denotes the three SM left handed charged lepton doublets and $\tilde{H}=i\tau H^*$ where $H$ is the standard model Higgs doublet. $M_{ij}'$ and $y^\prime_{ij}$ denote the mass matrix and 
the coefficient of the dim-5 operator of the neutrinos in the unphysical basis. The scale 
$\Lambda$ is the cut-off scale of the model. 
It is worth noting that the symmetry of the current framework 
allows for the inclusion of additional dimension-5 operators, such as the
Weinberg operator: $ {\cal O}_W \simeq \frac{C_{W}}{\Lambda_1} \overline{\tilde{L}} H
\tilde{H}^\dagger L$ and another operator involving right-handed neutrinos:
${\cal O}_{HN}\simeq \frac{C_{HN}}{\Lambda_2}(H^{\dagger}H) \overline{N^{'c}} N'$
where the suppression scales $\Lambda_1 $ and $\Lambda_2 $
correspond to the masses of two distinct heavy new particles arising from different underlying dynamics in their respective UV-complete models.
In our analysis, we assume that the  coefficients of these operators,
$C_W/\Lambda_1 $ and $C_{HN}/\Lambda_2 $ should satisfy
$C_W/\Lambda_1, C_{HN}/\Lambda_2 << y_{ij}^\prime/\Lambda $ (where various
Wilson coefficients ($C_W, C_{HN} $ and $y_{ij}^\prime$) are of similar order magnitude ), indicating a hierarchy among
the scales $\Lambda_1, \Lambda_2 $ and $\Lambda $.
This hierarchical structure reflects the idea that different operators can originate from distinct sources in the UV-complete model, leading to varying suppression factors. This assumption greatly simplifies the analysis by allowing us to ignore contributions from both ${\cal O}_W$ and ${\cal O}_{HN}$. 

The scalar potential $V(H,\Phi)$ reads as
\begin{eqnarray}
V(H,\Phi) &=& -\mu_{H}^{2} (H^{\dagger} H) + \lambda_{H} (H^{\dagger} H)^{2}
- \mu_{\Phi}^{2} (\Phi^{\dagger} \Phi) + \lambda_{\Phi} (\Phi^{\dagger} \Phi)^{2} \nonumber\\
&& +\lambda_{H\Phi} (\Phi^{\dagger} \Phi)(H^{\dagger} H)  + \frac{\mu_{3}}{2} (\Phi^{3}+\Phi^{\dagger 3}).
\label{Eq_potential}
\end{eqnarray} 
Note that the potential has a $U(1)$ global symmetry, which is softly broken by the explicit $\mu_3$ term. In its absence, the CP odd state of $\Phi$, denoted by $\chi$, would be a massless Nambu Goldstone Boson after the spontaneous breaking of $\Phi $. The presence of the $\mu_3$ term introduces a mass to $\chi$. 
As $\Phi$ gets a vev, $\mathcal{Z_{\rm 3}}$ has broken spontaneously \footnote{{The spontaneous $\mathcal{Z_{\rm 3}}$ symmetry breaking leads to degenerate vacuum states, where the field can settle into any of them. The boundaries are known as domain walls, where energy is stored due to a mismatched vacuum. If the domain walls are stable, they can influence the cosmological observations. To address this, one can introduce an explicit $\mathcal{Z}_3$ breaking term at higher order, which does not impact our analysis \cite{Dvali:1994wv, Wu:2022stu, Ghosh:2017fmr}.}}. Even though the $\mathcal{Z_{\rm 3}}$ has broken the Lagrangian still holds a $\Phi \rightarrow \Phi^{\dagger}$ symmetry (which is equivalent to $\chi \rightarrow -\chi$) due to which the $\chi$ is stable \cite{Kannike:2019mzk}. Therefore, $\mathcal{Z_{\rm 2}}$ is the remnant discrete symmetry under which $\chi$ transforms non-trivially, while the other fields remain unchanged.
The real component of the complex scalar field $\Phi$ acquires nonzero vev $v_\phi$ (at high temperature $T_{\Phi}\gg T_{EW}$) along the CP even field direction and the $\Phi$ field can be expanded around the vev as 
\begin{equation*}
    \Phi=\frac{1}{\sqrt{2}}(\phi+v_\phi+i \chi) .
\end{equation*}
Both the CP-even and the CP-odd states of $\Phi$ acquire non-degenerate masses, given by 
\begin{eqnarray}
    M_{\phi}^2= 2\lambda_\Phi v_{\phi}^2 + \frac{3\, \mu_3 v_\phi}{2\sqrt{2}}  ~~{\rm and}~~M_{\chi}^2=-\frac{9\, \mu_3 v_\phi}{2 \sqrt{2}}.
\end{eqnarray}
The right-handed neutrino mass matrix $M_{ij}^\prime $ receives an additional contribution from the vev $(v_\phi)$ of $\Phi$, {\it i.e., $(M_{ij}'+ v_\phi^2 y_{ij}'/(2\Lambda))$}.
After diagonalizing, the relevant Lagrangian can be 
written as:
\begin{eqnarray}
-\mathcal{L} &\supset &  \left[\lambda_{\alpha i}\, \overline{L_\alpha} \tilde{H} N_i + \frac{1}{2} (D_N)_{ij} \overline{N_{i}^{c}} N_{j} + \alpha_{ij} \,\phi \overline{N_{i}^{c}} N_{j} + h.c. \right]+V(H,\phi) ,
\end{eqnarray} 
where  $D_N \equiv \text{diag}(M_{N_1}, M_{N_2})$ is the mass matrix of the right-handed neutrinos in their physical basis.
Here, the dimensionless variable $\alpha_{ij}$ represents the strength of the trilinear interaction term: $\phi \overline{N_i^c} N_j$. The $\alpha_{ij}$ is a complex symmetric matrix because of the Majorana nature of the right-handed neutrinos. The $(D_N)_{ij}$ and $\alpha_{ij}$ do not diagonalize simultaneously, which allows flavor-changing neutral current interactions among the right-handed neutrinos. Both these terms violate the global lepton number.

After Electroweak Symmetry Breaking (EWSB) (the scale is substantially lower than the $\mathcal{Z_{\rm 3}}$ breaking scale), the SM Higgs doublet gets a non-zero vev along the CP even direction. The Higgs around the EW vev ($v \simeq 246$GeV) can be parameterized as,
\begin{equation*}
H=
\begin{pmatrix}
0 \\ \frac{1}{\sqrt{2}}(h +v)
\end{pmatrix}\,.
\end{equation*}
 
\noindent After EWSB, the low-energy Lagrangian obtains the following form \cite{Alanne:2018brf}:
\begin{eqnarray}
-\mathcal{L} &\supset & \left[\lambda_{\alpha i}\, \nu_{\alpha} N_i \left(\frac{v+h}{\sqrt{2}} \right) +\frac{1}{2} (D_N)_{ij} \overline{N_{i}^{c}} N_{j} + \alpha_{ij} \phi \overline{N_{i}^{c}} N_{j} + h.c.\right]+V(h,\phi)~.
\label{type-1_Lag}
\end{eqnarray}
This Lagrangian leads us to the Type-I Seesaw mechanism.

Minimizing the scalar potential $V(H,\Phi)$ at the vacuums ($v$ and $v_\phi$), one can obtain the following relations,
\begin{eqnarray}
    \mu_{_H}^2 &=& \lambda_{_H}v^2 + \frac{\lambda_{H\Phi} v_\phi^2}{2} ,\nonumber \\
    \mu_{_\Phi}^2 &=& \lambda_{_\Phi}v_\phi^2 + \frac{\lambda_{H\Phi} v^2}{2} +\frac{3\mu_3 v_\phi}{2 \sqrt{2}} .
\end{eqnarray}

\noindent The two CP even (CPE) states $h$ and $\phi$ are mixed up after the EWSB, and the  mass matrix reads as,
\begin{equation}
\mathcal{M}_{\rm CPE}^2=\begin{pmatrix}
2 \lambda_{_H} v^{2} & \lambda_{_{H\Phi}} \,v\, v_{\phi}  \\
\lambda_{_{H\Phi}} \,v \,v_{\phi}  & ~~2 \,\lambda_{\Phi} v_{\phi}^{2} + \frac{3\sqrt{2}}{4} \,v_{\phi} \,\mu_{3}
\end{pmatrix} =\begin{pmatrix}
A & C  \\
C  & B
\end{pmatrix} .
\end{equation} 
The eigenvalues of the aforementioned mass matrix associated with the two physical states $h_1$ and $h_2$ are as follows:
\begin{eqnarray}
    M_{h_1,h_2} = \frac{1}{2} \left((A+B) \mp \sqrt{(A-B)^2+4 C^2}\right) .
\end{eqnarray}
The mass eigenstates $h_1$ and $h_2$ are related to the flavor states $h$ and $\phi$ through the following orthogonal transformation, parameterized by the mixing angle $\theta$ :
\begin{equation}
\begin{pmatrix}
h_1 \\ h_2
\end{pmatrix}=
\begin{pmatrix}
\cos\theta & -\sin\theta \\ \sin\theta & \cos\theta
\end{pmatrix}\,
\begin{pmatrix}
h \\ \phi
\end{pmatrix}\, ~~~{\rm with}~~~ \tan2\theta= \frac{2 C}{A-B}.
\end{equation}
Here $h_1$ is identified as the SM-like Higgs with mass $M_{h_1} \simeq 125$ GeV and $h_2$ is the beyond the SM (BSM) scalar with a mass denoted as $M_{h_2}$. 
Following the above relations, we can express the quartic and cubic couplings in terms of various measurable physical quantities: heavy scalar masses ($M_{h_1}, M_{h_2}, M_\chi$), vevs ($v,v_\phi$), and the scalar mixing angle ($\sin\theta$). The relations are as follows:
\begin{eqnarray}
\lambda_{H} &=& \frac{1}{2 v^{2}} \left( M_{h_{1}}^{2} \cos^{2} \theta + M_{h_{2}}^{2} \sin^{2}\theta \right), \nonumber\\
\lambda_{\Phi} & = & \frac{1}{2 v_{\phi}^{2}} \left( M_{h_{2}}^{2} \cos^{2}\theta + M_{h_{1}}^{2} \sin^{2}\theta + \frac{1}{3} M_{\chi}^{2} \right) , \nonumber\\
\lambda_{H\Phi} &=& \frac{1}{v\,v_{\phi}} \left( M_{h_{2}}^{2}-M_{h_{1}}^{2} \right) \cos\theta \sin\theta , \nonumber\\
\mu_{3} & = & -\frac{2\sqrt{2}}{9} \frac{M_{\chi}^{2}}{v_{\phi}} \,.
\label{eq:lambda_para}
\end{eqnarray}

The phenomena of Leptogenesis, the Electroweak phase transition, and DM production via the freeze-out mechanism depend on the thermal history of the Universe, which we will explore in our discussion.
The phenomenon of Leptogenesis occurs at high temperatures ($T_{\Phi} > T > T_{\rm EW}$)\footnote{The generated lepton asymmetry transfers to the baryon asymmetry through EW sphaleron processes. The baryon asymmetry is conserved after the EW sphaleron processes decouple from the thermal bath at a temperature ($T_\text{sph}\sim 10^2$) GeV}. {Contrarily, the DM maintains thermal equilibrium even after the EWSB phase, {\it i.e.,} at a temperature around $T_{\rm{FO}}\sim \frac{M_{\chi}}{20}~ {\rm to} ~\frac{M_{\chi}}{30} < T_{\rm EW}$ \cite{Kolb:1990vq}}. The phenomenology of the model depends on the following independent parameters:
\begin{eqnarray}
    \{  M_{\phi},~v_{\phi},~M_{N_{1,2}},~\alpha_{ij}\} &&~~~~~{\rm for~ Leptogenesis}\nonumber \\
  {\rm and} ~~ \{ M_{h_2} ,~M_{\chi},~v_{\phi},~\sin\theta \} && ~~~~~{\rm for~ Dark ~Matter }.
\end{eqnarray}
In our discussion of DM, we consider that the right-handed neutrino masses ($M_{N_{1,2}}$) are much heavier than the DM mass $M_\chi$. Therefore, $M_{N_{1,2}}$ will not appear in the DM analysis.
The mass parameter $M_{\phi}$, which represents the mass of the $\phi$ state prior to 
the EWSB, is related to the mass parameter $M_{h_2}$, corresponding to 
the mass of the physical state $h_2$ in the following manner:
\begin{eqnarray}
    M_{\phi}^2=M_{h_{2}}^{2} \cos^{2}\theta + M_{h_{1}}^{2} \sin^{2}\theta ~ \xrightarrow{\sin\theta \to 0} ~ \simeq M_{h_2}^2~.
    \label{mphi_mh2_expr}
\end{eqnarray}
\subsection{Theoretical and Experimental constraints }\label{sec:const}
\noindent $\bullet$\textbf{ Stability of potential:}\\
The quartic terms of the scalar potential $V(H,\Phi)$ play an important role in ensuring the stability of the potential, followed by
the following co-positivity conditions \cite{Chakrabortty:2013mha}:
\begin{align}
 \lambda_{H} \geq 0,~~~~\lambda_{\Phi} \geq 0 ~~~{\rm and}~~~\lambda_{H \Phi} + 2\sqrt{\lambda_H \lambda_\Phi} \geq 0 .
\end{align}
\noindent $\bullet$\textbf{Perturbative Unitarity:} \\
The quartic couplings of the scalar potential can also be constrained from tree-level unitarity of the theory, considering all possible $2\to 2$ scattering amplitudes that contribute to the S matrix \cite{Hektor:2019ote,Bhattacharya:2017fid}. The eigenvalues of the $S$ matrix are bounded from above as: 
\begin{align}
   |\lambda_H| < 4 \pi,~~|\lambda_\Phi| < 4 \pi,~~|\lambda_{H\Phi}| < 8 \pi \nonumber \\
   |(3 \lambda_H + 2 \lambda_\Phi) \pm \sqrt{2 \lambda_{H \Phi}^2 + (3 \lambda_H - 2 \lambda_\Phi)^2 }| < 8 \pi~~.
\end{align}

\noindent $\bullet$\textbf{ Collider constraints:} \\
The presence of the BSM scalar can modify the tree-level interactions of the SM Higgs with other SM particles through the mixing ($\sin\theta$). Combining measurements of different final states ($\gamma \gamma, \gamma Z, WW, ZZ, bb, \mu \mu, \tau \tau$) by ATLAS \cite{ATLAS:2022vkf} and CMS \cite{CMS:2022dwd}, the Higgs signal strengths set an upper limit on the mixing angle at $95\%$ CL: $|\sin\theta| \lesssim 0.29$ \cite{Lane:2024vur}.
The $W$ mass correction at NLO imposes the most stringent constraint on the mass range of $M_{h_2}\sim\{250-1000\}$ GeV, with the mixing angle $\sin\theta \sim \{0.2-0.3\}$ \cite{Lopez-Val:2014jva}. On the other hand, the electroweak precision observables impose weaker constraints on $\sin\theta$ compared to those obtained from $W$-boson mass corrections \cite{Lopez-Val:2014jva}. 

If the DM mass is below $M_{h_1}/2$, the SM-like Higgs can decay to DM pairs ($h_1 \to \chi \chi$), contributing to the Higgs invisible decay width. The ATLAS collaboration has placed a strong constraint on the Higgs invisible branching ratio, Br($h_{\rm SM} \to {\rm inv}$), setting it below $13\%$ \cite{ATLAS:2015ciy}. The Higgs invisible branching ratio can be expressed as (considering $M_{h_2} > M_{h_1}$):
\begin{equation}
    {\rm Br}(h_1 \rightarrow {\rm inv})=  \frac{ \sin^2\theta ~\Gamma(\phi \rightarrow \chi \chi)}{\sin^2\theta ~\Gamma(\phi \rightarrow \chi \chi)+~\cos^2\theta~\Gamma (h \rightarrow {\rm SM}~{\rm SM})}
\end{equation}
with $\Gamma(h \rightarrow {\rm SM}~{\rm SM}) \simeq 4.1 $ MeV.

\section{Leptogenesis in presence of a singlet scalar} \label{sec:-Leptogenesis}
The additional singlet scalar $\Phi$ opens up a large CP-violation compared to standard thermal Leptogenesis in the Type-I seesaw model. Once $\Phi$ acquires a vacuum expectation value, the relevant couplings that appear for Leptogenesis are:
\begin{equation}
 \frac{y_{ij}'}{\Lambda} \Phi^\dagger \Phi  \overline{N_{i}^{'c}} N_{j}' + \lambda_{H \Phi} \Phi^\dagger \Phi H^\dagger H  \rightarrow 
    ~\alpha_{ij}\, \phi \,\overline{N_{i}^{c}} N_{j}+\xi \,\phi \,(H^{\dagger} H)\,,
    \label{Lagr_Lep}
\end{equation}
where we define $\xi=\lambda_{H\Phi}\, v_{\phi}$.

\begin{figure}[h!]
	\includegraphics[scale=0.4]{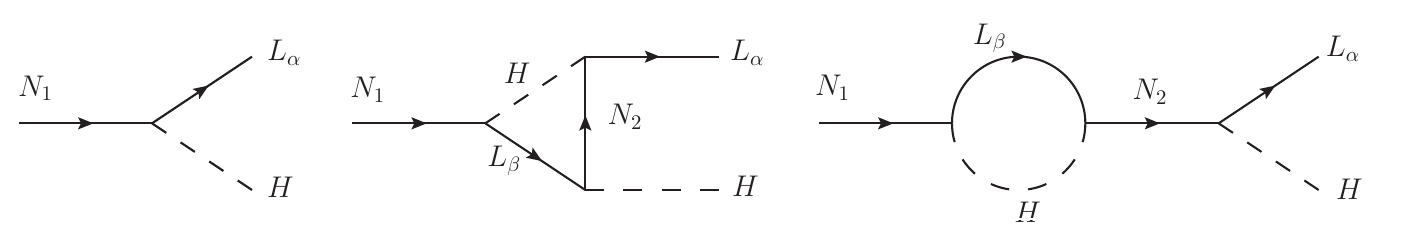}
	\includegraphics[scale=0.4]{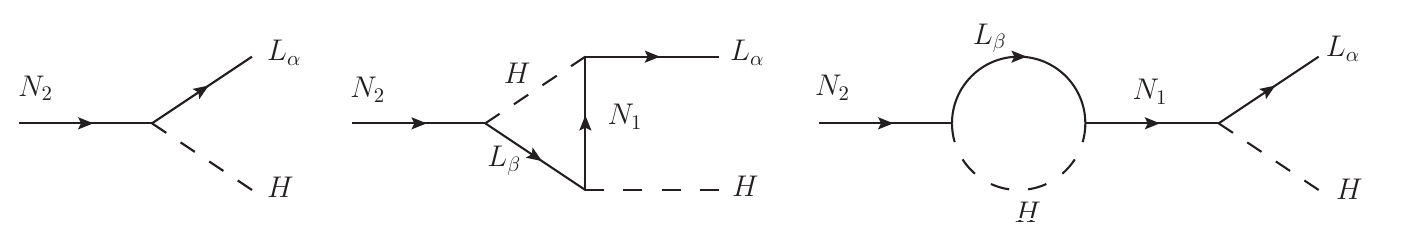}
 	\includegraphics[scale=0.4]{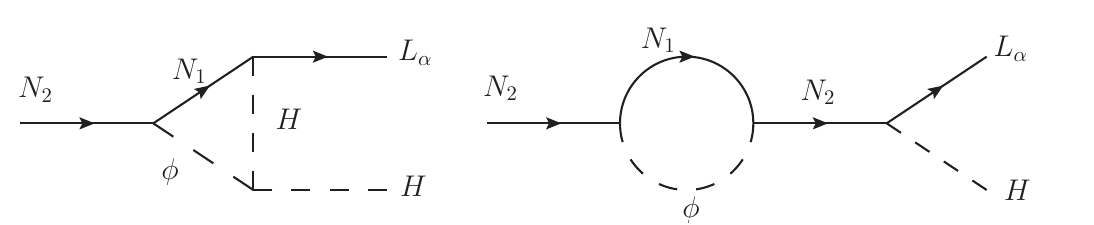}
	\caption{\it Contribution of the tree and loop level (vertex and self-energy) diagrams to the total CP asymmetry.}
	\label{CP_asy_Feydiag}
\end{figure}

The interference of tree and loop level diagrams gives a non-zero contribution to the CP asymmetry. In Fig.\,\ref{CP_asy_Feydiag}, we show the tree and loop-level (vertex and self-energy) Feynman diagrams that appear for the standard thermal Leptogenesis and the additional diagrams for the $N_2$-Leptogenesis. The additional loop diagrams that appear due to additional interaction terms, given in the third row of Fig.\,\ref{CP_asy_Feydiag}, play a crucial role in enhancing the CP violation in the $N_2$ Leptogenesis scenario\footnote{ $N_1 \rightarrow N_{2}+ \phi$ kinematically forbidden since we consider $M_{N_2} > M_{N_1} + M_\phi$\,\cite{LeDall:2014too}.}.

As we mentioned in Sec.\ref{sec:-model}, the Lagrangian given in Eq.\ref{type-1_Lag} sets the stage for the Type-I seesaw mechanism. After integrating out the heavy degrees of freedom, we get the light neutrino mass matrix $m_\nu$ as follows:
\begin{equation}
    m_\nu=-m_{D} D_N^{-1} m_D^T,
\end{equation}
 where $m_D=\lambda \,v/\sqrt{2}$ is the Dirac mass matrix. The light neutrino matrix can be diagonalized by a unitary transformation $U$, where $U$ coincides with the PMNS (Pontecorvo-Maki-Nakagawa-Sakata) lepton mixing matrix.  
\begin{eqnarray}
    D_\nu=U^{T} m_\nu U = \text{diag}(m_1,m_2,m_3).
\end{eqnarray}
We work in the flavor basis where the charged-lepton Dirac mass matrix is diagonal, and use the Casas-Ibarra parametrization\cite{Casas:2001sr} to re-express the neutrino Yukawa coupling matrix $\lambda$ in terms of low energy parameters as given below:
\begin{eqnarray}
    \lambda = \frac{i}{v} U^* D_{\nu}^{1/2} R D_{N}^{1/2} ,
    \label{CI}
\end{eqnarray}
where $R$ is a complex $3\times 2$ orthogonal matrix ($R^T R=\mathbb{1}$) which can be parametrized in terms of one complex angle, $z'$. The $U$ matrix contains three mixing angles ($\theta_{12}, \theta_{23} \,\, \text{and} \,\,\theta_{13}$), Dirac phase ($\delta$) and Majorana phase ($\gamma_1$). The $D_\nu$ depends on two mass-squared differences $\Delta m_{\text{sol}}^{2}$ and $\Delta m_{\text{atm}}^{2}$ in light neutrino mass spectrum\cite{deSalas:2020pgw, Esteban:2020cvm}.

In the standard thermal Leptogenesis, without additional loop contributions, the CP-asymmetry can be expressed as
\begin{eqnarray}
    \epsilon_{i=1,2}^{0}=\frac{1}{8 \pi (\lambda^{\dagger}\lambda)_{ii}} \sum_{j\neq i} \text{Im}\left[(\lambda^{\dagger}\lambda)_{ji}^{2}\right] \mathcal{F}\left(\frac{M_{N_j}}{M_{N_i}} \right)\,,
\end{eqnarray} 
where 
\begin{equation}
  \mathcal{F}(x) = x \left[ 1+(1+x^2) \text{ln}\left( \frac{x^2}{x^2+1}\right) -\frac{1}{x^2-1}\right].
\end{equation}

The CP-asymmetry $\epsilon_{i}^{0}$ consists of contributions from both the vertex and self-energy diagrams. In the present scenario, the standard CP asymmetry is further modified by additional contributions from the two newly introduced diagrams.
\begin{equation}
    \epsilon_{1}=\epsilon_{1}^{0}, ~~~~\epsilon_{2}=\epsilon_{2}^{0}+\epsilon_{2}^{v}+\epsilon_{2}^{s}
\end{equation}
where
\begin{eqnarray}
    \epsilon_{2}^v &=& \frac{1}{8\pi (\lambda^{\dagger}\lambda)_{22} M_{N_2}} \{ \text{Im}\left[(\lambda^{\dagger}\lambda)_{12}\, \xi \, \alpha_{21} \right] \mathcal{F}_{21,R}^v + \text{Im}\left[(\lambda^{\dagger}\lambda)_{12} \,\xi \,\alpha_{21}^* \right] \mathcal{F}_{21,L}^v \}, \nonumber\\
    \epsilon_{2}^s &= & \frac{1}{8\pi (\lambda^{\dagger}\lambda)_{22}} \{ \text{Im}\left[(\lambda^{\dagger}\lambda)_{12} \alpha_{21} \alpha_{11} \right] \mathcal{F}_{211,RR}^s + \text{Im}\left[(\lambda^{\dagger}\lambda)_{12} \alpha_{21}^* \alpha_{11} \right] \mathcal{F}_{211,RL}^s \nonumber\\
     && + \text{Im}\left[(\lambda^{\dagger}\lambda)_{12} \alpha_{21} \alpha_{11}^* \right] \mathcal{F}_{211,LR}^s + \text{Im}\left[(\lambda^{\dagger}\lambda)_{12} \alpha_{21}^* \alpha_{11}^* \right] \mathcal{F}_{211,LL}^s \}. 
     \label{CP_asy_exp}
\end{eqnarray}
The explicit expressions for the loop function $\mathcal{F}$ can be found in Appendix \ref{App_loop}, and greater details can be found in \cite{Alanne:2018brf, LeDall:2014too}.

To study the evolution of the number densities of the right-handed neutrinos
$N_i\, (i=1,2)$ and the amount of $B-L$ asymmetry $N_{B-L}$, we consider a set of coupled Boltzmann equations while taking care of their decay and inverse decay rates and scattering processes. In the present scenario, the decay of $N_2$ to $N_1$ and $\phi$, ({\it i.e.} $N_2\rightarrow N_1 \phi$) and the washout processes {\it i.e.,} $\Delta L=2$ scatterings $N_i N_j \rightarrow HH$ play a key role in addressing the low-scale Leptogenesis. The relevant Boltzmann equations for the number densities $N_{N_{1,2}}$ and $N_{B-L}$ can be expressed as \cite{Alanne:2018brf}
\begin{eqnarray}
\frac{dN_{N_{2}}}{dz} & = & -\left(D_{2}+D_{21} \right)\left( \frac{N_{N_{2}}(z)}{N_{N_{2}}^{\text{eq}}(z)}-1 \right)  + D_{21} \left( \frac{N_{N_{1}}(z)}{N_{N_{1}}^{\text{eq}}(z)}-1 \right)  \nonumber\\
&& - S_{N_{1}{N_{2}}\rightarrow H H} \left( \frac{N_{N_{1}} N_{N_{2}}}{N_{N_{1}}^{\text{eq}} N_{N_{2}}^{\text{eq}}}-1 \right)  - S_{N_{2}{N_{2}}\rightarrow H H} \left(\frac{N_{N_{2}} N_{N_{2}}}{N_{N_{2}}^{\text{eq}} N_{N_{2}}^{\text{eq}}}-1 \right) \nonumber\\
\frac{dN_{N_{1}}}{dz} & = & -(D_{1}+D_{21}) \left( \frac{N_{N_{1}}(z)}{N_{N_{1}}^{\text{eq}}(z)}-1 \right) + D_{21} \left( \frac{N_{N_{2}}(z)}{N_{N_{2}}^{\text{eq}}(z)}-1 \right) \nonumber\\
&& - S_{N_{1}{N_{2}}\rightarrow H H} \left( \frac{N_{N_{1}} N_{N_{2}}}{N_{N_{1}}^{\text{eq}} N_{N_{2}}^{\text{eq}}}-1 \right)  -S_{N_{1}{N_{1}}\rightarrow H H} \left(\frac{N_{N_{1}} N_{N_{1}}}{N_{N_{1}}^{\text{eq}} N_{N_{1}}^{\text{eq}}}-1 \right)  \nonumber\\
\frac{dN_{B-L}}{dz} & = & \epsilon_{1} D_{1} \left( \frac{N_{N_{1}}(z)}{N_{N_{1}}^{\text{eq}}(z)}-1 \right) +\epsilon_{2} D_{2} \left( \frac{N_{N_{2}}(z)}{N_{N_{2}}^{\text{eq}}(z)}-1 \right) - \left( W_{1} + W_{2} \right) N_{B-L},
\label{BEq}
\end{eqnarray}
where the $z_1 \equiv M_{N_1}/T$ (with $z \equiv z_1$) and $z_2 \equiv M_{N_2}/T= (M_{N_2}/M_{N_1}) z$ are the dimensionless parameters. The $N_{N_i}^{eq}$ are the equilibrium number densities,
\begin{eqnarray}
N_{N_i}^{\text{eq}}(z) & = & \frac{z_{i}^{2}}{2} \mathcal{K}_{2}(z_{i})\,.
\end{eqnarray}
The $D_{1,2}$, $D_{21}$ and $W$ are (function of $z$) the decay rate of right-handed neutrinos $N_{1,2} \rightarrow L H$, $N_2 \rightarrow N_1 \phi$ and washout from the inverse decays $L H \rightarrow N_{1,2} $, respectively,
\begin{eqnarray}
D_{i}(z) & = & K_{i} z \frac{\mathcal{K}_{1}(z_{i})}{\mathcal{K}_{2}(z_{i})} N_{N_{i}}^{eq}(z), \\
\label{D}
D_{21}(z) & = & K_{21} z \frac{\mathcal{K}_{1}(z_{2})}{\mathcal{K}_{2}(z_{2})} N_{N_{2}}^{eq}(z), \\
\label{D21}
W(z) & = & \sum_{i} \frac{1}{4} K_{i} z_{i}^{3} \mathcal{K}_{1}(z_{i}),
\label{Di_D21_W_inv}
\end{eqnarray}
where $\mathcal{K}_{1,2}(z)$ are the modified Bessel functions of the second kind. 
The decay parameters are
\begin{equation}
    K_i \equiv \frac{\Gamma (N_i \rightarrow L H)}{\mathcal{H}(T=M_{N_i})}~~~,~~~K_{21} \equiv \frac{\Gamma (N_2 \rightarrow N_1 \phi)}{\mathcal{H}(T=M_{N_2})} \,,
\end{equation}
where $\mathcal{H}=\sqrt{8 \pi^3 g_*/90}\,\, T^2/M_{\text{pl}}$ is the Hubble rate.
The decay widths are
\begin{equation}
    \Gamma_i=\Gamma (N_i \rightarrow L H)+\Gamma (N_i \rightarrow \bar{L} \bar{H})=\frac{(\lambda^{\dagger} \lambda)_{ii}}{8\pi} M_{N_i} \,,
\end{equation}
\begin{equation}
    \Gamma (N_2 \rightarrow N_1 \phi )=\frac{|\alpha_{12}|^2 M_{N_2}}{16 \pi} \left[\left(1+\frac{M_{N_1}}{M_{N_2}} \right)^2-\frac{M_\phi^2}{M_{N_2}^2} \right] \sqrt{\left(1-\frac{M_{N_1}^2}{M_{N_2}^2}-\frac{M_\phi^2}{M_{N_2}^2}\right)^2 -4 \frac{M_{N_1}^2}{M_{N_2}^2} \frac{M_\phi^2}{M_{N_2}^2}}\,.
    \label{gamma21}
\end{equation}

The scattering cross-section function for $N_i N_j \rightarrow H H$ can be expressed as,
\begin{eqnarray}
S_{N_{i} N_{j}\rightarrow H H} \equiv \frac{M_{N_i}}{64 \pi^{2} \mathcal{H}(T=M_{N_i})} \int_{w_{\text{min}}}^{\infty} \sqrt{w} K_{1}(\sqrt{w}) \hat{\sigma}_{N_{i} N_{j}\rightarrow H H}\left(\frac{w M_{N_i}^{2}}{z_{i}^{2}} \right) \,,
\label{Eq_scatterig}
\end{eqnarray}
where $w_{\text{min}}=(M_{N_i}+M_{N_j})^{2}$, and
\begin{equation}
    \hat{\sigma}=\frac{1}{s} \delta(s,M_{N_i},M_{N_j}) \sigma_{N_{i} N_{j}\rightarrow H H} \,,
\end{equation}
where $\sigma$ is cross-section for the scatterings $N_i N_j \rightarrow H H$,
\begin{eqnarray}
\sigma(N_{i}N_{j}\rightarrow H H)=\frac{|\alpha_{ij}|^{2} \xi^{2}}{32\pi} \frac{s-\left( M_{N_i}+M_{N_j} \right)^{2}}{\sqrt{\delta(s,M_{N_i},M_{N_j})} (s-M_{S}^{2})^{2}} \,,
\label{scattering_NNtoHH}
\end{eqnarray}

with $\delta(s,M_{N_i},M_{N_j})=(s-M_{N_i}^{2}-M_{N_j}^{2})^{2} - 4 M_{N_i}^{2} M_{N_j}^{2}$.

\begin{figure}[ht]
  \subfigure[]{
  \includegraphics[scale=0.5]{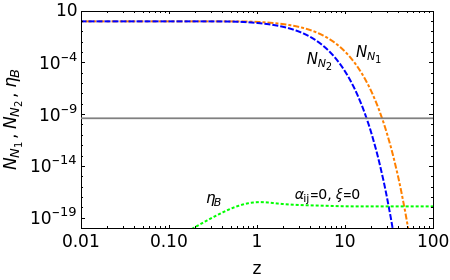}}
  \label{abund_z_a}
  \subfigure[]{
  \includegraphics[scale=0.5]{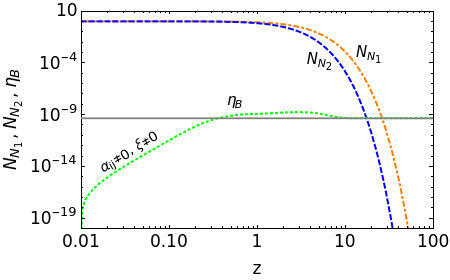}}
  \label{abund_z_b}
    \caption {\it
    The variation of the abundances of $N_1$, $N_2$, and the baryon asymmetry $\eta_B$ is shown as a function of the variable $z = M_{N_1}/T$. The dotted green line denotes the baryon asymmetry $\eta_B$, while the black horizontal line represents the observed value of $\eta_B$. For illustrative purposes, the parameters are fixed as follows: $M_{N_2} = 7.5 \times 10^3$ GeV, $M_{N_1} = 5 \times 10^3$ GeV, and $M_\phi = 500$ GeV. The left panel $(a)$ corresponds to the case without the additional loop diagrams involving the scalar $\Phi$ (cf. Fig.~\ref{CP_asy_Feydiag}), i.e., with $\alpha_{ij} = 0$ and $\xi = 0$. In contrast, the right panel $(b)$ depicts the scenario in which the complex scalar $\Phi$ significantly contributes to the generation of the observed baryon asymmetry, with parameters $\alpha_{ij} = 10^{-3}$ and $\xi = 6 \times 10^3$ GeV.}
    \label{abund_z}
\end{figure}

In our scenario, the contribution to the total $B-L$ asymmetry comes from the decays of both heavy right-handed neutrinos $N_2$ and $N_1$. As the Universe expands and cools down to a temperature $T \sim M_{N_2}$, {\it i.e., $z_2 \sim 1$}, the out-of-equilibrium decay $N_2 \to L H$ generates the primary lepton asymmetry. 
In the presence of the new interactions, $N_2$ can additionally decay to $N_1$ and $\phi$ through the coupling $\alpha_{12}$. It can increase the abundance of $N_1$. Subsequently, $N_1$ also decays in its out-of-equilibrium around the temperature $T \sim M_{N_1}$, {\it i.e., $z_1 \sim 1$}, analogous to the $N_2$ decay. Due to the small CP asymmetry, this contribution is very small to the total lepton asymmetry for $M_{N_1}\lesssim 10^{6}$ GeV. When $N_1$ is in equilibrium, its inverse decays can reduce the asymmetry generated earlier by the decay of $N_2$. To understand the dynamics, we numerically solve the Boltzmann equations given in Eq.\ref{BEq}. The estimated abundances of $N_{1}, N_{2}$ and $N_{B-L} (\eta_B)$ as a function of inverse temperature are depicted in Fig.\ref{abund_z}.

The generated $N_{B-L}$ asymmetry is converted to the baryon asymmetry $(\eta_B)$ via sphaleron processes. The predicted $N_{B-L}$ is related to the measured $\eta_B$ at the time of recombination in the following way:
\begin{equation}
    \eta_B= \left(\frac{a_{sph}}{f} \right)  N_{B-L} \,,
    \label{etaB}
\end{equation}
where $a_{sph}=28/79$ is the fraction of $B-L$ asymmetry converted into the baryon asymmetry by sphaleron processes, and $f=N_\gamma^{\text{rec}}/N_\gamma^{\text{*}}=2387/86$ is the dilution factor calculated assuming standard photon production from the onset of Leptogenesis till recombination \cite{Buchmuller:2004nz}.

The Green dotted line in Fig.\ref{abund_z} shows the abundance of the baryon asymmetry, $\eta_B$. Here we fix the parameters: $M_{N_2}=7.5\times 10^{3} $ GeV, $M_{N_1}=5\times 10^{3}$ GeV, $M_\phi=500$ GeV, $\alpha_{ij}=10^{-3}$, and $\xi=6 \times 10^3$ GeV. 
We illustrate our results in terms of $\xi (=\lambda_{H\Phi} v_\phi)$ only, where $\lambda_{H\Phi}$ plays a key role in DM and FOPT. Whereas $\alpha_{ij}$, apart from its magnitude, won't play much phenomenology. As $\alpha_{ij}$ is a complex matrix, it can be an additional source of CP violation; however, we ignore its effect in our analysis for simplicity. The CP-violation appears due to $\lambda_{ij}$ (see Eq.\,\ref{CP_asy_exp}). It can be parametrized through the Casas-Ibarra parametrization given in Eq.\,\ref{CI}. 
Since the CP-asymmetry ($\epsilon_2$) is proportional to $\alpha_{ij}$, it can maximize the CP-asymmetry, but it can also impact the washouts from the scattering processes, given in Eq.\,\ref{scattering_NNtoHH}. For large values of $\alpha_{ij}$, the CP-asymmetry ($\epsilon$) can be large, but the washout is also high. It suppresses the RHN abundance and, hence, the lepton asymmetry. The washout may be smaller for small values of $\alpha_{ij}$, but the CP-asymmetry is suppressed, hence the lepton asymmetry. So we fix it with an appreciable value of $\alpha_{ij}=10^{-3}$. Throughout our analysis, we fix the complex angle $z'=0.01+ i\, 0.8$. The horizontal Gray line represents the observed baryon asymmetry of the Universe. In the left panel of Fig.\ref{abund_z}, we display the baryon asymmetry abundance without the complex singlet scalar field $\Phi$. When $\Phi$ is included, new processes emerge, as depicted in Fig.\ref{CP_asy_Feydiag}, significantly enhancing $\eta_B$. This enhancement is described by parameters $\alpha_{ij}$ and $\xi$, and Fig.\ref{abund_z} illustrates how
non-zero values of those new parameters can affect $\eta_B$.  

\begin{figure}[ht]
  \subfigure[]{
  \includegraphics[scale=0.48]{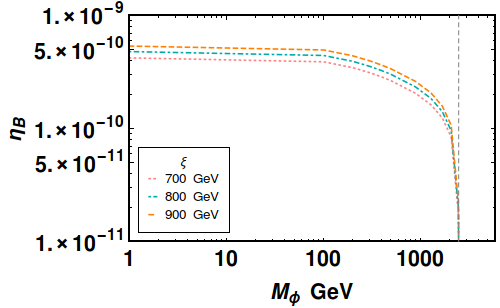}}
  \label{mphi_etaB}
  \subfigure[]{
  \includegraphics[scale=0.45]{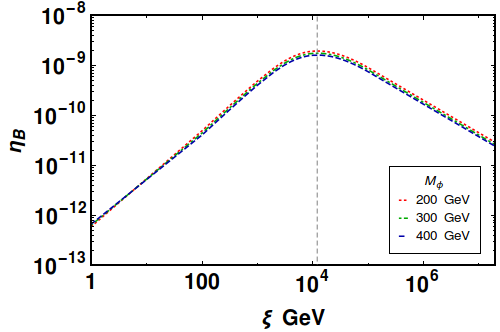}}
  \label{beta_etaB}
    \caption {\it Figure shows the variation of the baryon asymmetry, $\eta_B$ with respect to $M_{\phi}$ $(a)$ and $\xi$ $(b)$ for three representative values of $\xi$ and $M_\phi$ respectively.  The vertical dashed line in the left plot represents the kinematically allowed maximum value of $M_{\phi}$, while in the right plot, it represents the $\xi_{\rm Thres}$. The values of
    $M_{N_1}, M_{N_1}$ and $\alpha_{ij}$ are set identical values to those in Fig.\ref{abund_z}.}
\label{mphibeta_etaB}
\end{figure}

In Fig.\ref{mphibeta_etaB}, we show the variation of $\eta_B $ with 
$M_\phi $ and $\xi$. In Fig.\ref{mphibeta_etaB}(a), we fix the $\xi=700 \,\text{GeV}\,\,{\rm (pink ~dotted)}, 800\,\text{GeV} \,\,{\rm (cyan ~dashed)},$ and $900\,\text{GeV} \,\,{\rm (orange ~dotdashed)}$ and vary $M_\phi$. 
The baryon asymmetry remains almost constant until $M_\phi$ nears the mass difference between the heavy neutrinos, $M_{N_2}-M_{N_1} \sim M_{\phi}$.
When $M_\phi$ becomes comparable to this mass difference, the decay width $\Gamma_{21}$ decreases significantly, see Eq.\ref{gamma21},
leading to a reduction in the baryon asymmetry. In this plane, we can also observe that as we increase $\xi$, the $\eta_B $ increases. In Fig.\,\ref{mphibeta_etaB} (b), we illustrate the variation of $\eta_B$ with $\xi$. Here we fix $M_{\phi}=200\,\text{GeV} \,\,{\rm (Red, dotted)}, 300\,\text{GeV} \,\,{\rm (Green, dotdashed)},$ and $400\,\text{GeV} \,\,{\rm (Blue, dashed)}$. Since $M_{\phi}$ is far from the mass difference of heavy neutrinos, we see a mild variation in $\eta_B$ with different choices of $M_\phi$. Fixing $M_\phi$, if we increase $\xi$, the $\eta_B$ rises until it reaches a threshold 
value $\xi_{\rm Thres}$. Beyond this threshold, 
scatterings of the form $N_i N_j \to H H $, given in Eq.\ref {Eq_scatterig}, begin to dominate over the decay rates $D_1, D_2$ and $D_{21}$. This dominance leads to an increase in the washout
process that can suppress the abundance of $N_{1,2}$, causing $\eta_B$ to decrease. To minimize the 
washout effects, we restrict our analysis to the region where $\xi < \xi_{\rm Thres}$ in the rest of the paper.
This ensures that our numerical estimations align well with the analytical predictions, 
as discussed in \cite{Alanne:2018brf}.

\begin{figure}[ht]
  \subfigure[]{
  \includegraphics[scale=0.3]{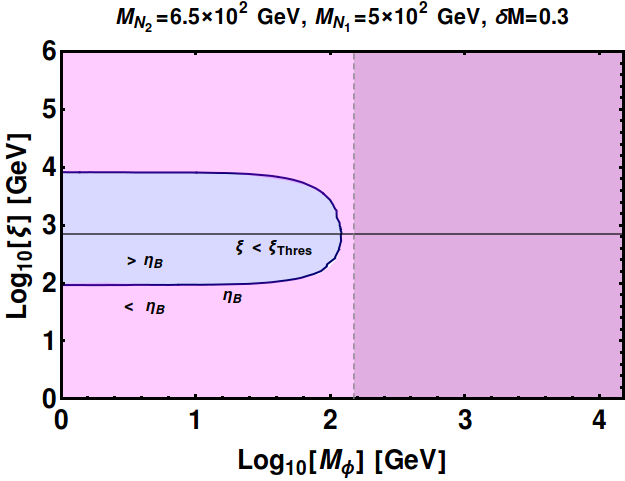}}
  \label{mphi_beta_a}
  \subfigure[]{
  \includegraphics[scale=0.3]{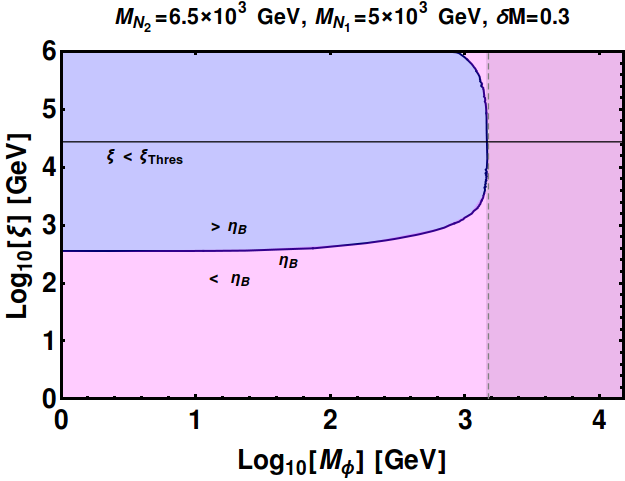}}
  \label{mphi_beta_b}\\
  \subfigure[]{
  \includegraphics[scale=0.37]{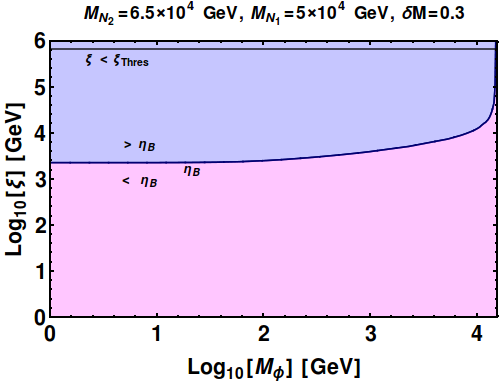}}
  \label{mphi_beta_c}
  \subfigure[]{
  \includegraphics[scale=0.37]{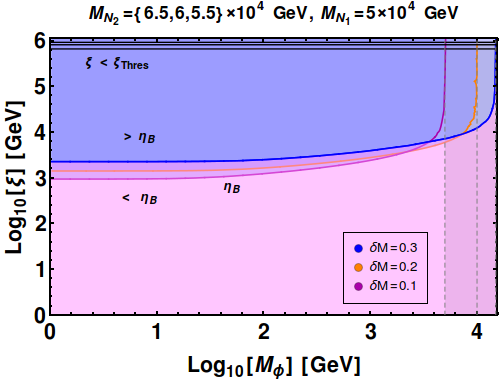}}
  \label{mphi_beta_d}
    \caption {\it The behavior of the observed baryon asymmetry of the Universe at different scales, $M_{N_2}=6.5 \times 10^{\{2,3,4\}}$ GeV with $\delta M=0.3$. 
    To avoid large suppression from scattering processes, the parameter $\alpha_{ij} $ is set to $10^{-3}$. The horizontal line corresponds to the threshold limit, where the washout from the scattering rate dominates over the decay rate. }
    \label{mphi_beta}
\end{figure}

In Fig.\ref{mphi_beta}, we show the allowed parameter space of the observed baryon asymmetry of the Universe in the plane of $M_{\phi}$ and $\xi$. The Blue line successfully explains the observed baryon asymmetry of the Universe. The Pink and Blue regions show the under and over-abundant baryon asymmetry, respectively. We choose the parameters for Fig.\,\ref{mphi_beta} (a): $M_{N_2}=6.5 \times 10^{2}$ GeV and $M_{N_1}=5 \times 10^{2}$ GeV, Fig.\,\ref{mphi_beta} (b): $M_{N_2}=6.5 \times 10^{3}$ GeV and $M_{N_1}=5 \times 10^{3}$ GeV, Fig.\,\ref{mphi_beta} (c): $M_{N_2}=6.5 \times 10^{4}$ GeV and $M_{N_1}=5 \times 10^{4}$ GeV by keeping $\delta M=(M_{N_2}-M_{N_1})/M_{N_1}=0.3$. Here, we show the variation baryon asymmetry parameter space in the $M_{\phi} - \xi$ plane for different mass scales of the right-handed neutrinos. Until $M_{\phi}$ value approaches the kinematic limit, the baryon asymmetry remains almost constant with 
the variation of $\xi$ to satisfy the observed BAU. Once the $M_\phi$ reaches near the kinetic limit, the baryon asymmetry starts to decrease; hence, the $\xi$ begins to increase so that the observed BAU can be obtained. The process continues until $\xi$ approaches the threshold $\xi_{Thres}$, represented by the horizontal line. Near the threshold, the washout processes due to scatterings become dominant, leading to a reduction in the baryon asymmetry. To compensate and achieve the observed BAU, the $M_{\phi}$ has to be reduced by keeping $\xi$ constant. That results in two allowed values for the $\xi$ with a single value of $M_\phi $, see Fig.\ref{mphi_beta} (a). We also notice that as we increase the mass scale of the right-handed neutrinos, the allowed $\xi$ shifts towards larger values, see Fig.\ref{mphi_beta} (b) and (c). In Fig.\ref{mphi_beta} (d), we consider $\delta M=0.3,0.2,0.1$ corresponding to $M_{N_2}=(6.5,6,5.5) \times 10^4$ GeV and $M_{N_1}=5\times 10^4$ GeV respectively. We see that for a particular mass scale of right-handed neutrinos, if we reduce the $\delta M$, the BAU allowed $\xi$ decreases.

In the following section, we explore the phenomenology of dark matter (DM) within this framework. We analyze the role of the parameters $\alpha_{ij}, \xi $, and $ M_\phi(M_{h_2})$ in determining the dynamics of DM in this scenario. 
Furthermore, in the results and analysis section, we highlight the correlations between Leptogenesis and DM phenomenology.

\section{Pseudo Scalar Dark Matter Phenomenology}\label{sec:-DM}
This section focuses on the phenomenology of DM $(\chi )$, which is the CP-odd state of the complex singlet scalar $\Phi$. The residual $\Phi \rightarrow \Phi^{\dagger}$ symmetry ($\chi \rightarrow -\chi$) ensures the stability of DM $\chi$. The DM communicates with the visible sector (SM) through the portal interaction $\lambda_{H\Phi}(H^\dagger H)(\Phi^\dagger \Phi)$. In the early Universe, $\chi$ maintained thermal equilibrium with the bath particles through the Higgs portal interaction. This equilibrium condition was determined by the inequality between the interaction rate $\Gamma_{\rm DM -SM} \equiv n_\chi^{\rm eq.} {\langle {\sigma v} \rangle}_{\chi \chi \to XY}$ ($X, Y$ represent the thermal bath particles)  and the Hubble expansion rate $\mathcal{H}$ as  $\Gamma_{\rm DM -SM} > \mathcal{H}$. As the Universe expanded, the rate of interaction diminished with decreasing temperature. When the temperature reached a point where $\Gamma_{\rm DM -SM}$ went below $\mathcal{H}$, DM froze out of the thermal bath, resulting in today's observed DM density. {This type of DM is commonly known as WIMP-like (Weakly Interacting Massive Particle) candidate} \cite{Kolb:1990vq}. Note that $\chi$ maintains thermal equilibrium even after the EWSB at temperatures around $T_{\rm FO} \sim M_\chi /{20} < T_{\rm EW}$.
Before DM freezeout ($T> T_{\rm FO}$), the number density of DM follows the equilibrium density, denoted as $n_\chi^{\rm eq}$. After EWSB, it turns out that the SM Higgs $(h)$ and the CP-even 
component of the BSM singlet $(\phi)$ mix to form two physical states $h_1$ (SM-like) and $h_2$. Therefore both $h_1$ and $h_2$ mediated scattering processes between the DM $(\chi)$ and the bath particles ($X, Y=\{{\rm SM}, h_2\}$) are responsible for the number density of $\chi$. The Feynman diagrams of the number-changing processes of DM are shown in Fig.\ref{Feyn_diag1} and Fig.\ref{Feyn_diag2}. 
\begin{figure}[htb!]
  \begin{center}
    \begin{tikzpicture}[line width=0.6 pt, scale=1.1]
        \draw[dashed] (-6,1)--(-5,0);
	\draw[dashed] (-6,-1)--(-5,0);
	\draw[dashed] (-5,0)--(-4,1);
	\draw[dashed] (-5,0)--(-4,-1);
	\node at (-6.2,1.1) {$\chi$};
	\node at (-6.2,-1.1) {$\chi$};
	\node at (-3.8,1.1) {$h_1$};
	\node at (-3.8,-1.1) {$h_1$};
	\draw[dashed] (-1.8,1.0)--(-0.8,0.5);
	\draw[dashed] (-1.8,-1.0)--(-0.8,-0.5);
	\draw[dashed] (-0.8,0.5)--(-0.8,-0.5);
	\draw[dashed] (-0.8,0.5)--(0.2,1.0);
	\draw[dashed] (-0.8,-0.5)--(0.2,-1.0);
	\node at (-2.0,1.1) {$\chi$};
	\node at (-2.0,-1.1) {$\chi$};
	\node at (-1.0,0.07) {$\chi$};
	\node at (0.4,1.1) {$h_1$};
	\node at (0.4,-1.1) {$h_1$};
	%
        \draw[dashed] (2.4,1)--(3.4,0);
	\draw[dashed] (2.4,-1)--(3.4,0);
	\draw[dashed] (3.4,0)--(4.4,0);
	\draw[solid] (4.4,0)--(5.4,1);
	\draw[solid] (4.4,0)--(5.4,-1);
	\node  at (2.2,-1) {$\chi$};
	\node at (2.2,1) {$\chi$};
	\node [above] at (3.9,0.05) {$h_{1,2}$};
	\node at (5.8,1.0){A};
	\node at (5.8,-1.0) {B};
     \end{tikzpicture}
 \end{center}
\caption{\it{Feynmann diagrams for DM annihilation to SM: $\chi ~\chi \rightarrow $ A ~B with $\{A,B\}=\{W,Z,h_1(M_{h_1}=125 \,{\rm GeV}),f ({\rm SM \, fermions})\}$ }.}
\label{Feyn_diag1}
 \end{figure}
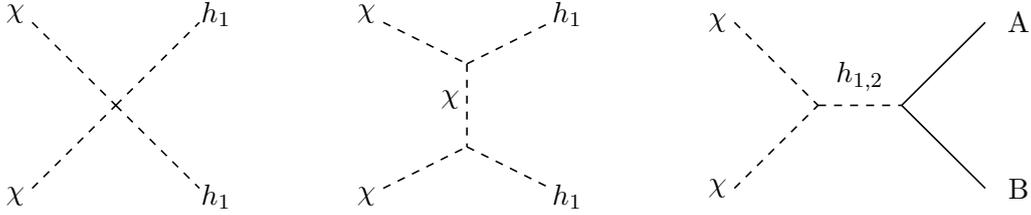
\begin{figure}[htb!]
  \begin{center}
    \begin{tikzpicture}[line width=0.6 pt, scale=1.1]
        \draw[dashed] (-6,1)--(-5,0);
	\draw[dashed] (-6,-1)--(-5,0);
	\draw[dashed] (-5,0)--(-4,1);
	\draw[dashed] (-5,0)--(-4,-1);
	\node at (-6.2,1.1) {$\chi$};
	\node at (-6.2,-1.1) {$\chi$};
	\node at (-3.8,1.1) {$h_2$};
	\node at (-3.7,-1.16) {$h_{1,2}$};
	\draw[dashed] (-1.8,1.0)--(-0.8,0.5);
	\draw[dashed] (-1.8,-1.0)--(-0.8,-0.5);
	\draw[dashed] (-0.8,0.5)--(-0.8,-0.5);
	\draw[dashed] (-0.8,0.5)--(0.2,1.0);
	\draw[dashed] (-0.8,-0.5)--(0.2,-1.0);
	\node at (-2.0,1.1) {$\chi$};
	\node at (-2.0,-1.1) {$\chi$};
	\node at (-1.0,0.07) {$\chi$};
	\node at (0.4,1.1) {$h_2$};
	\node at (0.48,-1.1) {$h_{1,2}$};
	%
        \draw[dashed] (2.4,1)--(3.4,0);
	\draw[dashed] (2.4,-1)--(3.4,0);
	\draw[dashed] (3.4,0)--(4.4,0);
	\draw[dashed] (4.4,0)--(5.4,1);
	\draw[dashed] (4.4,0)--(5.4,-1);
	\node  at (2.2,-1) {$\chi$};
	\node at (2.2,1) {$\chi$};
	\node [above] at (3.9,0.05) {$h_{1,2}$};
	\node at (5.8,1.0){$h_2$};
	\node at (5.8,-1.0) {$h_{1,2}$};
     \end{tikzpicture}
 \end{center}
\caption{\it{Feynmann diagrams for DM annihilation to $h_{1,2}$: $\chi ~\chi \rightarrow h_2~ h_{1,2} $ }.}
\label{Feyn_diag2}
 \end{figure}
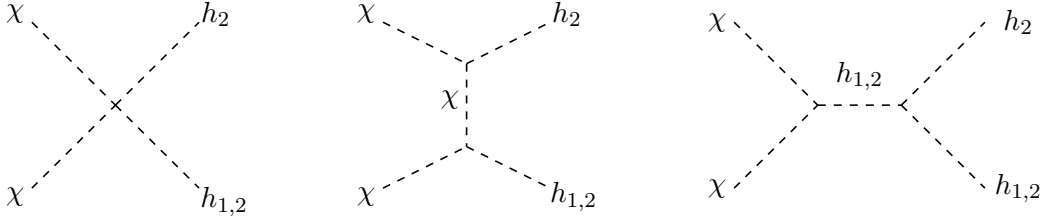
 
\noindent  The evolution of DM number density can be described by solving the Boltzmann equation, which is given by \cite{Kolb:1990vq,Bhattacharya:2016ysw}:
\begin{eqnarray}
\frac{dn_{\chi}}{dt}+3\mathcal{H} n_{\chi}&=&- \sum_{\rm SM}\langle \sigma v \rangle_{\chi\chi\to {\rm SM~SM}}(n_{\chi}^2-{n_{\chi}^{eq}}^2) \,\Theta(M_\chi-M_{\rm SM}) \nonumber \\
&& -\langle \sigma v \rangle_{\chi\chi\to h_1\,h_2}(n_{\chi}^2-{n_{\chi}^{eq}}^2) \,\Theta(2 M_\chi-M_{h_1}-M_{h_2})\nonumber \\
&& -\langle \sigma v \rangle_{\chi\chi\to h_2\,h_2}(n_{\chi}^2-{n_{\chi}^{eq}}^2) \,\Theta(M_\chi-M_{h_2}) \nonumber \\
&=& -{\langle \sigma v \rangle}_{\rm eff} (n_{\chi}^2-{n_{\chi}^{eq}}^2).
\end{eqnarray}
Here $n_\chi^{eq}=\frac{g_\chi}{2\pi^2} M_\chi^2 T K_2[\frac{M_\chi}{T}]$ \cite{Kolb:1990vq} is the equilibrium density where $g_\chi=1$ and $K_2$ is the modified Bessel function of the second kind. 
The ${\langle {\sigma v} \rangle}_{\chi\chi \to a b} $ is the thermal average cross-section for the number changing process $\chi \chi \to a~b$ defined in Ref.\cite{Kolb:1990vq}. These thermal average annihilation cross-sections of DM depend on the model parameters $\{ M_{\chi},\, M_{h_2}, \,v_{\phi}, \,\sin\theta\}$. In this thermal freeze-out scenario, the relic density of DM and the total effective thermal-averaged cross-section are related as \cite{Kolb:1990vq,Bhattacharya:2016ysw}:
\begin{eqnarray}
    \Omega_\chi h^2 &\propto  &\frac{1}{{\langle \sigma v \rangle}_{\rm eff}}~~,
    \label{eq:raprx}
\end{eqnarray}
where 
\begin{eqnarray}
   {\langle \sigma v \rangle}_{\rm eff}&  =&\sum_{\rm SM} {\langle \sigma v \rangle_{\chi\chi\to {\rm SM~SM}} \Theta(M_\chi-M_{\rm SM}) } + \langle \sigma v \rangle_{\chi\chi\to h_1\,h_2} \Theta(2 M_\chi-M_{h_1}-M_{h_2})   \nonumber \\ 
   &&~~~~~~~~~~ + \langle \sigma v \rangle_{\chi\chi\to h_2\,h_2} \Theta(M_\chi-M_{h_2}) .
   \label{eq:sigeff}
\end{eqnarray}
 Here, $\Theta$ is the Heaviside theta function, representing the kinematics of the number-changing process. Depending on DM mass, different number-changing processes open up and contribute to the relic density. The approximate relation in Eq.\ref{eq:raprx}, will help us to understand the behavior of DM density as a function of model parameters. Note we use the publicly available package MicrOmegas \cite{B_langer_2015} for relic density computation, after generating the model files using FeynRule \cite{Alloul_2014}.

\begin{figure}
  \includegraphics[scale=0.25]{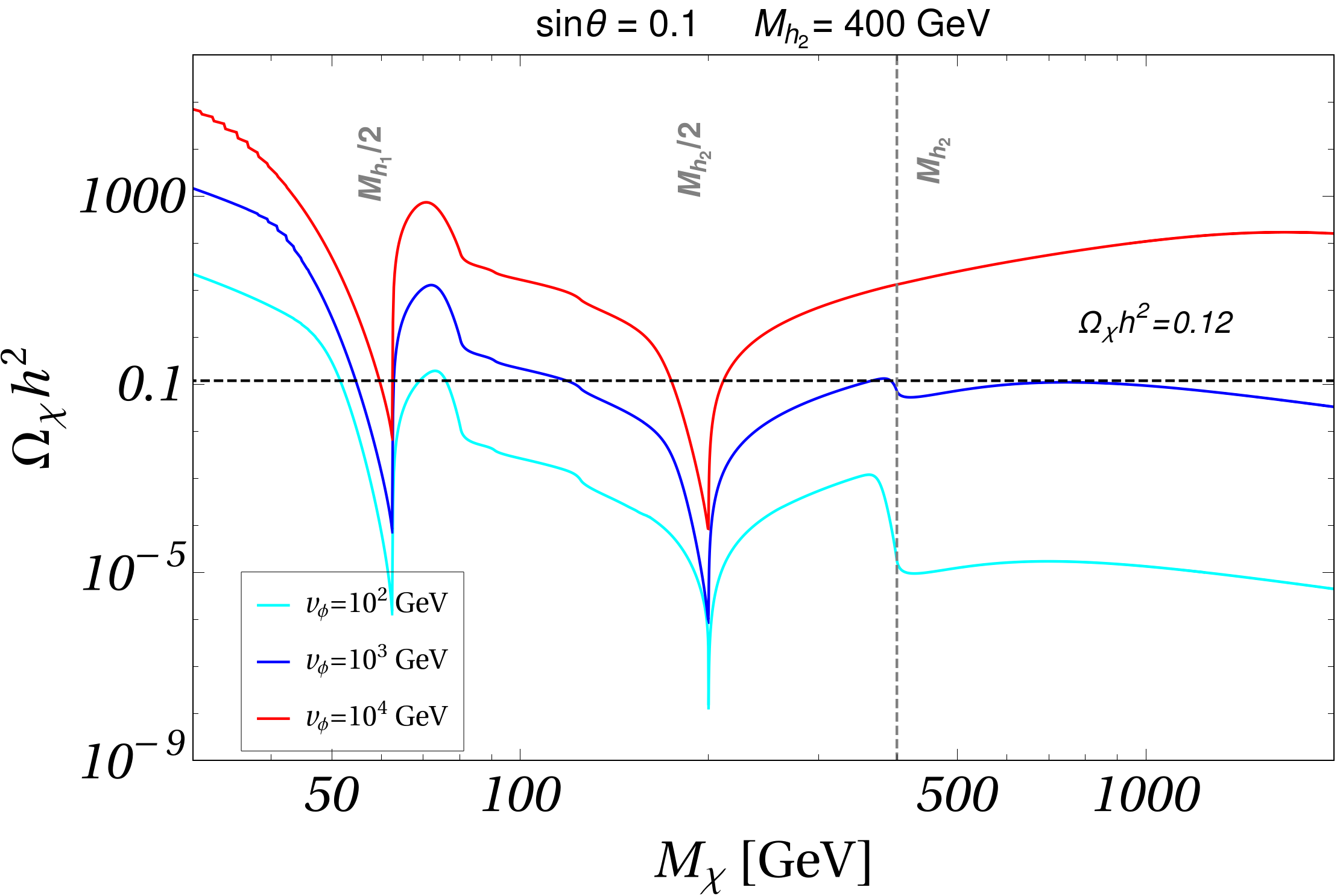}~~
  \includegraphics[scale=0.25]{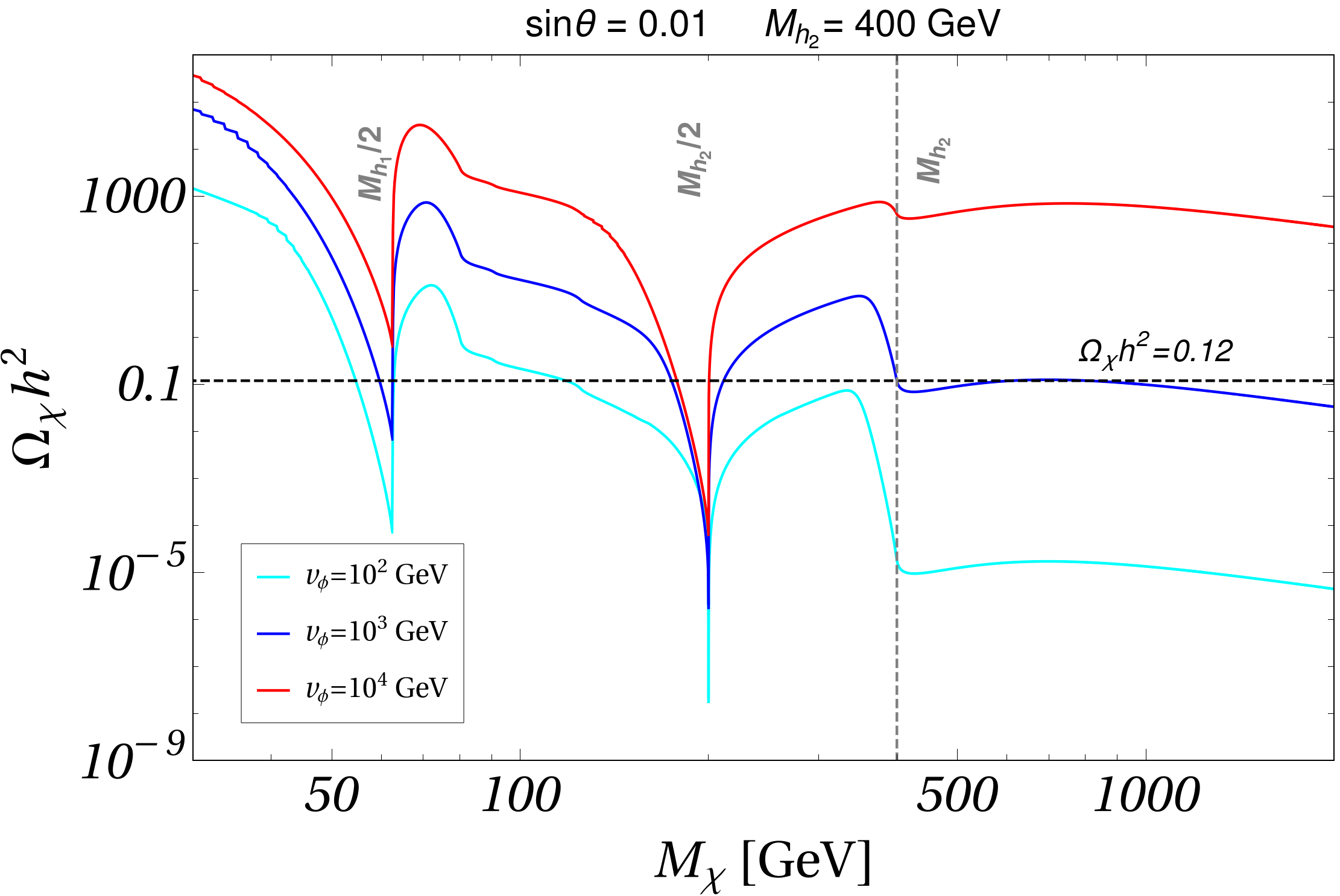}
  \caption{\it Variation of relic density as a function of DM mass $M_\chi$ 
  for three values of $v_\phi = 10^2, 10^3$ and  $10^4 $ {\rm GeV} corresponding to cyan, blue, and red lines respectively. $M_{h_2} = 400 $ {\rm GeV} is fixed for both the panels, while the left (right) panel corresponds to $\sin\theta = 0.1(0.01)$ respectively.}
	\label{fig:Relic1}
\end{figure}

In Fig.\ref{fig:Relic1}, we show the variation of DM relic density ($\Omega_\chi h^2$) as a function of $M_\chi$ for three different values of $v_\phi$ in GeV: $10^2$ (cyan line), $10^3$ (blue line) and $10^4$ (red line). For demonstration, we kept fixed $\sin\theta=0.1$ in the left panel and $\sin\theta=0.01$ in the right panel, and $M_{h_2}=400$ GeV for both figures. The black dotted horizontal line in each figure indicates the observed DM relic density measure by PLANCK $\Omega_{\chi} h^2=0.12$ \cite{Planck:2018vyg}. 
As stated earlier, the DM is connected to the thermal bath particles via the portal coupling $\lambda_{H \Phi}$. Therefore, $\lambda_{H\Phi}$ plays a crucial role in deciding the abundance of DM ($\Omega_\chi h^2$).  For a fixed  $M_\chi$ and  $M_{h_2}$, the portal coupling $\lambda_{H \Phi}$ varies as:  $\lambda_{H\Phi} \propto \frac{\sin\theta}{v_\phi}$ followed by Eq.\ref{eq:lambda_para}. 
For a fixed value of $\sin\theta$, with an increase in $v_{\phi}$, $\lambda_{H\Phi}$ decreases, and as a result, ${\langle \sigma v \rangle}_{\rm eff}$ decreases. Therefore, relic density increases with the increase of $v_\phi$, as depicted in Fig.\ref{fig:Relic1}. On the contrary, for a fixed value of $v_\phi$, as $\sin\theta$ decreases, ${\langle \sigma v \rangle}_{\rm eff}$ decreases, increasing relic density. 
The dependence on $\sin\theta$ can be understood from the left ($\sin\theta=0.1$) and right ($\sin\theta=0.01$) panels of Fig.\ref{fig:Relic1} for a fixed value of $v_\phi$. 

Now we will demonstrate the variation of relic density as a function of DM mass $M_\chi$, keeping $v_\phi, \sin\theta$ and $M_{h_2}$ constant, thereby fixing the value of $\lambda_{H\Phi}$. One can observe two dips in relic density: one
at $M_\chi \sim M_{h_1}/2~ (\sim 62.5~{\rm GeV})$ and another at $M_{h_2}/2 ~(\sim 200~{\rm GeV})$ due to resonance enhancement in the cross-sections at $M_{h_1}$ and $M_{h_2}$ poles respectively. Depending on $M_{\chi}$, different final states are opened up, adding their contribution to ${\langle \sigma v \rangle}_{\rm eff}$. Therefore, the total effective thermal-averaged cross-section increases with the increase of $M_\chi$ as follows in Eq.\ref{eq:sigeff}. Hence, relic density drops with the increase of  $M_\chi$. The active DM annihilation processes vary across different $M_\chi$ regions, as follows.\\
$\ bullet~~M_\chi < M_{h_1}:$ $\chi~\chi \to {\rm SM~SM}$ with $M_\chi > M_{\rm SM}$ are the dominant number-changing processes, which are mediated by both the CP-even physical states $h_{1,2}$. As already shown, the relic density varies with both the $\sin\theta$ and $v_\phi$. \\
%
$\bullet~ M_{h_1} < M_\chi < M_{h_2}:$ New annihilation channels contribute to relic density depending on $M_\chi$ as $\chi \chi \to h_1 h_1 $  with $M_\chi > M_{h_1}$, $\chi \chi \to ~t ~\overline{t}$  with $M_\chi > M_t$  and   $\chi \chi \to h_1 h_2$ with $M_\chi > (M_{h_1}+M_{h_2})/2 ~ (\sim 262.5)$. \\
$\bullet~  M_\chi > M_{h_2}:$ In addition to the two aforementioned annihilation processes, the following new process: $\chi \chi \to h_2 h_2$ also contributes to ${\langle \sigma v \rangle}_{\rm eff}$. Consequently, there is a suppression in relic density near $M_\chi \sim M_{h_2}$.

\begin{figure}[h!]
$$
 \includegraphics[scale=0.25]{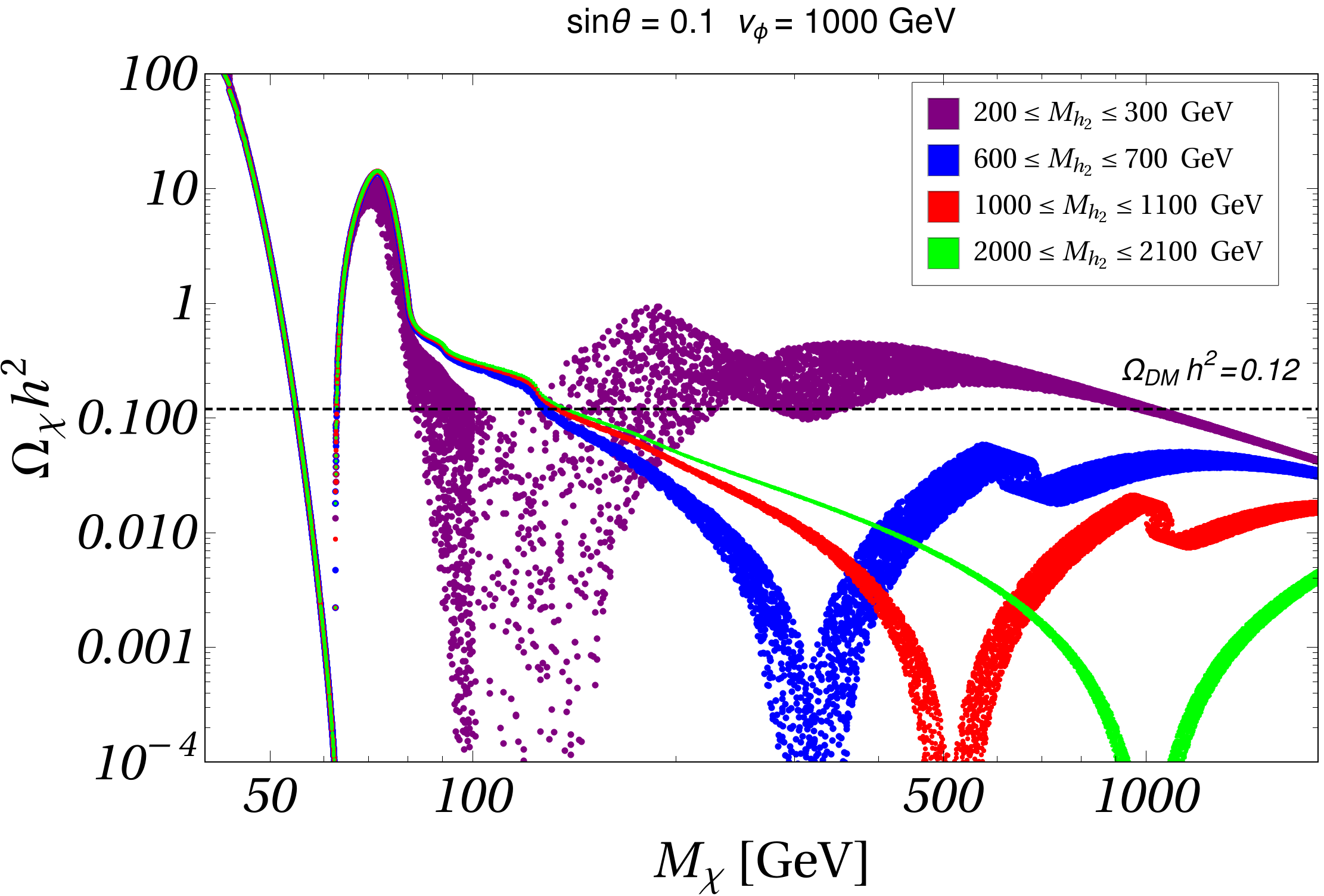}~~
  \includegraphics[scale=0.25]{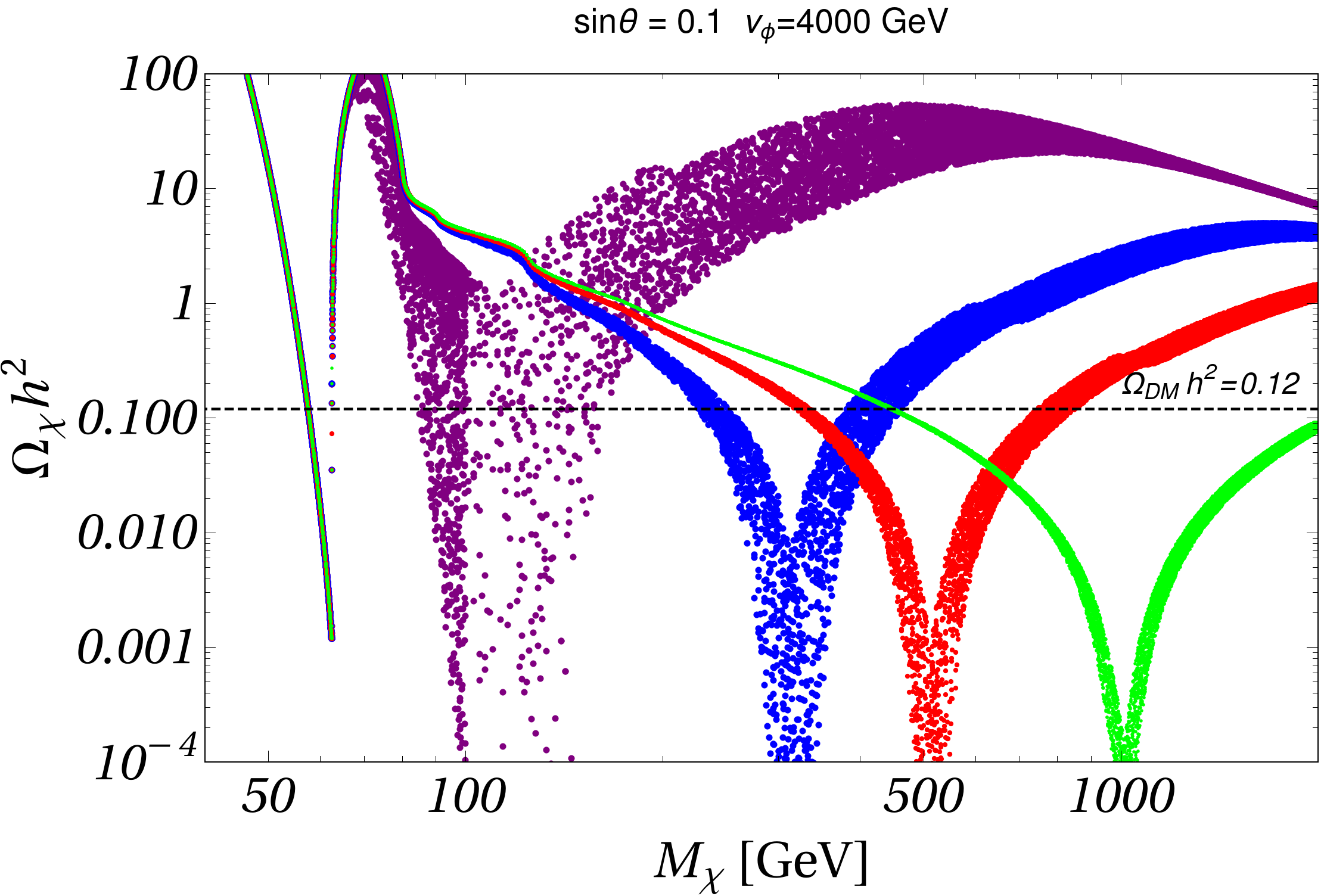}
 $$
\caption{\it  Variation of relic density as a function of DM mass with different ranges of $M_{h_2}$. $\sin\theta = 0.1 $ is fixed for both the panels, while the left (right) panel corresponds to $v_\phi = 1000 ~(4000)$ ${\rm GeV}$ respectively.}
\label{fig:Relic2}
\end{figure}

Next, we show the variation of the DM relic density with $M_\chi$ for four representative regions of $M_{h_2}$ shown in both panels of Fig.\ref{fig:Relic2}. The relic density is almost independent of $M_{h_2}$ when DM mass is below $M_W$, as the coupling strengths between SM Higgs and light fermions are suppressed. In contrast, when $M_\chi > M_W$, $M_{h_2}$ turns crucial and significantly affects the relic density as the DM annihilation into the gauge and scalar final states becomes available. In the presence of the new annihilation processes ($\chi ~\chi \to h_1 h_2, h_2 h_2$) and the resonance-induced drop in relic density (discussed before ), the $M_{h_2}$ substantially influences the relic density of DM. This is due to the dependence of the quartic couplings ($\lambda$) on $M_{h_2}$ (see Eq.\ref{eq:lambda_para}).  In Fig.\ref{fig:Relic2}, we see that for fixed values of $v_\phi$ and $\sin\theta$, the relic density decreases with the increase of $M_{h_2}$ (for $M_\chi > M_{W}$). This can be attributed to the fact that the portal coupling $\lambda_{H\Phi}$  increases with the increase of $M_{h_2}$ as illustrated in Eq.\ref{eq:lambda_para}. As discussed before, with the increase of $v_\phi$, the relic density of DM increases, as shown in the right panel of Fig.\ref{fig:Relic2}. Note that the relic density drops near $M_\chi \sim M_{h_2}$ due to the opening of a new annihilation process $\chi~\chi \to h_2~h_2$, and it becomes prominent for the lower value of $v_\phi$ as $\lambda_{H\Phi} \propto 1/v_\phi$. For this reason, we do not observe any noticeable relic density reduction near $M_\chi \sim M_{h_2}$ for $v_\phi=4000$ GeV in the right panel of Fig.\ref{fig:Relic2}. 
\begin{figure}[h!]
$$
  \includegraphics[scale=0.3]{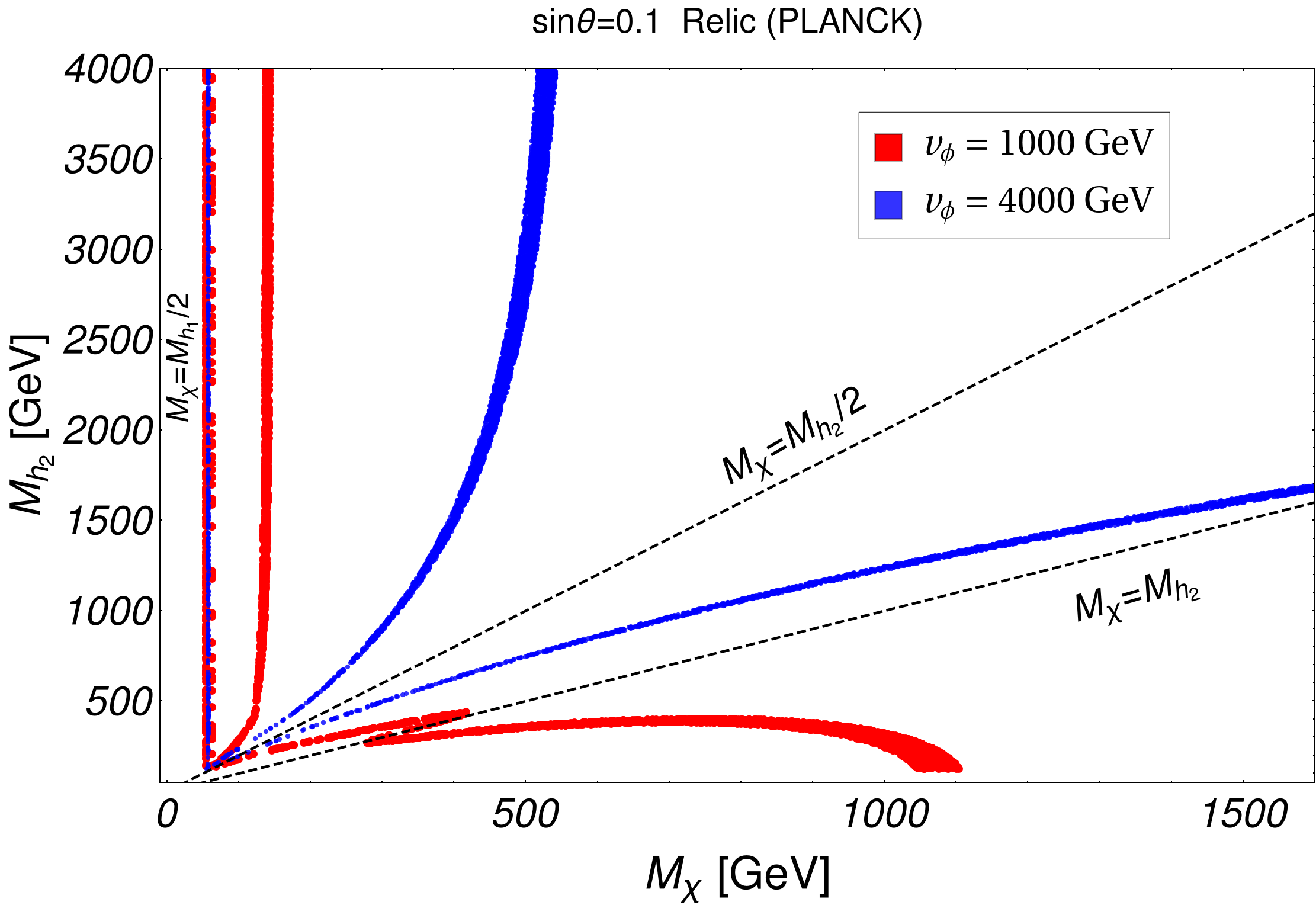}
 $$
\caption{\it  The allowed parameter space for the relic density of DM ($\Omega_\chi h^2 =0.120 \pm 0.001$ \cite{Planck:2018vyg}) in the plane of $M_\chi$ vs $M_{h_2}$ for fixed $\sin\theta=0.1$ and $v_{\phi}=\{1000 \,{\rm (Red)}, 4000 \,{\rm (Blue)} \}$ {\rm GeV}. The dashed black lines correspond to $M_{\chi}=M_{h_2}$ and $M_{\chi}=M_{h_2}/2$ as mentioned on the representative lines of the figures. }
\label{fig:Relic3}
\end{figure}

In Fig.\ref{fig:Relic3}, we show the relic allowed parameter space in $M_\chi$ vs $M_{h_2}$ plane for the same values of $v_\phi$ and $\sin\theta$, considered in Fig.\ref{fig:Relic2}. The vertical region around $M_\chi \sim M_{h_1}/2$ satisfies the observed DM abundance, which is independent of $M_{h_2}$, as mentioned earlier. For $M_\chi > M_W$, the observed relic density parameter space looks like a {\bf V} shape in the plane of $M_\chi-M_{h_2}$ for a fixed value of $v_\phi$ (1000 GeV (red region) and 4000 GeV (blue region)). The region within each {\bf V} shape corresponding to a fixed $v_\phi$ represents the under-abundance ($\Omega _\chi h^2 < \Omega_{\rm DM} h^2$), while the region outside it represents over-abundance ($\Omega _\chi h^2 > \Omega_{\rm DM} h^2$). These regions can be understood from both the figures in Fig.\ref{fig:Relic2}. The diagonal dotted lines represent $M_{\chi}=M_{h_2}$ and $M_{\chi}=M_{h_2}/2$, as depicted in the figure. 

 \underline{$v_\phi=1000$ GeV (red region)}:  
 First, we focus on the lower mass region of $h_2$, where $M_{h_1} < M_{h_2} \lesssim 500$ GeV. The observed DM density is satisfied around $M_\chi \sim M_{h_2}/2$ due to the resonance effect. The opening of new annihilation processes $\chi \chi \to h_2 h_2$ with $M_\chi > M_{h_2}$, the DM relic density falls in the correct ballpark near DM mass $M_\chi \sim M_{h_2}$. At the same time with an increase in $M_\chi$, the coupling strengths $\lambda_{h_i \chi \chi}$ and $\lambda_{h_i h_j \chi \chi}$ also increase, resulting in a parameter space that satisfies the relic density for $M_\chi \approx 500-1100$ GeV with $M_{h_1} \lesssim M_{h_2} \lesssim 500$ GeV. 
 \noindent In the heavier mass region of $h_2$ with $M_{h_2} \gtrsim 500$ GeV, the portal coupling $\lambda_{H\Phi}$ gets enhanced, leading to under-abundance. However, a vertical region $M_\chi \sim M_{h_1}$ satisfies the observed DM density independently of $M_{h_2}$. In this mass region, the DM density decreases because of new number-changing processes and the enhanced cross-section near the $h_2$ pole. These phenomena can be understood from the left panel of Fig.\ref{fig:Relic2}, around the mass region $M_{\chi} \sim M_{h_1}$. 

\underline{$v_\phi=4000$ GeV (blue region)}: Similar feature can also be observed in this case. With an increase in $v_\phi$, the portal coupling is suppressed as $\lambda_{H\Phi} \propto 1/v_\phi$, resulting in a higher DM density. To satisfy relic density in this case, we rely on resonance enhancement in the ${\langle \sigma v \rangle}_{\rm eff}$ near the $h_2$ pole. Therefore, it satisfies the observed DM abundance on both sides of $M_\chi = M_{h_2}/2$ line with $M_\chi > M_{h_1}/2$ and $M_{h_1} < M_{h_2} \lesssim 2000$ GeV. Beyond $M_{h_2} \gtrsim 2000$ GeV, the portal coupling $\lambda_{\Phi H}$ is further enhanced with $M_{h_2}$ as followed by Eq.\ref{eq:lambda_para}. This leads to an under-abundance for $M_{\chi} \gtrsim 500$ GeV, which can be understood from the right panel of Fig.\ref{fig:Relic2}. Therefore, there is a vertical region around $M_\chi \sim 500$ GeV with $M_{h_2} \gtrsim 2000$ GeV, regardless of $M_{h_2}$,  which meets the observed relic density.  In this region, the relic density falls due to resonance enhancement in the annihilation cross-section.

\subsection{Direct detection}

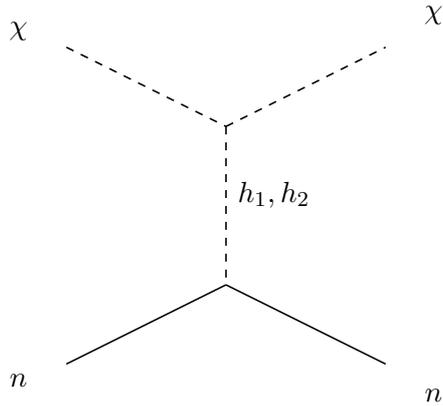
\begin{figure}[ht]
   \begin{center}
    \begin{tikzpicture}[line width=0.6 pt, scale=2.1]
\draw[dashed] (-1.8,1.0)--(-0.8,0.5);
\draw[solid] (-1.8,-1.0)--(-0.8,-0.5);
\draw[dashed] (-0.8,0.5)--(-0.8,-0.5);
\draw[dashed] (-0.8,0.5)--(0.2,1.0);
\draw[solid] (-0.8,-0.5)--(0.2,-1.0);
\node at (-2.1,1.1) {$\chi$};
\node at (-2.1,-1.1) {$n$};
\node at (-0.5,0.07) {$h_1,h_2$};
\node at (0.5,1.2) {$\chi$};
\node at (0.5,-1.2) {$n$};
     \end{tikzpicture}
 \end{center}
\caption{\it Feynman diagrams for spin-independent DM-nucleon scattering process for DM ($\chi$).} 
\label{fig:DD}
 \end{figure}

We shall now move to the DM-nucleon scattering process relevant to direct detection (DD). In direct detection experiments, the flux of DM may scatter with the nuclei in the target crystals, and the recoil rate of the target nucleus can be searched for as a signal of the DM. In this case, the spin-independent (SI) $\chi-n$ scattering cross-section occurs via two CP even scalars ($h_1$ and $h_2$) exchange t-channel diagrams as shown in Fig.\ref{fig:DD}. The corresponding spin-independent $\chi-n$ scattering cross-section with  the fractional DM density $f_\chi ~(\equiv \Omega_{\chi}/\Omega_{\rm DM})$ is given by \cite{Ghosh:2017fmr}
\begin{eqnarray}
    \sigma_{\rm DD}^{\rm SI} &=& f_\chi ~\frac{1}{4\pi} \bigg(  \frac{f_n \mu_n}{M_\chi}    \bigg)^2 \bigg( \frac{m_{n}}{v}\bigg)^{2} ~\bigg[\frac{\lambda_{h_1\chi\chi} \cos\theta }{t-m_{h_1}^{2}}+\frac{\lambda_{h_2 \chi\chi} \sin\theta}{t-m_{h_2}^{2}}\bigg]^{2} \nonumber \\
    &\overset{t \to 0}{=}&  f_\chi ~ \frac{1}{4\pi} \bigg(  \frac{f_n \mu_n}{M_\chi}    \bigg)^2 \bigg( \frac{m_{n}}{v}\bigg)^{2} ~\bigg[\frac{\lambda_{h_1\chi\chi} \cos\theta }{m_{h_1}^{2}}+\frac{\lambda_{h_2 \chi\chi} \sin\theta}{m_{h_2}^{2}}\bigg]^{2} ,
    \label{eq:ddX}
\end{eqnarray}
where 
\begin{eqnarray}
\lambda_{h_1\chi\chi} &=& - \lambda_{H\Phi}\, v \cos\theta + 2  \lambda_{\Phi} v_\phi \sin\theta - \frac{3 \mu_3 \sin\theta}{\sqrt{2}}  \nonumber \\
~~{\rm and}~~\lambda_{h_2\chi\chi} &=& - \lambda_{H\Phi} \,v \sin\theta - 2 \lambda_{\Phi} v_\phi \cos\theta 
+ \frac{3 \mu_3 \cos\theta}{\sqrt{2}}. \nonumber
\end{eqnarray}
\noindent Here $\mu_n=\frac{m_n ~M_\chi}{m_n+M_\chi}$ is the reduced mass of DM-nucleon system with $m_n=0.946$ GeV (neutron mass) and $f_n=0.28$ is the nucleon form factor \cite{Alarcon:2012nr}. For small $\sin\theta$ limit the cross-section turns out to be $\sigma_{\rm DD}^{\rm SI} \propto \frac{\lambda_{H\Phi}^2}{{M_\chi}^2}$ where the expression of $\lambda_{H\Phi}$ is given in Eq.\ref{eq:lambda_para}. Non-observation of DM at direct search experiments such as XENON-1T \cite{XENON:2018voc}, PANDAX-4T \cite{PandaX-4T:2021bab} and the most recent LUX-ZEPLIN (LZ) 2024 \cite{LZ:2024zvo} put stringent constraints on the $M_\chi-\sigma_{\rm DD}^{\rm SI}$ plane, which can be translated in terms of the model parameters.

\begin{figure}[ht]
  \subfigure[]{
  \includegraphics[scale=0.27]{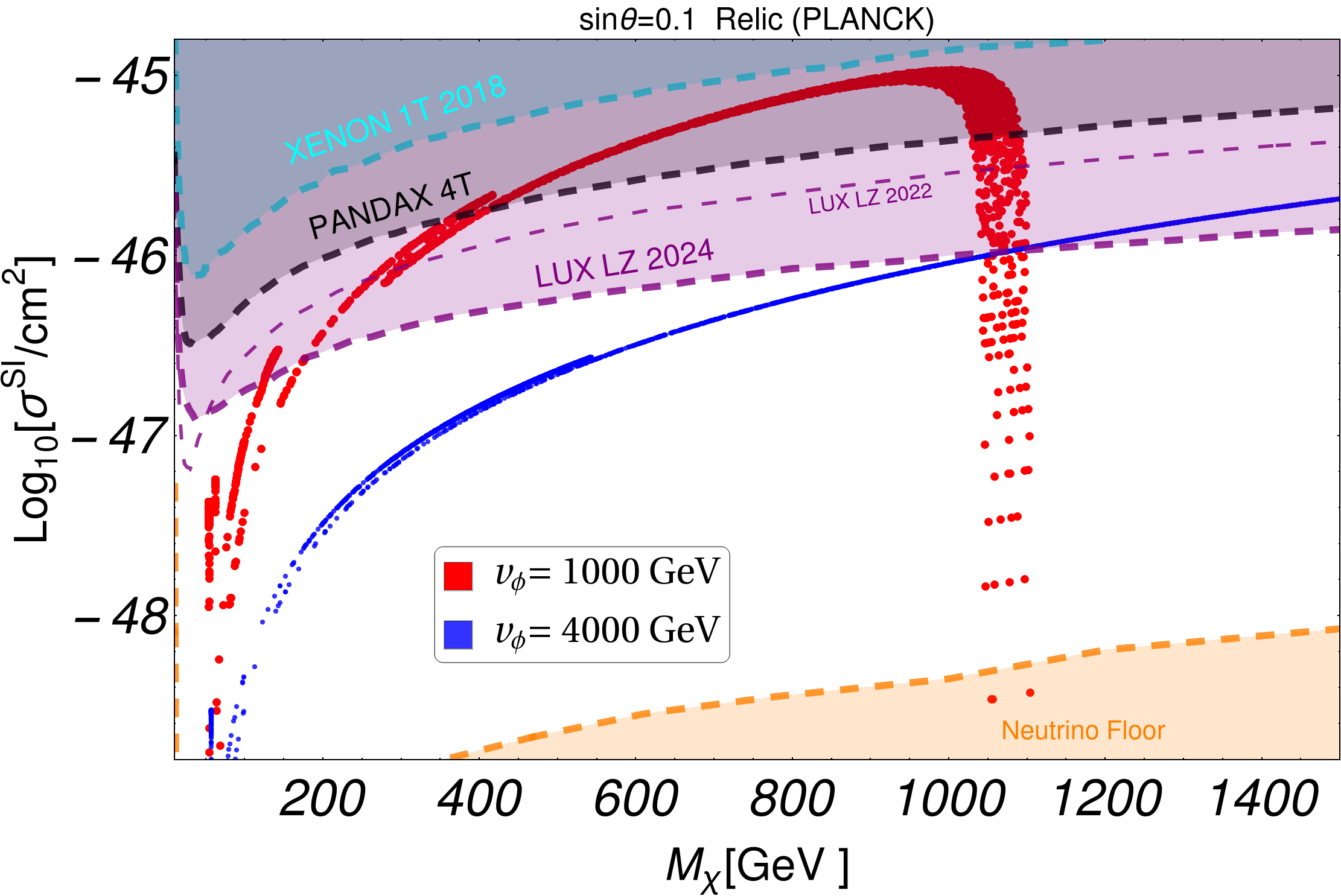}}
  \subfigure[]{
  \includegraphics[scale=0.28]{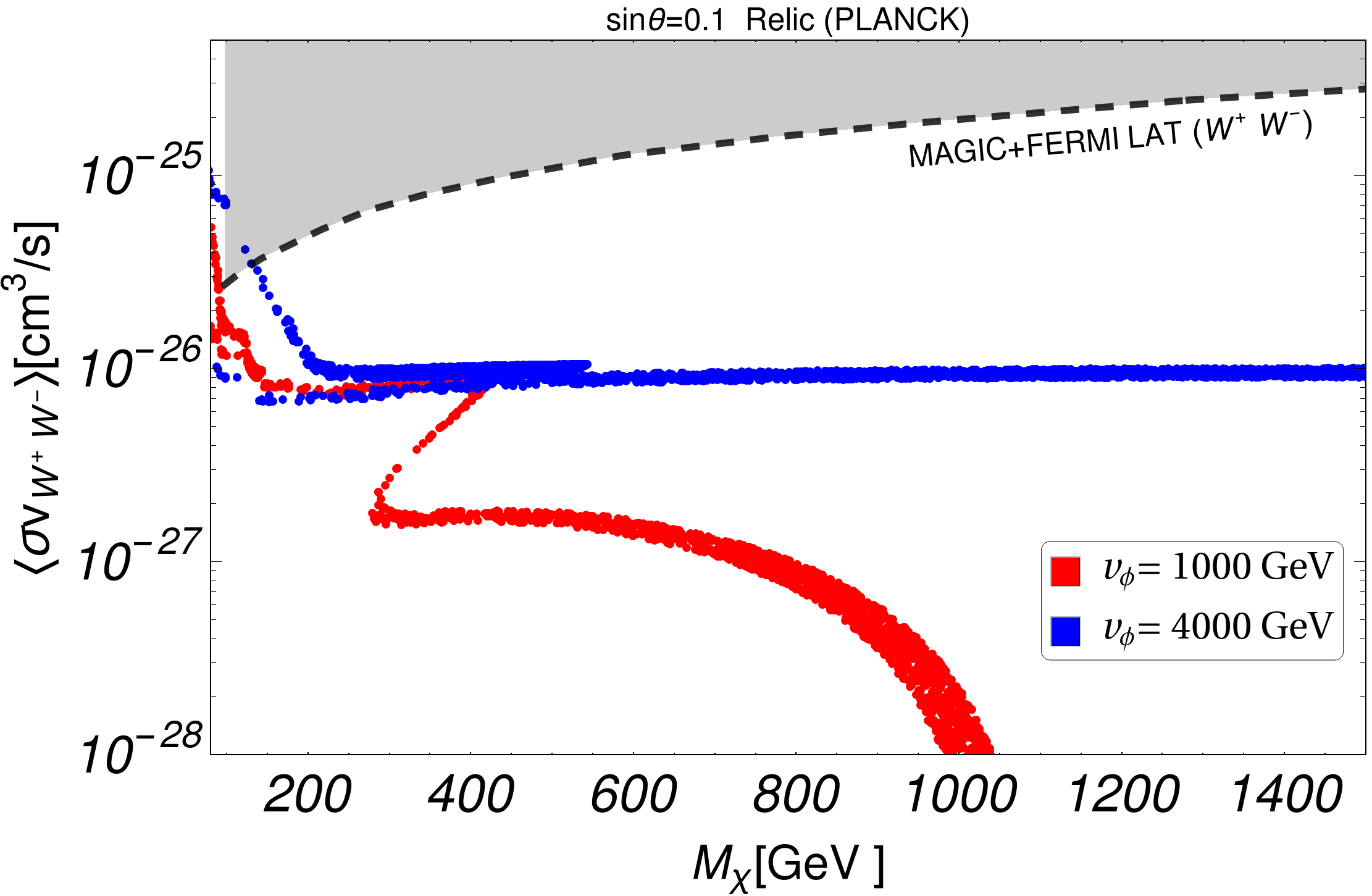}}
    \caption {\it (a) Relic density allowed parameter space is plotted in the plane of DM mass versus SI DM-nucleon cross-section for $v_\phi=1000$ GeV (red region) and $v_\phi=4000$ GeV (blue region) while keeping $\sin\theta=0.1$. The parameter space is compared with the experimental upper bounds from XENON-1T\cite{XENON:2018voc}, PANDAX 4T\cite{PandaX-4T:2021bab}, and LZ 2024 \cite{LZ:2024zvo} in the same plane. The orange shaded region represents the neutrino floor. (b) The thermally averaged cross-section of the $\chi \chi \to W^+ W^-$ process for the parameter space allowed by observed DM density (PLANCK) is plotted as a function of $M_\chi$. The combined exclusion bound from indirect search experiments by Fermi-LAT \cite{Fermi-LAT:2015att} and MAGIC \cite{MAGIC:2016xys} is shown in the gray region for the same DM annihilation channel. Note that for both plots, $\chi$ represents $100\%$ ($f_\chi =1 $) of the observed DM density. }
    \label{mchi_DD_ID}
\end{figure}

We plot the relic density allowed parameter space ($f_\chi=1$ i.e. $100\%$ of the observed relic density) in $M_\chi $ vs $\sigma_{\rm DD}^{\rm SI}$ plane in Fig.\ref{mchi_DD_ID} (a) to compare with the current upper bounds from XENON-1T\cite{XENON:2018voc}, PANDAX 4T\cite{PandaX-4T:2021bab}, and LZ 2024 \cite{LZ:2024zvo}. The red region corresponds to $v_{\phi}=1000$ GeV, and the blue region corresponds to $v_\phi=4000$ GeV, both with $\sin\theta=0.1$. The increase in $v_{\phi}$ leads to a decrease in the DM-nucleon scattering cross-section due to suppression of the portal coupling $\lambda_{H\Phi}$ with higher $v_\phi$, which is depicted in Fig.\ref{mchi_DD_ID} (a). {Therefore, for $v_\phi=4000$ GeV, the SI DD cross-section becomes smaller compared to $v_\phi=1000$ GeV.} The current LZ 2024 \cite{LZ:2024zvo} data excludes our parameter space in the intermediate-mass region $M_\chi \sim \{300-1000\}$ GeV with $M_{h_2} \lesssim 500$ GeV for $v_\phi=1000$ GeV. There is a drop in $\sigma_{\rm DD}^{\rm SI}$ near $M_\chi \sim 1000$ GeV with $v_\phi=1000$ GeV. This phenomenon occurs because the observed DM relic density in this region demands a lower $M_{h_2}$, as discussed earlier. As a result, the cross section decreases since $\lambda_{H \Phi}$ diminishes with $M_{h_2}$. {On the other hand, for $v_\phi = 4000$ GeV, the upper bound on SI DM-nucleon cross section excludes $M_{\chi} \gtrsim 1100$ GeV}.

\subsection{Indirect detection}

DM can also be detected at various indirect search experiments, including space-based observatories like the Fermi-LAT \cite{Fermi-LAT:2015att} and ground-based counterparts like MAGIC \cite{MAGIC:2016xys} telescopes. These telescopes detect gamma rays produced via DM annihilation or decay in the local Universe. In our discussion, the gamma-ray flux can be produced when DM $\chi$ annihilates into SM-charged particle pairs ($\psi^+\psi^-$), followed by their subsequent decay. The total gamma-ray flux for a given mode $\chi \chi \to \psi^+\psi^-$ ($\psi=\{\mu, \tau, b, W\}$)in a specific energy range is given by \cite{MAGIC:2016xys}
\begin{eqnarray}
    \Phi^\gamma_{\psi^+ \psi^-}&=&\frac{1}{4\pi} \frac{\langle \sigma v \rangle_{\chi\chi \to \psi^+ \psi^- }}{2 M_{\chi}^2} \int_{E_{\rm min}}^{E_{\rm max}} \frac{dN_\gamma}{dE_\gamma} dE_{\gamma} \int dx ~\rho_\chi^2\big(r(b,l,x)\big)
    \label{eq:id}~.
\end{eqnarray}
The notation follows standard conventions as ref.~\cite{MAGIC:2016xys}. The indirect search experiments like Fermi-LAT and MAGIC \cite{Fermi-LAT:2015att, MAGIC:2016xys} collectively put an upper bound $\langle \sigma v \rangle_{\chi\chi \to \psi^+ \psi^-}$ from the non-observation of gamma-ray flux produced from DM. It is evident from the above Eqn.\ref{eq:id}, to compare the experimental bounds with the theoretical $\langle \sigma v \rangle_{\chi\chi \to \psi^+ \psi^-}$, one must scale the cross-section by the fractional DM abundance as: ${\langle \sigma v \rangle^{\rm ID}_{\chi\chi \to \psi^+ \psi^-}}=f_\chi^2 \langle \sigma v \rangle_{\chi\chi \to \psi^+ \psi^-}$ with $f_{\chi}=\Omega_\chi/\Omega_{\rm DM}~ (\leq 1)$.

The most stringent constraint is found to come from the DM annihilation mode $\chi \chi \to W^+ W^-$ compared to other modes, due to the gauge coupling. In Fig.\ref{mchi_DD_ID} (b), we show $\langle {\sigma v}_{W^+ W^-} \rangle$ as a function of $M_\chi$ for all relic satisfied points with $f_\chi=1$ ($\chi$ contributes $100\%$ of the observed relic density), and compare it with the combined Fermi-LAT and MAGIC exclusion bound \cite{Fermi-LAT:2015att, MAGIC:2016xys}, shown in the gray region. Similar to the DD cross-section, the $\langle \sigma v_{W^+ W^-} \rangle$ decreases with an increase in $v_\phi$ (shown in red for $1000$ GeV and in blue for $4000$ GeV) and a decrease in $M_{h_2}$, which is influenced by the coupling $\lambda_{H \Phi}$ (see Eq.\ref{eq:lambda_para}). From the plot, it turns out that apart from lower $M_\chi$, most of the parameter space lies below the combined Fermi-LAT and MAGIC exclusion bound.
Note that the other DM annihilation modes $\psi^+\psi^-:\{b \overline{b}, ~\tau^+\tau^-,~\mu^+\mu^- \}$ are well below the upper bound set by indirect searches due to the relatively suppressed SM Yukawa coupling of these fermions. 

\begin{figure}[h!]
  \includegraphics[width=.65\linewidth]{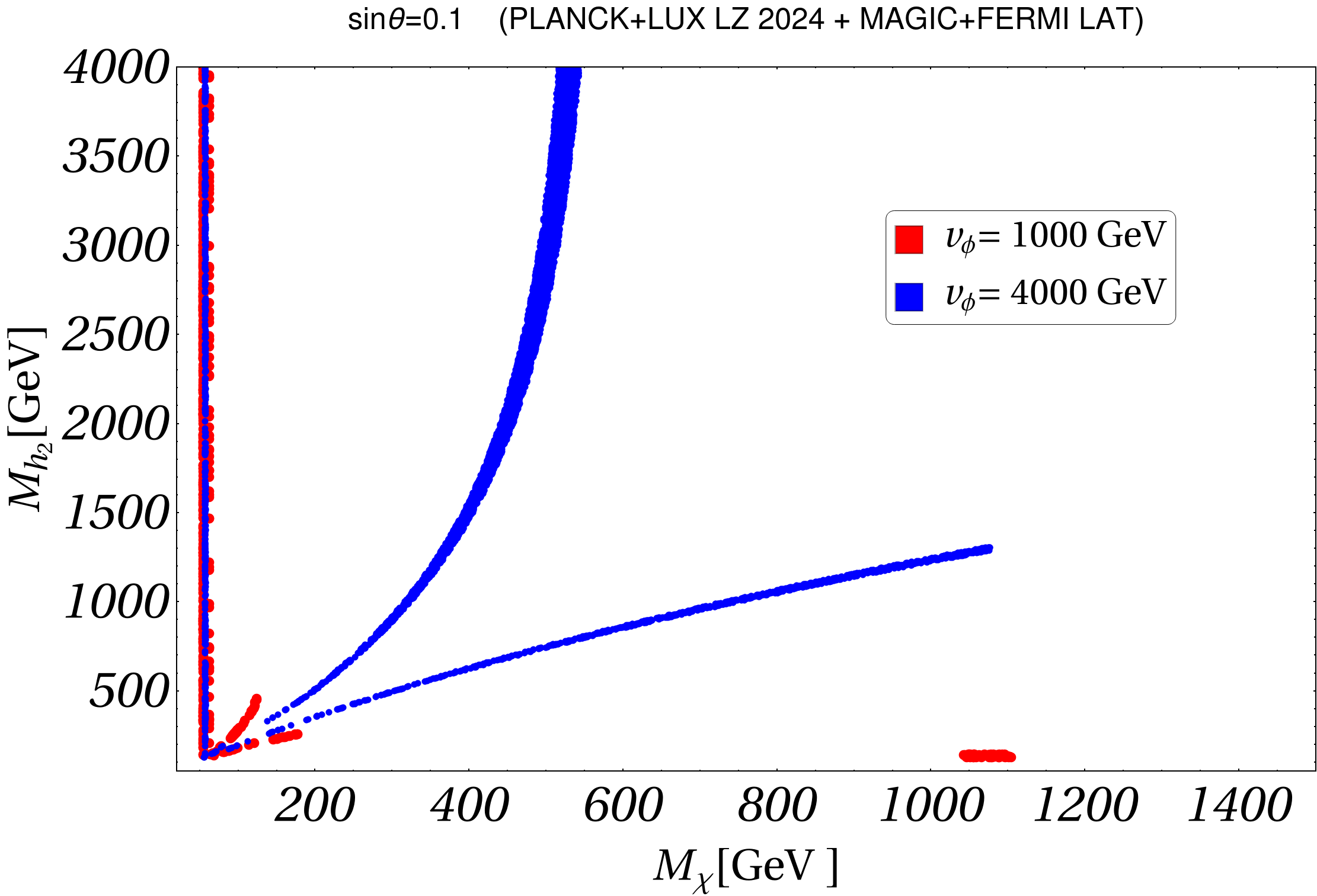}
\caption{\it Relic (PLANCK) +DD (LZ 2024)+ ID (Fermi LAT+ MAGIC ) allowed parameter space is shown in the plane of $M_\chi$ and $M_{h_2}$ for the same parameters. Here DM ($\chi$) contributes $100\%$ of the observed relic density ($f_\chi=1$). }
\label{fig:RDDID}
\end{figure}

Finally in Fig.\ref{fig:RDDID}, we show the parameter space in the $M_{\chi}-M_{h_2}$ plane which collectively satisfies Relic (PLANCK \cite{Planck:2018vyg}) + DD (LZ 2024 \cite{LZ:2024zvo})+ ID (Fermi LAT+ MAGIC \cite{Fermi-LAT:2015att,MAGIC:2016xys}) constraints. The red and blue regions correspond to the $v_\phi=1000$ GeV and $v_\phi=4000$ GeV, respectively, with $\sin\theta=0.1$. {Note that the intermediate DM mass region, $M_{\chi} \sim \{300-1000\}$ GeV for $v_{\phi}=1000$ GeV and the higher mass region, $M_\chi \gtrsim 1100$ GeV for $v_{\phi}=4000$ GeV are excluded from the upper bound on DM-nucleon cross-section by LZ 2024 \cite{LZ:2024zvo}.}

\section{Phase Transition} \label{sec:-PT}
We now discuss the possibility of a strong first-order phase transition in the parameter space relevant to both Leptogenesis and DM phenomenology. Our main objective for studying FOPT is that it can give rise to stochastic GWs, which can be detected by experiments in the future. The physics comprising low-scale Leptogenesis and DM phenomenology leaves its imprints on the GW spectrum, which can be detected in GW detectors.  To study the phase transition, we consider finite temperature corrections to the effective potential.

The Coleman Weinberg effective  potential (or the quantum corrections to the tree level potential) at one loop level in the $ \overline{\rm MS} $ renormalization scheme at zero temperature is given by \cite{PhysRevD.7.1888} 
  \begin{equation}
      V_{\rm cw}(h_{i}) = \sum_{j=W^{\pm},Z,h_1,h_2,\chi,t} (-1)^{S_{j}}\frac{n_{j}m^4_{j}(h_{i})}{64\pi^2}
\left[\log\frac{m_{j}^{2}(h_{i})}{4\pi \mu^{2}} - C_{j}  \right] ~,
\label{Vcw}
  \end{equation}
  where $h_{i},\,\{i=1,2\}$ are the scalar fields in physical basis, $m_{j}$ is the mass of the $j^{\rm th}$ particle and $n_{j}$  is the number of degrees of freedom of the $j^{\rm th}$ particle. $S_{j}$ has the value 0 for bosons and 1 for fermions. $\mu$ is the renormalisation energy scale which is taken to be $m_{t}$. $C_{j}$ are the constants which have the value $\frac{3}{2}$ for scalars and fermions, and $\frac{5}{6}$ for gauge bosons. Considering thermal effects, the temperature-dependent part of the effective potential at one loop level can be expressed as \cite{PhysRevD.9.3320}
  \begin{eqnarray}
 V_{T}(h_{i},T)=\frac{T^4}{2\pi^2}\Bigg [\sum_{B}n_{B}J_{B}(m_{B}^2(h_{i})/T^2) + \sum_{F}n_{F}J_{F}(m_{F}^2(h_{i})/T^2) \Bigg ].
   \label{finiteT}\end{eqnarray}
   The $n_{B/F}$ are the degrees of freedom of bosons/fermions respectively, and the $J_{B/F}$ are Bosonic and Fermionic functions, which are represented as 
    \begin{equation}
        J_{B/F}(x^2) = \int_{0}^{\infty} y^2 \log[1\mp e^{-\sqrt{x^2+y^2}}]dy.
        \label{JbJf}
    \end{equation}
At high temperatures, the perturbative expansion of the effective potential may lose validity due to the emergence of infrared divergences, leading to the breakdown of the loop expansion of the effective potential \cite{PhysRevD.9.3320, Linde:1980ts}. These divergences arise from the Matsubara modes, which account for periodicity in the imaginary time direction in finite-temperature calculations \cite{Matsubara:1955ws}.
{To account for the infrared divergences in the bosonic sector at finite temperature, we include the daisy resummation following the Arnold-Espinosa approach \cite{Arnold:1992rz}. In this method, only the zero Matsubara modes of bosonic fields receive thermal mass corrections, which are incorporated into the daisy (ring) term. The potential due to such ring diagrams can be written as \cite{quirós1993daisy},
\begin{equation}
    V_{\rm ring}(h_{i},T)=-\sum_{j}\frac{n_{j}T}{12\pi}\Big(\big[M_{j}^2(h_{i},T)]^{\frac{3}{2}}-m_{j}^3(h_{i})\Big )~.
    \label{Vring}
\end{equation}

\noindent The quantities $M_{j}(h_{i},T)$ and $m_{j}(h_{i})$ are the eigenvalues of the mass matrix at finite and zero temperature, respectively. The mass terms at finite temperature are given by,
\begin{equation}
-\mu_{H/\phi}^2(T)=-\mu_{H/\phi}^2+\Pi_{H/\phi}T^2 \nonumber ~.
\end{equation} 
The quantities $\Pi_{H/\phi}$ are called the Daisy Coefficients, which are obtained from the coefficients of $T^2$ in the expression of finite temperature correction to the effective potential in the high-temperature limit.

\noindent The Daisy coefficient matrix $\Pi$ for the CP-even scalar fields is given by
\begin{equation}
   \begin{bmatrix}
    \frac{\lambda_{H}}{2}+\frac{\lambda_{H\Phi}}{12}+\frac{3g^2}{16}+\frac{g'^2}{16}+\frac{y^2}{4} &0\\
    0& \frac{\lambda_{\Phi}}{3}+\frac{\lambda_{H\Phi}}{6} \\
    \end{bmatrix} T^2 .
\end{equation}
}
While taking into account the Coleman Weinberg at zero temperature corrections, generally, the tree level vevs and the masses get changed. To avoid that, we have to add a zero temperature counter term $\delta V_{\rm ct} (h_{i})$ to the effective potential, which is given by,
\begin{eqnarray}
\delta V_{\rm ct}(h_{i}) &=& -\delta\mu_{H}^{2} (H^{\dagger} H) + \delta\lambda_{H} (H^{\dagger} H)^{2}
- \delta\mu_{\Phi}^{2} (\Phi^{\dagger} \Phi) + \delta\lambda_{\Phi} (\Phi^{\dagger} \Phi)^{2} \nonumber\\
&& +\delta\lambda_{H\Phi} (\Phi^{\dagger} \Phi)(H^{\dagger} H)  + \frac{\delta\mu_{3}}{2} (\Phi^{3}+\Phi^{\dagger 3}).
\label{ctpot}
\end{eqnarray}
To find out the expressions of the counter terms corresponding to each parameter, we use the following conditions,
\begin{eqnarray}
    \partial_{h_{a}}(\delta V_{\rm ct}+\Delta V)=0 , \nonumber\\
     \partial_{h_{a}}\partial_{h_{b}}(\delta V_{\rm ct}+\Delta V)=0 , 
     \label{ctcon}
\end{eqnarray}
where the partial derivatives are taken with respect to $h$ and $\phi$ fields expressed as $h_{a(b)}$. The derivatives are evaluated at $h=v$ and $\phi=v_{\phi}$. The $\Delta V$ is the effective potential at zero temperature, excluding the tree-level part of the potential. The expressions of the counter term containing $\delta V_{\rm ct}$ corresponding to each parameter in the tree-level potential are given in \ref{cterms}. The total effective potential can be written as the sum of the contribution that comes from the Eqs.\,\ref{Vcw},\,\ref{finiteT},\,\ref{Vring},\,\ref{ctpot}, as:
\begin{equation}
    V_{\rm eff}(h_{i},T) = V_{0}(h_{i})+ V_{\rm cw}(h_{i})+V_{T}(h_{i},T)+V_{\rm ring}(h_{i},T)+\delta V_{\rm ct}(h_{i}) ,
    \label{vtot}
\end{equation}
 where $V_{0}(h_{i})$ is the tree level potential in terms of physical scalar basis and is given in Eq.\ref{Eq_potential}. $V_{cw}(h_{i})$ and $V_{T}(h_{i},T)$ are the one loop corrections to the potential at zero and finite temperature, respectively. $V_{ring}(h_{i},T)$ are the daisy corrections and $
 \delta V_{ct}(h_{i})$ are the zero temperature counter terms to the effective potential.
In general, a phase transition involves an important quantity that characterizes the transition between two phases and is called the critical temperature. In the context of FOPT, the critical temperature $T_{c}$ is determined by equating the potential values at the two vevs, corresponding to the high vev and the low vev, respectively, which is given by \cite{Patel:2011th}
\begin{equation}
\label{tc}
    V(h_{i}^{\rm High},T_{c})= V(h_{i}^{\rm Low},T_{c}) .
\end{equation}
The strong first-order phase transition will generate Gravitational waves with high amplitudes that can have a significant overlap with the sensitivity regions of upcoming GW detectors.
The condition for strong first-order phase transition is $\zeta_{c} \geq 1$, where the quantity $\zeta_{c}$ is called the order parameter and is defined as 
\begin{equation}
    \zeta_{c} = \frac{\Delta h_{i}}{T_{c}} ,
\end{equation} with $\Delta h_{i}$ is the difference between high and low vevs of the SM/BSM scalar field. 

It is important to note that the total effective potential given in Eq.\ref{vtot} depends on the choice of gauge explicitly. Thus, the order parameter $\zeta_{c}$, and the extremas of the effective potential are gauge dependent, which are the important ingredients required for the study of phase transition \cite{PhysRevD.9.3320,Laine_1995,Garny_2012,Espinosa_2017,Patel:2011th}.  In our case, all the finite temperature calculations are done in Landau gauge ($\xi^{'}$),\,\footnote{As the effective potential calculations depend on the gauge choice explicitly, a gauge-independent detailed analysis is beyond the scope of this paper. 
Further details can be found in \cite{NIELSEN1975173,PhysRevD.13.3469}. }   where $\xi^{'}$ is a gauge-fixing parameter.

We generate the results of the phase structures of the scalar fields using the publicly available CosmoTransition package \cite{Wainwright_2012}. In our analysis, we obtain two main phase transition patterns. We characterize them as Type I and Type II phase transitions. 
\begin{itemize}
\item {\bf Type I}: single-step, first-order phase transition. 
\item {\bf Type II}: two-step, the first step is first-order while the second step is second-order.

\end{itemize}

\subsection{Gravitational Wave Spectrum} \label{sec:-GW}

Cosmological phase transitions in the early Universe can give rise to stochastic Gravitational waves. The generation of such waves necessitates a first-order phase transition. These GWs originate from the release of energy of the colliding bubbles of the true vacuum as they propagate throughout the entire plasma. Such bubble formation can only take place in first-order phase transitions. FOPT can be analyzed by two main temperatures. They are the critical temperature $T_{c}$, which is defined in Eq.\ref{tc}, and the nucleation temperature $T_{n}$. The $T_{n}$  is  defined as the temperature that satisfies the  following condition given by
\cite{PhysRevD.45.3415},
\begin{equation}
    \frac{S_{3}(T_{n})}{T_{n}}=140 \label{nuctemp}~.
\end{equation}
where $S_{3}$ represents the 3 dimensional Euclidean Action and is given by \cite{1983544}:
\begin{equation}
\label{action}
    S_{3}=4\pi \int  r^2 dr\left[\frac{1}{2}\left(\frac{d h_{i}}{dr}\right)^2 + V_{\rm eff}(h_{i},T)\right]~,
\end{equation}
where $V_{\rm eff}(h_{i},T)$ is defined in Eq.\ref{vtot}.
The FOPT proceeds via bubble nucleation at $T_{n}$, which 
is, in general, slightly below $T_{c}$. During nucleation, the tunneling probability per unit volume from the false vacuum to the true vacuum at finite temperature $T$ \cite{Grojean_2007} can be calculated from the following expression:
\begin{equation}
    \Gamma(T) = T^4 \left(\frac{S_{3}}{2\pi T}\right)^{\frac{3}{2}} e^{-\frac{S_{3}}{T}}.
\end{equation}
The differential equation satisfied by the scalar fields $h_{i}$ where $i=\{1,2\}$ is obtained from extremizing the Euclidean Action of Eq.\ref{action} and thus given by \cite{1983544,LINDE1977306,PhysRevLett.46.388}
\begin{equation}
    \frac{d^2h_{i}}{dr^2} + \frac{2}{r}\frac{dh_{i}}{dr}=\frac{dV_{\rm eff}(h_{i},T)}{dh_{i}}~,
\end{equation}
with the boundary conditions $h_{i}=0$ as $r\rightarrow\infty$ and $\frac{dh_{i}}{dr}=0$ at $r=0$.

There are three main sources of the generation of the stochastic Gravitational waves, which are:
\begin{itemize}
\item The Bubbles of the true vacuum collide with each other, and the energy of the collision is propagated in the form of Gravitational Waves.
\item Sound waves are generated in the plasma when the bubbles are propagating through it.
\item Magnetohydrodynamic turbulence forming in the plasma after the collision of the bubbles.
\end{itemize}

In general, these three sources co-exist, and the total Gravitational Wave energy spectrum can be expressed as \cite{Caprini_2016,Ellis_2020}
\begin{equation}      
    \Omega_{\rm GW} h^2 \simeq \Omega_{\rm col} h^2+ \Omega_{\rm sw} h^2+ \Omega_{\rm turb}h^2 .
\end{equation}

The GW spectrum depends upon four important parameters. They are: \\ (i) $\alpha$: A quantity that is proportional to the latent heat corresponding to the phase transition and indicates the strength of the phase transition.\\ (ii) $\frac{\beta}{H_{n}}$: A quantity that is inversely proportional to the time taken for the phase transition to complete. \\(iii) $T_{n}$: The Nucleation temperature. \\(iv) $v_{w}$: Velocity of the Bubble Wall.

Among these parameters, $\alpha$ signifies the strength of the phase transition \cite{PhysRevD.49.2837}\begin{equation}
    \alpha = \frac{\epsilon(T_{n})}{\rho_{R}(T_{n})} ,
\end{equation}
where $\epsilon$ is expressed as \cite{Kehayias_2010}
\begin{equation}
    \epsilon(T_{n})= \Delta V_{\rm eff} - T\frac{d\Delta V_{\rm eff}}{dT}\bigg|_{T=T_{n}} .
\end{equation}
The $\Delta V_{\rm eff}$ is the difference between the effective potentials at false and true vacuum, and $\rho_{R}(T_{n})$ is the energy density of radiation given by,
    \begin{equation}
        \rho_{R}(T_{n})=\frac{\pi^2 g_{*}T_{n}^{4}}{30} ,
    \end{equation}
with $g_{*}$ representing the relativistic degrees of freedom at $T_{n}$. The parameter $\frac{\beta}{H_{n}}$ denotes the ratio of the inverse time duration of the phase transition to the Hubble parameter value at $T_{n}$  \cite{Nicolis_2004}
\begin{equation}
    \frac{\beta}{H_{n}}= T_{n}\frac{d(S_{3}/T)}{dT}\bigg|_{T=T_{n}} .
\end{equation}

The part of the GW spectrum resulting from bubble collisions, redshifted to today, can be expressed as  \cite{Jinno_2017}
\begin{equation}
    \Omega_{\rm col} h^2 = 1.67\times10^{-5}\left(\frac{\beta}{H_{n}}\right)^{-2}\left(\frac{\kappa_{\rm col}  \alpha}{1+\alpha}\right)^{2}\left(\frac{100}{g_{*}}\right)^{\frac{1}{3}}\left(\frac{0.11 v^{3}}{0.42 + v^{2}}\right)\left(\frac{3.8\left(\frac{f}{f_{\rm col}}\right)^{2.8}}{1+2.8\left(\frac{f}{f_{\rm col}}\right)^{3.8}}\right) ,
\end{equation}
where we have the peak frequency $f_{\rm col}$ red-shifted to today as\cite{Jinno_2017}
\begin{equation}
    f_{\rm col}=1.65\times10^{-5}\left(\frac{0.62}{1.8-0.1 v_{w}+v_{w}^{2}}\right)\left(\frac{\beta}{H_{n}}\right)\left(\frac{T_{n}}{100}\right)\left(\frac{g_{*}}{100}\right)^{1/6} .
\end{equation}
 The efficiency factor for the bubble collision is \cite{Borah:2023zsb},
 \begin{equation}
     \kappa_{\rm col}=\frac{0.715 \alpha +\frac{4}{27} \sqrt{\frac{3 \alpha }{2}}}{0.715 \alpha +1} .
 \end{equation}
As the bubble of true vacuum propagates through the plasma, it produces sound waves. The part of the GW spectrum resulting from these sound waves red-shifted to today \cite {Hindmarsh_2014,Hindmarsh_2018,Hindmarsh:2017gnf,Guo_2021},
\begin{equation}
     \Omega_{\rm sw} h^2 =2.65\times10^{-6} \Gamma_{\rm sw}\left(\frac{\beta}{H_{n}}\right)^{-1}v_{w}\left(\frac{\kappa_{\rm sw}  \alpha}{1+\alpha}\right)^{2}\left(\frac{g_{*}}{100}\right)^{\frac{1}{3}}\left(\frac{f}{f_{\rm sw}}\right)^{3}\left(\frac{4}{7}+\frac{3}{7}\left(\frac{f}{f_{\rm sw}}\right)^{2}\right)^{-\frac{7}{2}} ,
\end{equation}
with
\begin{align}
    \Gamma_{\rm sw}=\left(1-\frac{1}{\sqrt{1+2\tau_{\rm sw}H_{n}}}\right),\,\,
    \tau_{\rm sw}=\frac{(8\pi)^{\frac{1}{3}}}{\beta   ~U}\,\,,
    U=\sqrt{\frac{3}{4} \,\alpha \, \kappa_{\rm sw}} ,
\end{align}
where $\Gamma_{\rm sw}$ is the suppression factor arising from the finite lifetime of the sound waves, generated with $\tau_{\rm sw}$ being the lifetime of the sound waves, and $U$ is the root mean square velocity of the sound waves. 

The peak frequency of the sound waves redshifted to today \cite{Hindmarsh:2017gnf},
\begin{equation}
    f_{\rm sw}=1.9\times10^{-5}\left(\frac{1}{ v_{w}}\right)\left(\frac{\beta}{H_{n}}\right)\left(\frac{T_{n}}{100}\right)\left(\frac{g_{*}}{100}\right)^{1/6} .
\end{equation}

The efficiency factor corresponding to the contribution of the sound waves \cite{Borah:2023zsb},
\begin{equation}
    \kappa_{\rm sw}=\frac{\alpha }{\alpha +0.083 \sqrt{\alpha }+0.73} .
\end{equation}

The part of the GW spectrum resulting from magnetohydrodynamic turbulence generated within the ionized plasma red-shifted to today \cite{Caprini_2009},
\begin{equation}
    \Omega_{\rm turb} h^2 =3.35\times10^{-4} \left(\frac{\beta}{H_{n}}\right)^{-1}v_{w}\left(\frac{\kappa_{\rm turb}  \alpha}{1+\alpha}\right)^{\frac{3}{2}}\left(\frac{100}{g_{*}}\right)^{\frac{1}{3}}\left(\frac{\left(\frac{f}{f_{\rm turb}}\right)^{3}}{\left(1+\left(\frac{f}{f_{\rm turb}}\right)^{\frac{11}{3}}\right)\left(1+\frac{8\pi f}{h_{*}}\right)}\right)
\end{equation}
where $h_{*}$ is the inverse Hubble time during the production of Gravitational Waves,
\begin{equation}
    h_{*}= 16.5\times\left(\frac{T_{n}}{100}\right)\left(\frac{g_{*}}{100}\right)^{1/6} .
\end{equation}
The peak frequency due to turbulence generated in the ionized plasma due to the magnetic fields in the plasma, redshifted to today\cite{Caprini_2009},
\begin{equation}
    f_{\rm turb}=2.7\times10^{-5}\left(\frac{1}{ v_{w}}\right)\left(\frac{\beta}{H_{n}}\right)\left(\frac{T_{n}}{100}\right)\left(\frac{g_{*}}{100}\right)^{1/6} .
\end{equation}

The $\kappa_{\rm turb}$ represents the efficiency factor corresponding to the contribution of MHD turbulence and is generally given in terms of a small fraction of $\kappa_{\rm sw}$. We consider  $\kappa_{turb}= 0.1 \kappa_{sw}$ as suggested by simulations \cite{Borah:2023zsb}.

The above expressions of the efficiency factors $\kappa_{\rm col}$ and $\kappa_{\rm sw}$ are valid for relativistic bubble wall velocity $v_w$. We work in the limit where $v_{w}\rightarrow1$ \footnote{This choice of wall velocity corresponds to the so-called runaway regime, which arises in scenarios where the vacuum energy released during the phase transition is sufficient to continuously accelerate the bubble wall \cite{Espinosa:2010hh, Caprini:2019egz}. The detailed discussions on the conditions leading to $v_w \simeq1$, are shown in \cite{Athron:2023mer}.}.

To determine the detectability of any signal from the background, the most commonly used quantity is the signal-to-noise ratio (SNR), which is defined as \cite{Ellis_2020}
\begin{equation}
    \text{SNR} \equiv \sqrt{\tilde{T} \int_{f_{min}}^{f_{max}} \left[\frac{h^2 \,\Omega_{\text{GW}}(f)}{h^2 \,\Omega_{\text{Sens}}(f)} 
    \right]^{2} df}~~ .
    \label{SNR_expr}
\end{equation} 

 We have considered $\tilde{T}$ to be of 5 years' duration for all the relevant detectors. The $h^2 \,\Omega_{\text{Sens}}(f)$ corresponds to the experimental sensitivity of a given experimental configuration to cosmological sources obtained from the power spectral density (PSD) $S_{h}(f)$ \cite{Kuroda_2015}.

\section{Results and Analysis} \label{sec:-results}

\begin{figure}[htb!]
  \subfigure[]{
  \includegraphics[scale=0.3]{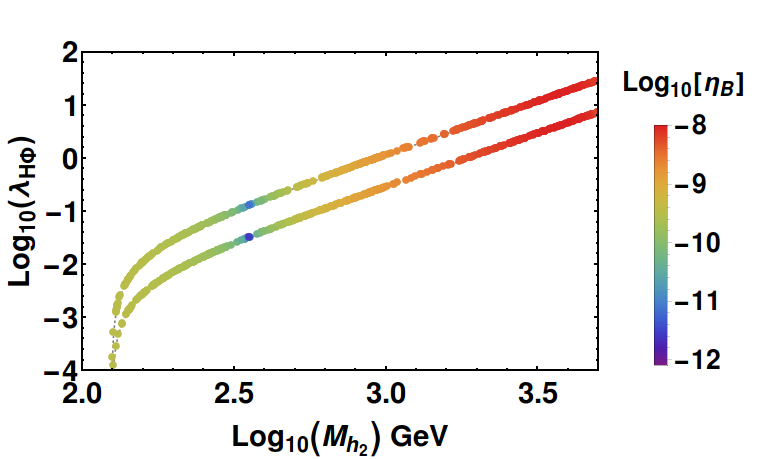}}
   \subfigure[]{
  \includegraphics[scale=0.35]{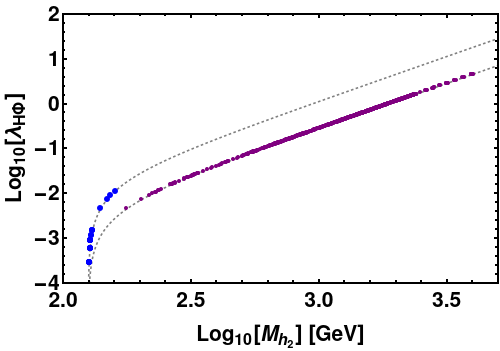}}
  \subfigure[]{
  \includegraphics[scale=0.3]{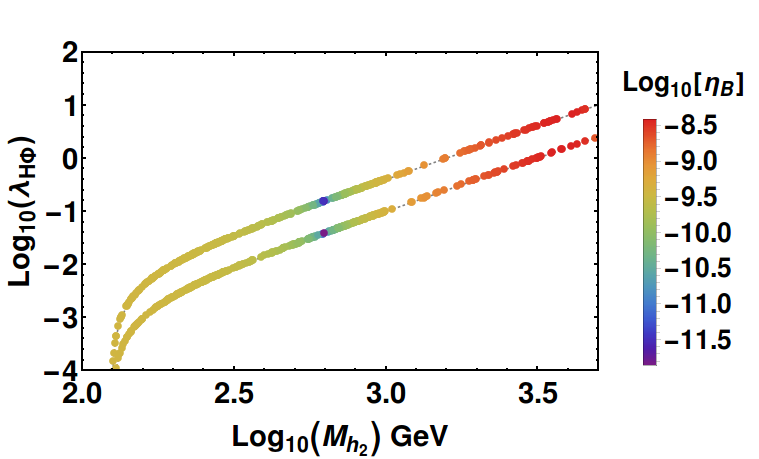}}
  \subfigure[]{
  \includegraphics[scale=0.35]{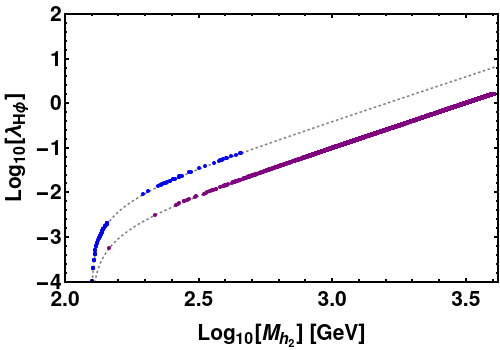}}
   \subfigure[]{
  \includegraphics[scale=0.3]{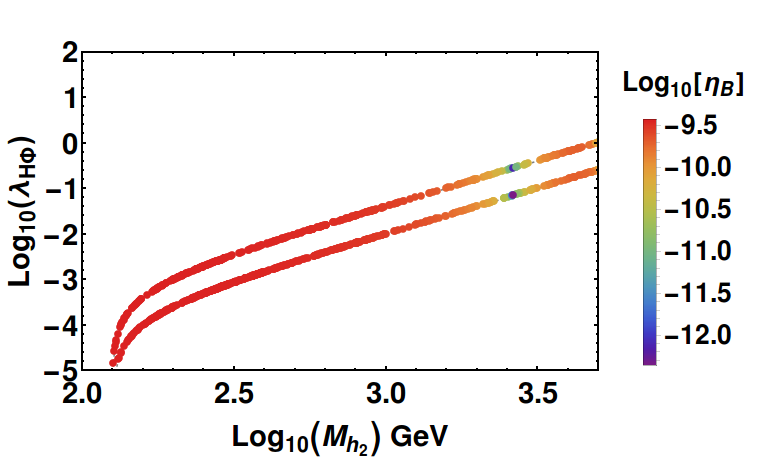}}
   \subfigure[]{
  \includegraphics[scale=0.35]{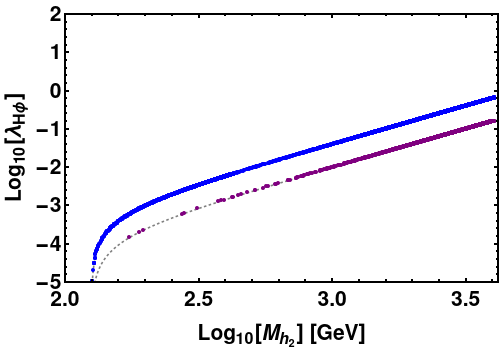}}
  
    \caption {\it The left column shows the variation of $\eta_B$, indicated by the color bar, in the $M_{h_2}$–$\lambda_{H \Phi}$ plane. The right column presents the allowed dark matter parameter space in the same plane, constrained by the observed relic density, direct detection limits (LZ 2024), and indirect detection bounds (FERMI-LAT + MAGIC).
 $M_{N_2}=5.5 \times 10^4$ {\rm GeV}, $M_{N_1}=5 \times 10^4$ {\rm GeV}. The top, middle, and bottom row panels correspond to $\sin\theta=0.3, 0.1, {\rm and \,} 0.01$, respectively. Each plot's top and bottom lines correspond to $v_\phi$=1000, 4000 {\rm GeV} respectively. The Grey dashed line represents the Eq.\ref{eq:lambda_para} for $\lambda_{H\Phi}$.}
\label{res_mh2_lmbd_etaB}
\end{figure}

In this section, we first identify a common parameter space that can explain both the observed BAU and 
the abundance of DM. Next, we examine the parameter space for Gravitational waves 
generated during the first-order phase transition. 
Here we present our findings in terms of the mass eigenstate $M_{h_2}$ as defined in Eq.\ref{mphi_mh2_expr} instead of $M_\phi$. 
As discussed earlier, at the Leptogenesis scale, the $\xi$ is a function of $\lambda_{H \phi}$ and $v_\phi$. Following scalar mixing, the coupling $\lambda_{H\Phi}$ can be rewritten in terms of low energy parameters $v_{\phi},\, \sin\theta$ and $M_{h_2}$ as
shown in Eq.\ref{eq:lambda_para}. We fix our choice for the $\sin\theta=\{0.3, 0.1, 0.01\}$ and for the $v_\phi=\{1000, 4000 \}$ GeV. 
Fig.\ref{res_mh2_lmbd_etaB} shows the common and compatible parameter space for the observed baryon asymmetry of the Universe (left column) and the allowed abundance of the DM (right column) in the plane of $M_{h_2}$ - $\lambda_{H\Phi}$. In the left panel, the
color bar represents the variation of $\eta_B$.

To illustrate the dynamics of the baryon asymmetry of the universe (BAU) and dark matter abundance, as well as their dependence on model parameters, we consider representative values of right-handed neutrinos masses $M_{N_2}=5.5 \times 10^{4}$ GeV and $M_{N_1}=5 \times 10^{4}$ GeV corresponding to $\delta M=0.1$ and $\alpha_{ij}=8 \times 10^{-3}$. Furthermore, the   
top, middle, and bottom panels correspond to $\sin\theta=0.3, 0.1, 0.01 $, respectively. In each plot, the top and bottom curves correspond to $v_\phi$=1000 GeV and 4000 GeV, respectively. 
In contrast to the behavior of $\eta_B$ with $M_\phi$
in Fig.\ref{mphi_beta}, it increases with increasing $M_{h_2}$, as 
$M_{h_2}$ is proportional to $\lambda_{H\Phi}$.
Consequently, an increase in the scalar self-coupling
$\lambda_{H\Phi}$ leads to an enhanced BAU.

In all three left panels of Fig.\ref{res_mh2_lmbd_etaB}, one can see that a drop
$(\lesssim {\cal O}(10^{-{11}}))$ occurs in baryon asymmetry (blue region) for certain values of $M_{h_2}$ and $\lambda_{H\Phi}$ corresponding to $v_\phi = 1000$ GeV and 4000 GeV, respectively. For these values, the 
vertex (being negative) contribution and self-energy contribution become equal. Initially, the self-energy contribution dominates over the vertex contribution, and as the $\lambda_{H\Phi}$ increases with the $M_{h_2}$, the vertex contribution dominates over the self-energy contribution, and when these parameters attain equal values, the $\eta_B$ drops. Additional details will be addressed later.
In each of the left three panels, the red regions of the curves indicate the model predicts 
the excessive baryon asymmetry $\eta_B$. 
Conversely, the greenish-yellow regions along each curve denote the values of $M_{h_2}$ and $\lambda_{H\Phi}$ that yield $\eta_B$ in agreement with the observed baryon 
asymmetry of the Universe.


\begin{table}[ht]
\begin{center}
\begin{tabular}{|c|c|c|c|c|c|c|c|c|c|}
\hline
Parameters  & $\sin\theta$ & $M_{h_{2}}$& $v_{\phi}$ & $\lambda_{H \Phi}$  & $\eta_B$ & $M_{\chi}$& $\Omega_{\chi}h^2 $& $ \sigma^{\rm SI}_{\rm DD}$  &  $\langle {\sigma v}\rangle $    \\
&&(GeV)&(GeV)&&&(GeV)&&($cm^{2}$)&($cm^{3}/s$)\\\hline 
1.   & 0.3 & 594.58  & 1000  & 0.3931 & $6.1 \times 10^{-10}$ &&&&  \\ 
\hline 
2.   & 0.3 & 594.18  & 4000  & 0.0981 & $6.1 \times 10^{-10}$&154&0.11&$1.63\times10^{-47}$&$2.03\times10^{-26}$ \\ 
\hline
3.   & 0.1 & 1128.31 & 1000  & 0.5086 & $6.1 \times 10^{-10}$&137&0.11&$2.65\times10^{-47}$&$2.04\times10^{-26}$   \\ 
\hline 
4.   & 0.1 & 1129.41 & 4000  & 0.1274 & $6.1 \times 10^{-10}$&873&0.11&$6.82\times10^{-47}$&$2.45\times10^{-26}$    \\
\hline
5.   & 0.01 & 801.98 & 1000  & 0.0255 & $3.1 \times 10^{-10}$&455&0.12&$2.91\times10^{-48}$&$2.51\times10^{-26}$   \\ 
\hline 
6.   & 0.01 & 802.02 & 4000  & 0.0064 & $3.1 \times 10^{-10}$&355&0.09&$1.1\times10^{-49}$&$1.76\times10^{-27}$  \\ 
\hline 
\end{tabular}
\end{center}
\caption{\it BAU-compatible parameter space consistent with relic density and current dark matter direct and indirect detection bounds. We omit the dark matter observables in the first row, as for the given $M_{h_2}$, no DM mass satisfies both the relic density and direct detection constraints. This is clear from Fig. \ref{res_mh2_lmbd_etaB}(b).}
\label{table_etaB}
\end{table}

The Blue and Purple color points that appear in the right column of Fig.\ref{res_mh2_lmbd_etaB} correspond to the regions of parameter space consistent with dark matter relic density, as well as current direct and indirect detection constraints for $v_{\phi}=1000 \,\,{\rm and \,}\, 4000$ GeV, respectively. We observe that the parameter space corresponding to DM phenomenology is consistent with the observed baryon asymmetry of the Universe for $\sin\theta=0.3, v_{\phi}=4000$ GeV (top panel), and for $\sin\theta=0.1, v_{\phi}=1000 \,\,{\rm and \,}\, 4000$ GeV (middle panel). In contrast, the baryon asymmetry produced insufficient for $\sin\theta=0.01$ (bottom panel). It is noteworthy that the observed BAU can be satisfied for the large value of $\sin\theta=0.3$ with two choices of $v_\phi$ (1000 $\rm{GeV}$ and 4000 $\rm{GeV}$), but all the DM phenomenology constraints is allowed only with the larger value of $v_\phi $ (4000 GeV). The dark matter phenomenology constraints may be permitted for small $\sin\theta=0.01$ with two $v_\phi$ choices, but the produced baryon asymmetry is insufficient. We observe that the parameter choices around $\sin\theta=0.1$ and $v_\phi $ (1000 and 4000 GeV) are the most permissible choices to see the common parameter space for the observed BAU along with all the dark matter phenomenology constraints. 
 The  points listed in Table \ref{table_etaB} are consistent with the observed dark matter relic density, satisfy current limits on the spin-independent direct detection cross-section, and simultaneously reproduce the observed baryon asymmetry of the universe (BAU).

\begin{figure}[ht]
  \subfigure[]{
  \includegraphics[scale=0.3]{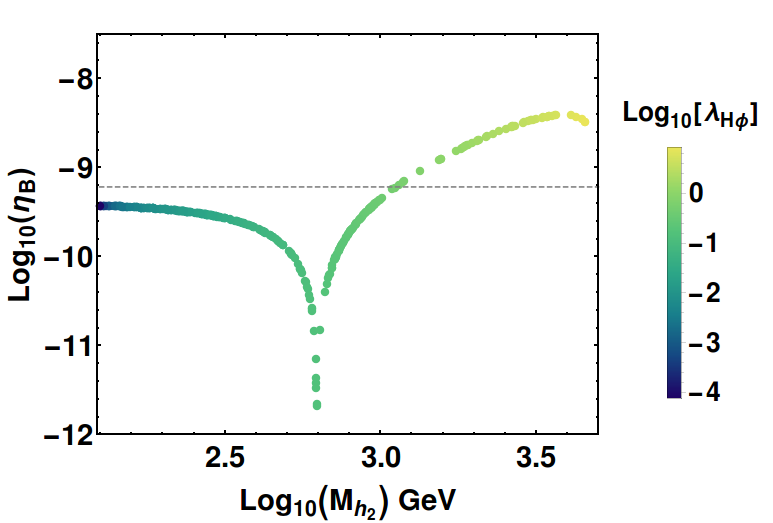}}
   \subfigure[]{
  \includegraphics[scale=0.3]{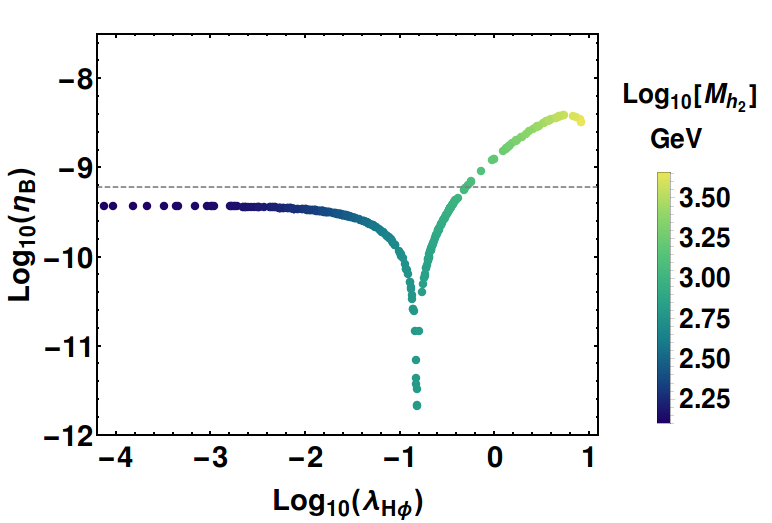}}
    \caption {\it Figure shows the drop in the baryon asymmetry in the plane of $M_{h_2}$, $\lambda_{H\Phi}$ with the $\eta_B$ for $\sin\theta=0.1$ and $v_{\phi}=1000$ GeV. The horizontal Grey line represents the observed BAU.}
\label{mh2_lmbd_etaB}
\end{figure}

As previously stated, the physical mechanisms responsible for the suppression of the baryon asymmetry of the universe (BAU) at specific values of $ M_{h_2} $ and $\lambda_{H\Phi}$ merit a detailed explanation, which we provide in Fig.\ref{mh2_lmbd_etaB}.
The left and right panels of Fig.\ref{mh2_lmbd_etaB} present the variation of the baryon asymmetry of the universe (BAU) as a function of the Higgs mass parameter $M_{h_2}$
and the scalar self-coupling $\lambda_{H\Phi}$, respectively. In the left panels, 
the color bar encodes the values of $\lambda_{H\Phi}$, while in the right panels, 
it represents the variation in the mass of the second scalar eigenstate $M_{h_2}$.
These plots correspond to a representative benchmark point in the parameter 
space, chosen with $\sin\theta=0.1 $ and $v_\phi = 1000 $ GeV, values which are consistent 
with both theoretical constraints and current experimental bounds.
The gray horizontal line in each panel indicates the observed value of the BAU, providing a direct visual comparison between the model predictions and 
experimental data. These figures are intended to offer further insight into the physical mechanism underlying the suppression of the BAU observed at 
certain parameter values in the previous figure (Fig.\ref{res_mh2_lmbd_etaB}). In particular, they illustrate how variations in $M_{h_2}$ and $\lambda_{H\Phi}$ affect the model predicted
BAU by modifying the inherent dynamics. Through these plots, we aim to clarify 
why certain regions of the parameter space, especially those corresponding to $v_\phi  =1000$ GeV and 4000 GeV yield BAU values significantly below the observed 
level, typically by many orders of magnitude.
The regions shaded in color, from light green to yellow in both panels of Fig.\ref{mh2_lmbd_etaB} represent areas of the parameter space where the model predicts an enhanced baryon asymmetry of the universe (BAU). Specifically, the portions of these colored patches that lie above the gray horizontal line correspond to parameter combinations for which the predicted BAU exceeds the observed value. 
This behavior is evident in both $M_{h_2}$ and $\lambda_{H\Phi}$-dependence plots, and reflects the sensitivity of the generated BAU to variations in the scalar sector parameters.

After successfully explaining the model parameter space that is suitable for the study of the DM phenomenology and the leptogenesis, we now investigate whether the model can support SFOPT within a region of parameter space that is simultaneously compatible with the above parameter space. 
For this study, we perform an independent random scan over the following parameter ranges: $M_{h_2}=\{500~{\rm GeV}-2000~{\rm GeV}\}, \, |\sin\theta|=\{0.1-0.3 \}, \,v_{\phi}=\{100~{\rm GeV}-1000~{\rm GeV} \}, \, M_{\chi}=\{100~{\rm GeV}-1000~{\rm GeV} \}$ using the CosmoTransition package. From the scan results, we get a few points that exhibit an SFOPT along the SM Higgs direction, while a slightly larger number of points show SFOPT along the BSM scalar direction. 
A substantial scalar self-coupling is necessary to initiate the SFOPT process, which, on the other hand, results in a very large DM direct detection cross-section that exceeds the current LZ limit. 
{We identify seven  benchmark points that exhibit SFOPT and list them in Table \ref{table_BP}.
We then evaluate the compatibility of these points with the observed BAU and dark matter phenomenology. In Table \ref{table_BP}, we show the model parameters 
associated with SFOPT for these benchmark points. Additionally, in 
Table \ref{table_BP}, we provide the dark matter relic density 
($ \Omega_{\chi}h^2 $), the spin-independent direct detection 
cross-section ($ \sigma^{\rm SI}_{\rm DD} $), the thermal averaged 
annihilation cross-section ($ \langle \sigma v \rangle^{\rm ID} $), and 
the baryon asymmetry parameter ($ \eta_B $) for four of these benchmark 
points, which we discuss in detail later in this section. 
In Table~\ref{table_BP}, we use the symbols \checkmark and \xmark ~ to indicate whether a given benchmark point satisfies a particular criterion. Specifically, a \checkmark indicates that the condition strong first-order phase transition (SFOPT), dark matter phenomenology (relic abundance, direct and indirect detection), or successful baryogenesis is fulfilled, while a \xmark~ indicates that it is not.


 In Table \ref{table_PT_op}, we present the output parameters of SFOPT corresponding to the first four benchmark points of Table \ref{table_BP}. 
It is worth noting that benchmark points BP3 - BP7 exhibit significant similarities in several key parameters, resulting in comparable behavior in the context of our study. To streamline our numerical analysis and focus on representative cases, we select BP3 and  BP4 among these points as the primary points for detailed investigation. This selection allows us to effectively capture the essential features and trends of the SFOPT results.}

The $(<h_1>,<h_2>)$ correspond to the vacuum expectation values for the SM Higgs and BSM Higgs direction, respectively. 
{For the benchmark points BP1 and BP2 the order parameter $\zeta_c > 1 (< 1)$ along the SM (BSM) Higgs direction respectively. On the other hand, for the BP3 and BP4, $\zeta_c > 1 (=0)$ along the BSM (SM) Higgs direction, respectively, as there is no FOPT along the SM Higgs direction. Thus, based on the behaviour of $\zeta_c$, we can infer that BP1 and BP2 exhibit a strong first-order phase transition (SFOPT) primarily along the direction of the Standard Model (SM) Higgs field. In contrast, BP3 and BP4 demonstrate SFOPT predominantly along the direction of the Beyond Standard Model (BSM) Higgs field. }

\begin{figure}[ht]
  \subfigure[]{
  \includegraphics[scale=0.4]{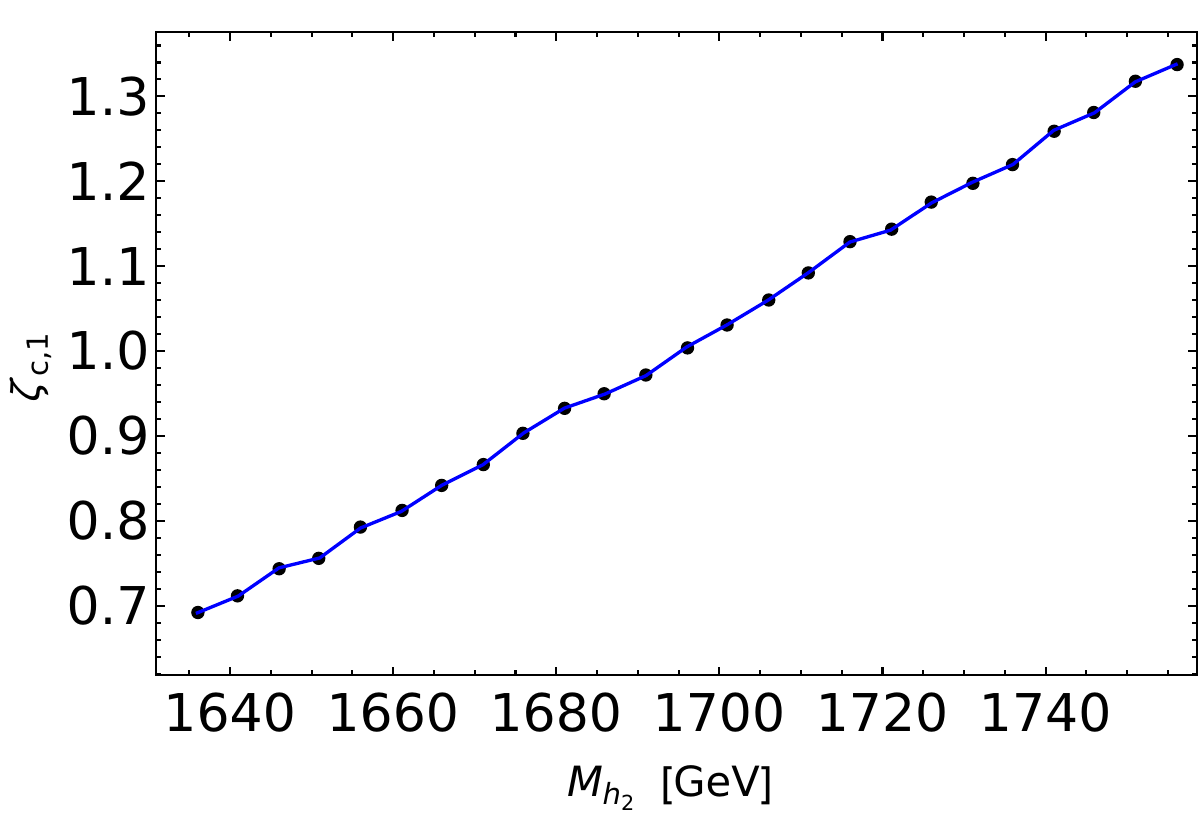}}
   \subfigure[]{
  \includegraphics[scale=0.4]{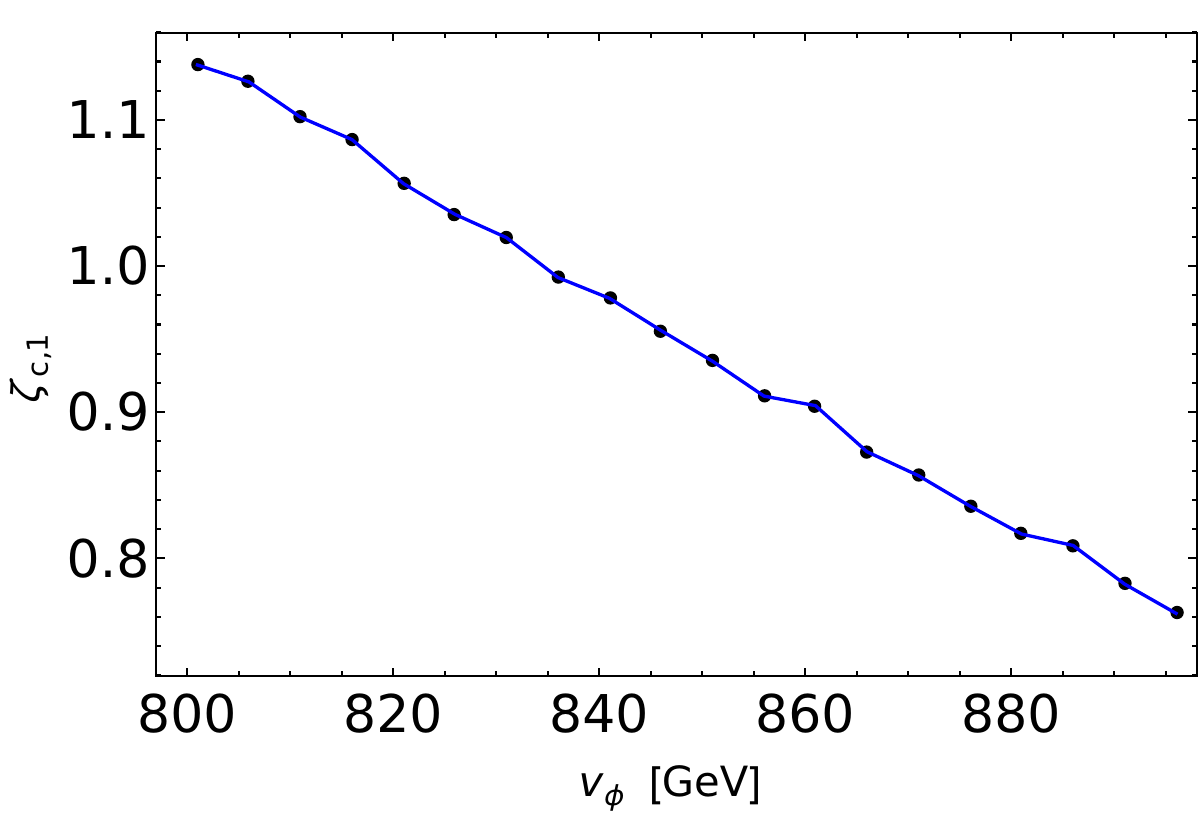}}
    \caption {\it The role of a few model parameters in determining the Phase Transition strength along the SM Higgs field direction. We utilize BP1 for this purpose. Dots represent the exact value of the order parameter obtained at the respective input variable.}
\label{paraPT}
\end{figure}

The order parameter $\zeta_c$, which quantifies the strength of the electroweak phase transition (EWPT), is strongly sensitive to the Higgs-portal coupling $\lambda_{H\Phi}$—the interaction strength between the Standard Model Higgs doublet $H$ and the additional complex scalar field $\Phi$ introduced in the model. Notably, $\lambda_{H\Phi}$ varies proportionally with the square of the mass of the BSM scalar $M_{h_2}$ and inversely with the singlet VEV $v_{\phi}$, assuming all other parameters are held fixed. As a result, both $M_{h_2}$ and $v_{\phi}$ can have a significant impact on the value of $\zeta_c$. To illustrate this sensitivity, in Fig.~\ref{paraPT}, we show how $\zeta_c$ changes as a function of $M_{h_2}$ and $v_{\phi}$, respectively, while keeping all other parameters constant in each case for BP1, where the strong first-order phase transition (SFOPT) occurs along the Standard Model Higgs direction.

 \begin{table}[ht]
\begin{center}
\begin{tabular}{|c|c|c|c|c|c|c|c|c|}
\hline
Parameters  & BP1 & BP2 & BP3 & BP4 & BP5&BP6&BP7  \\
\hline
$M_{h_{2}}$ (GeV)  & 1712.11 &1542.39 &398.11&400.16&401.08&398.45&395.26 \\\hline
$\sin\theta$    & -0.34& -0.28 & 0.087&0.082&0.080&0.081&0.080\\
\hline
$v_{\phi}$ (GeV)  &859.89&493.67&551.55&547.36&548.48&544.15&549.05\\\hline
$M_{\chi}$ (GeV) &144.28&670.33&879.77&880.87&876.36&877.43&880.5\\\hline
$\lambda_{H}$&2.915&1.735&0.138&0.137&0.137&0.137&0.136\\\hline
$\lambda_{H \Phi}$&-4.408&-5.347&-0.092&0.088&0.086&0.086&0.083\\\hline
$\lambda_{\Phi}$&1.758&4.789&0.682&0.697&0.692&0.699&0.686\\\hline
$\mu_{3}$ (GeV) &-7.608&-286.054&-441.01&-445.51&-440.05&-444.64&-443.75\\\hline
$\Omega_{\chi}h^2 (\times10^{-2})$   &  $1.02$& $0.00226$   &  $1.13$      & $1.09$ &$1.12$      & $1.12$&$1.13$       \\ \hline 
$ \sigma^{\rm SI}_{\rm DD}$ ($\times10^{-46} cm^2$)    &  $0.348 $ & $0.0359 $  &$2.125 $       &   $1.86 $ &$1.79$  &$1.82 $       &   $1.79 $      \\ \hline 
$ \langle {\sigma v}\rangle^{\rm ID}$ ($\times10^{-27}cm^3$/s)    &  $64$ &$0.005$  &$2.35 $       &   $2.27 $  &$2.37 $  &$2.36$       &   $2.33 $     \\ \hline 

$\eta_B (\times 10^{-10})$  &  $56.6$ & $39.5$ & $2.29$ &  $2.36$ &$2.36$ & $2.29$ & $2.28$   \\ \hline 
$SFOPT$            &  \checkmark & \checkmark & \checkmark & \checkmark & \checkmark & \checkmark & \checkmark \\ \hline
$DM~constraints$   &  \xmark     & \checkmark & \xmark     & \xmark     & \xmark     & \xmark     & \xmark     \\ \hline
$BAU$              &  \xmark     & \xmark     & \checkmark & \checkmark & \checkmark & \checkmark & \checkmark \\
\hline
\end{tabular}
\end{center}
\caption{\it The model parameters that give FOPT corresponding to the benchmark points. The \checkmark and \xmark ~
 represents the possibility and the non-possibility, respectively, of satisfying the criterion of SFOPT, DM constraints, and BAU.}
\label{table_BP}
\end{table}

\begin{table}[ht]
\begin{center}
\begin{tabular}{|c|c|c|c|c|c|}
\hline
Parameters  & BP1& BP2 & BP3 & BP4 \\\hline
$T_{c}$   &106.9  &106.5  &430.27&489.94    \\\hline
$(<h_1>,<h_2>)|_{\rm High~ 
 T_{c}}$     & (0, 813.88)& (0, 450.21)  & (0, 0)&(0, 0) \\
\hline
$(<h_1>,<h_2>)|_{\rm Low~ 
 T_{c}}$    &(115, 824.3)& (111.3, 452.3) &(0, 494.76) &(0, 499.92) \\\hline
  $\zeta_{c}$&(1.075, 0.097)&(1.045, 0.019) &(0, 1.149)&(0, 1.02)\\\hline
$T_{n}$&106.47&106.06 &359.09 &432.8 \\\hline
$(<h_1>,<h_2>)|_{\rm High~ 
 T_{n}}$    &(0, 813.74)&(0,453.98)&(0, 0)&(0, 0)\\\hline
$(<h_1>,<h_2>)|_{\rm Low~ 
 T_{n}}$   &(125.43, 826.19)&(114.91,463)&(0, 515.93)&(0, 519.04)\\\hline
$\alpha$&0.0122&0.01&0.0069&0.0045\\\hline $\frac{\beta}{H}$&71452.8 &146712&1595.25 & 2602.53\\\hline
\end{tabular}
\end{center}
\caption{\it Phase transition output parameters for SM and BSM Higgs directions corresponding to the first four BPs. The values of high and low vevs $(<h_1>,<h_2>)$ of each of the scalar fields are given at all steps for all the benchmark points for both $T_{c}$ and $T_{n}$.  Temperature and vevs are in the {\rm GeV} unit. }
\label{table_PT_op}
\end{table}

\begin{figure}[ht]
  \subfigure[]{
  \includegraphics[scale=0.4]{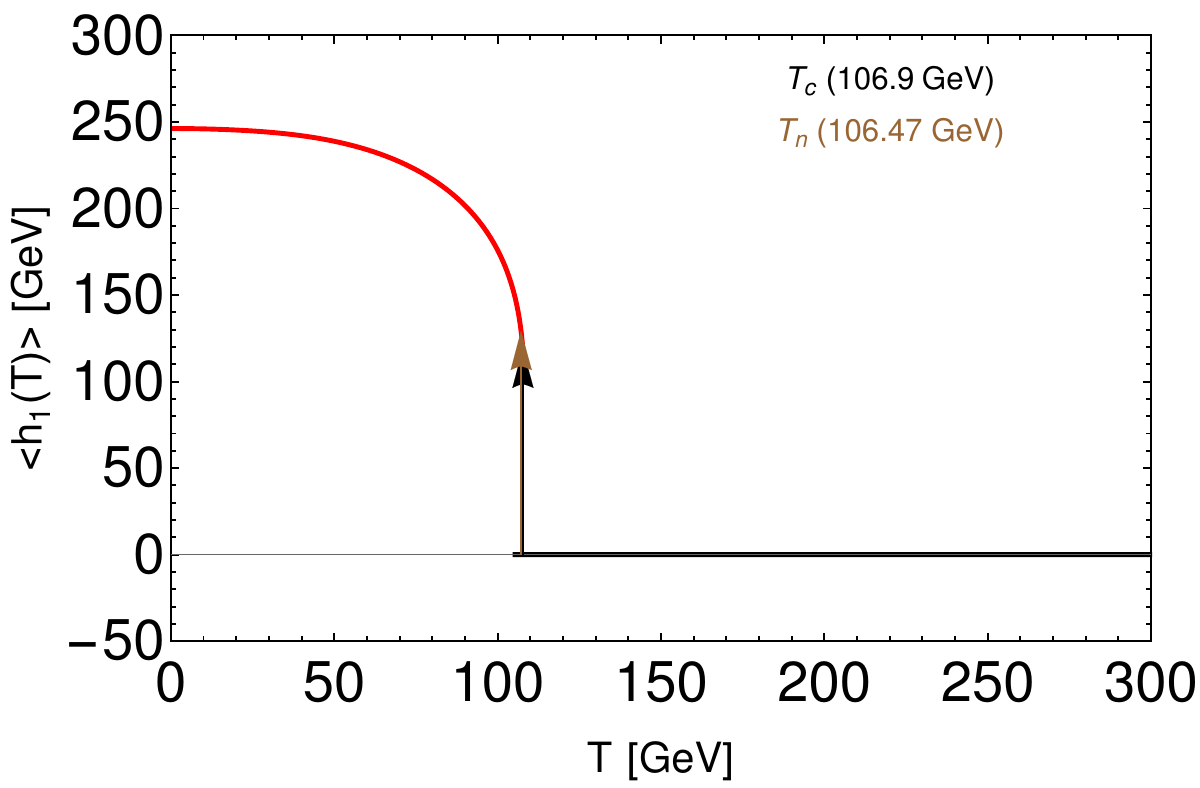}}
   \subfigure[]{
  \includegraphics[scale=0.4]{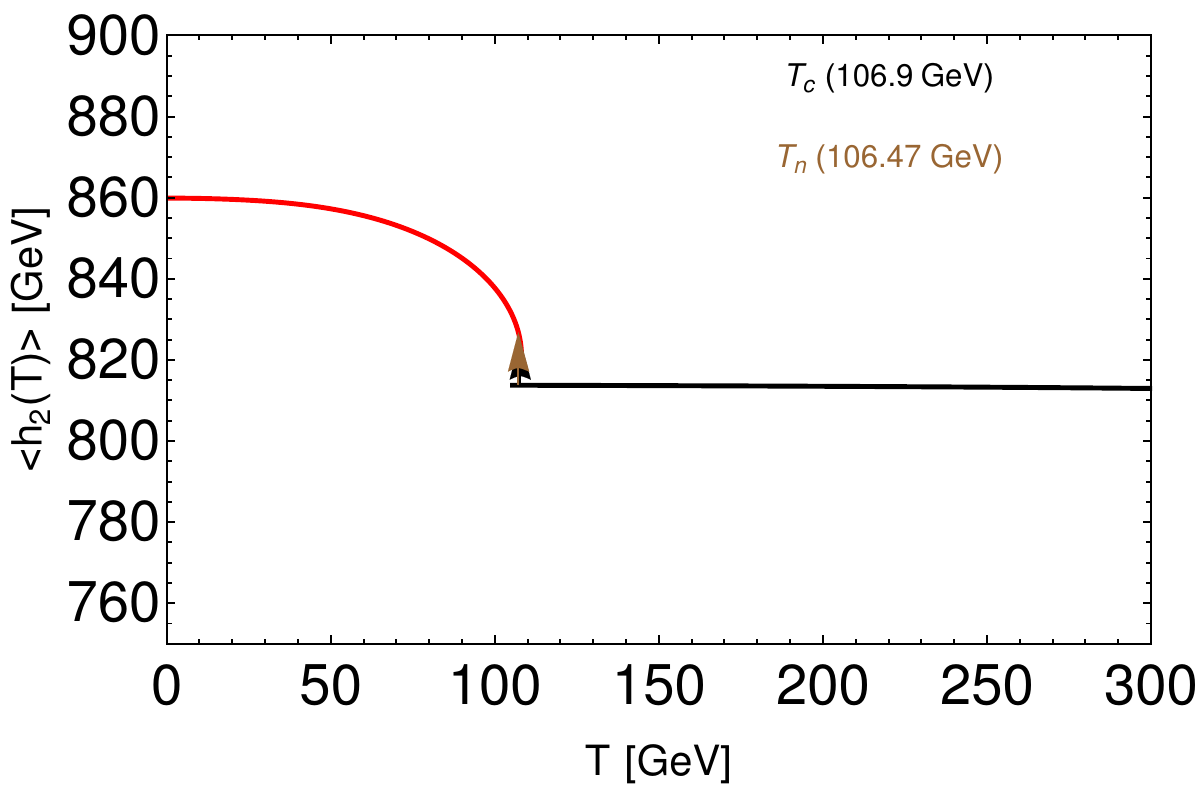}}
  \subfigure[]{
  \includegraphics[scale=0.4]{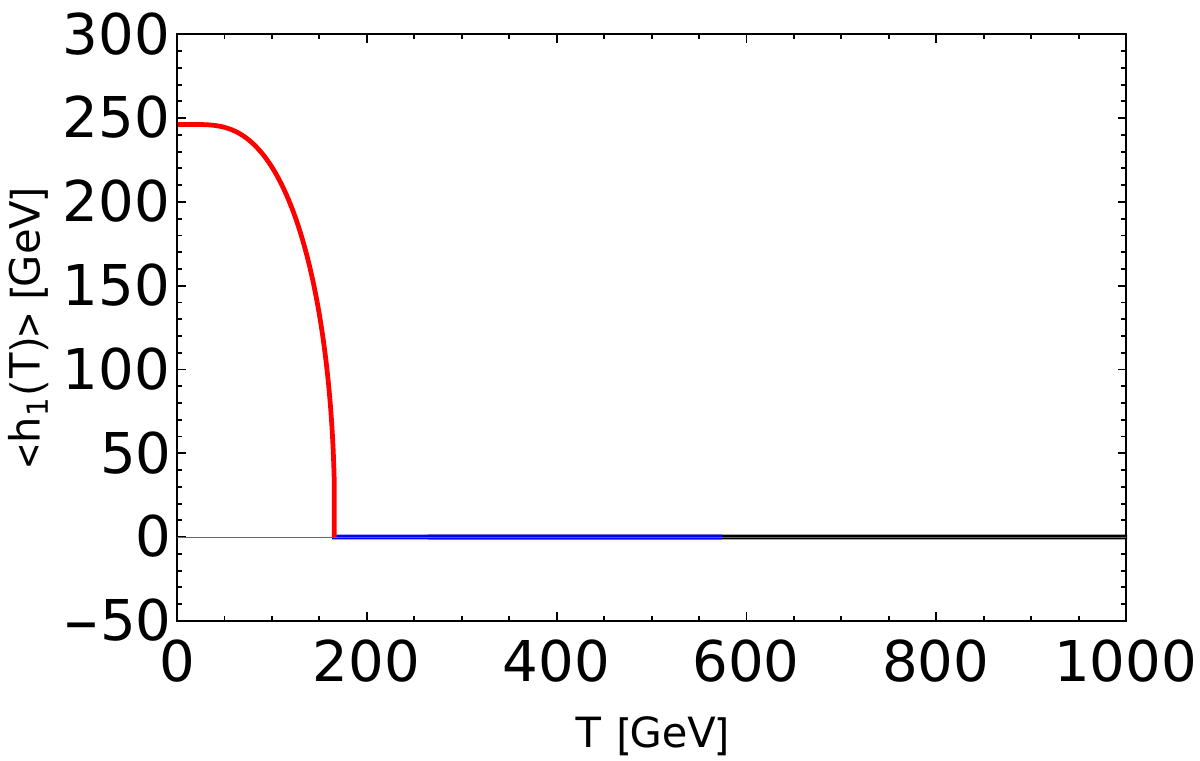}}
   \subfigure[]{
  \includegraphics[scale=0.4]{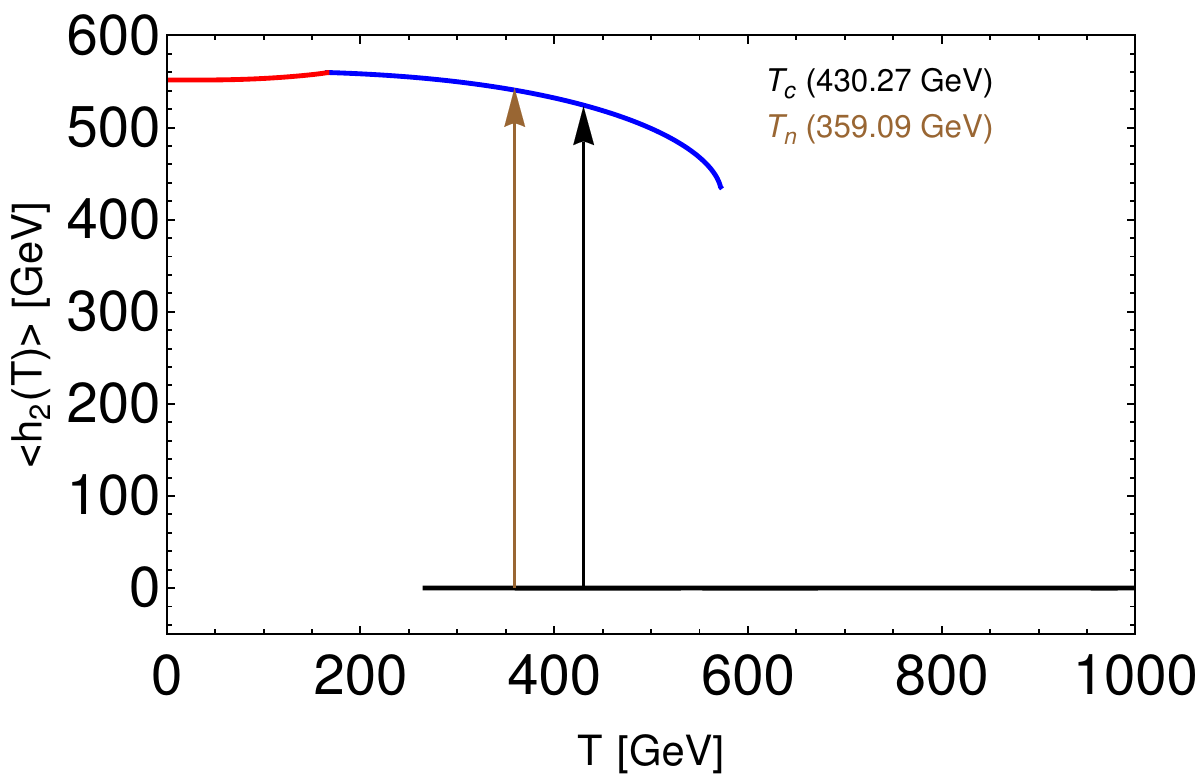}}
    \caption {\it Top panels: (a), (b) represent the phase structure of the fields $h_{1,2}(T)$ corresponding to BP1. Bottom panels: (c), (d) represent the phase structure of the fields $h_{1,2}(T)$ corresponding to BP3. } \label{figPT}
\end{figure}

 We notice that BP1 and BP2 fall in the Type I category, while BP3 and BP4 fall in the Type II category. In Fig.\ref{figPT}, we represented the phase structure of the scalar fields $h_{1,2}(T)$ as a function of the temperature. Here, we have shown Type I and Type II phase transitions, considering BP1 and BP3 as examples.  Different colors represent different phase transitions. Color change with (without) an arrow indicates the possibility of a first (second) order phase transition. The black arrow corresponds to the critical temperature, and the brown arrow corresponds to the nucleation temperature. 

In Fig.\ref{figPT} (a) and (b), the phase structure of SM Higgs and BSM Higgs fields, respectively, corresponding to BP1, is shown as an example of a Type I phase transition. From Fig.\ref{figPT}, we see that there is a single step first order phase transition at the critical temperature $T_{c}$=106.9 GeV along both $h_{1}$ and $h_{2}$ field directions. The transition is strong along SM Higgs direction with the order parameter $\zeta_{c,1}$ = 1.075, and the transition is weak along BSM Higgs direction with the order parameter $\zeta_{c,2}$ = 0.097.  In Fig.\ref{figPT} (c) and (d), the phase structure of SM Higgs and BSM Higgs fields, respectively, corresponding to BP3, is shown as an example of a Type II phase transition. A type II phase transition is a two-step phase transition. In the first step, there is a phase transition along BSM Higgs field direction at the critical temperature $T_{c}$ = 430.27 GeV.   But there is no phase transition along the SM Higgs direction. In the second step, there is a second order phase transition in both $h_{1}$ and $h_{2}$ field directions.

{In Table \ref{table_BP}, we observe that for benchmark points BP1 and BP2, the model predicted baryon asymmetry is excessively large, and neighbourhood scans
of those benchmark points also yield the 
baryon asymmetry that significantly exceeds the observed 
limit. For benchmark points BP3 to BP7, the baryon asymmetry aligns with the observed BAU, and their neighborhood scans can reproduce the observed BAU. 
All benchmark points show under-abundant relic density, but only BP2 satisfies both the direct and indirect DM  detection constraints.}  

\begin{figure}[h]
    \centering
    \includegraphics[width=13cm]{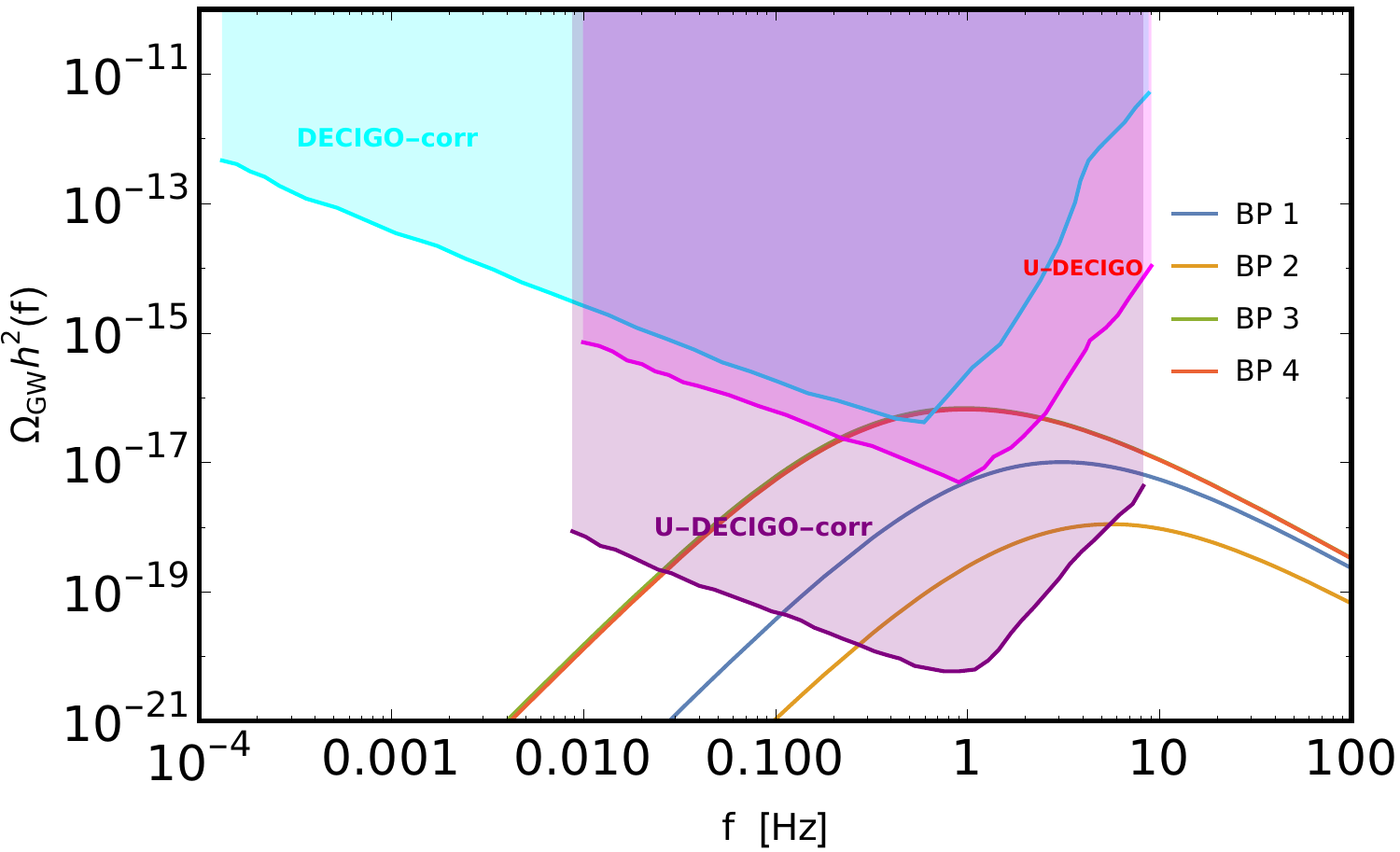}
    \caption{\it Gravitational Wave Spectrum corresponding to the BPs 1-4.}
    \label{GW}
\end{figure}
\begin{table}[ht]
    \centering
    \begin{tabular}{|c|c|c|c|}\hline
         BP&DECIGO-corr&U-DECIGO&U-DECIGO-corr  \\\hline
          1&-&-&$9.97\times10^{6}$\\\hline
          2&-&-&528234\\\hline
          3&10945.8&132896&$1.32\times10^{8}$\\\hline
          4&10287.2&126854&$1.26\times10^{8}$\\\hline
    \end{tabular}
    \caption{\it SNR values corresponding to the first four Benchmark Points.
    The dashed lines indicate that the corresponding detector will not detect the Gravitational wave spectrum associated with the benchmark point.}
    \label{SNRtab}
\end{table}

SFOPT provides a background for the formation of stochastic gravitational waves. The GW spectrum corresponding to the benchmark points is given in Fig.\ref{GW}, where the proposed sensitivities of the upcoming detectors DECIGO-corr, U-DECIGO, U-DECIGO-corr are depicted \cite{PhysRevD.73.064006,Yagi_2011}.

An important measure to detect the GW signal from its background is the SNR. We estimated the SNR values corresponding to all the BPs in Table.\ref{SNRtab} using Eq.\ref{SNR_expr}. For the detection of GW in each relevant detector, the SNR value must exceed a threshold value for a particular setup, which we considered SNR $\geq$ 10 in our analysis.  BP1 and BP2 SNRs in Table.\ref{SNRtab} are less than the threshold value of DECIGO-corr and U-DECIGO detectors, while for BP1, the SNR value is higher than that corresponding to BP2 for U-DECIGO-corr. GW Spectrum corresponding to BP3 and BP4 can be detected in the DECIGO and U-DECIGO detectors.  SNR values of BP3 are slightly higher than those of BP4, as from the Table \ref{table_PT_op}, we can see that the order parameter $\zeta_c$ of BP3 is always higher than that of BP4, resulting in greater signal strength. 
Thus, we verify that, higher the order parameter value in a particular field direction, the higher is the SNR of the GW signal corresponding to a particular detector. 

{Finally, we observe that among the seven benchmark points (BPs) listed in Table~\ref{table_BP}, all of which exhibit a strong first-order phase transition (SFOPT), only BP2 is consistent with the latest dark matter (DM) direct detection constraints. All points except BP1 and BP2 can account for the observed baryon asymmetry of the universe (BAU) within an order of magnitude. However, the majority of the points that exhibit SFOPT fail to satisfy the DM direct detection bounds, even in regions where the BAU is successfully explained. We did not find any point that simultaneously satisfies the requirements of SFOPT, BAU generation, and all DM phenomenology constraints.}

\section{Summary and Conclusion}\label{sec:-summary}
We considered an interesting scenario in which the Standard Model particle content is extended with a singlet scalar to probe leptogenesis at a low scale. The singlet scalar couples with the right-handed neutrino pair, which enhances additional CP violation while achieving the observed BAU at a low scale. 
Here, we attempted to examine the allowed parameter space of the relic density of DM without adding an extra particle content, but with the simple extension of the SM symmetry. We have shown a common parameter space that provides a low-scale Leptogenesis, which is consistent with low-energy neutrino data through the Casas-Ibarra parametrization. Furthermore, this parameter space can explain the observed relic density of DM while satisfying constraints from direct and indirect DM searches like  XENON-1T, LZ 2022, Fermi-LAT, and MAGIC {\it, etc}.
Our final observations are displayed in the plane of $\lambda_{H\Phi}-M_{h_2}$. For the parameter choice, we consider indicating that the $\sin\theta=0.1$ and $v_\phi=1000, 4000$ GeV are more appropriate parameter regions to look for the observed BAU and DM simultaneously. 

{
After identifying the allowed regions of parameter space consistent with the DM phenomenology and BAU through low-scale leptogenesis, we attempted to investigate whether one could also achieve the SFOPT in the same parameter space. In our analysis, we found seven benchmark points that exhibited the SFOPT. However, only one of them (BP2) satisfied all the
relevant constraints associated with the dark matter phenomenology, particularly, the latest LZ bound on DM direct detection appeared to 
be the most stringent one. The observed BAU can be obtained for all 
benchmark points except for BP1 and BP2.

Our detailed analysis showed that the SFOPT along the SM-like Higgs direction typically failed to satisfy the baryon asymmetry condition.
In contrast, the SFOPT along the BSM Higgs field direction could 
potentially provide the correct BAU. However, the corresponding
benchmark points exhibit an under-abundant dark matter relic density 
and marginally defy the latest direct detection constraints from the LZ collaboration. Furthermore, we studied the stochastic gravitational wave background generated by these phase transitions, focusing on the first four representative benchmark points. The resulting spectra lie within the sensitivity reach of future detectors such as DECIGO-corr, U-DECIGO, and U-DECIGO-corr.

Our results suggest that while the model can realize a successful electroweak phase transition, baryogenesis, and dark matter individually, achieving full compatibility among all three remains challenging. Notably, it satisfies both the BAU and DM constraints simultaneously. A more exhaustive scan of the parameter space may reveal regions where all these phenomena can simultaneously be satisfied.}

\vspace{5mm}
\noindent \textbf{Acknowledgments}

KM would like to thank Nandini Das for valuable discussions and acknowledges 
the financial support provided by the Indian Association for the Cultivation of Science (IACS), Kolkata. PG gratefully acknowledges the support from IACS, Kolkata, India, where most of this work was carried out.

\section{Appendices}\label{sec:-Appedix}
\subsection{Vertex factors}

The following are the vertex factors for vertices relevant to the calculation of the Relic density of DM given in Figs.\ref{Feyn_diag1}, \ref{Feyn_diag2},
\begin{eqnarray}
    h_1 \chi \chi &:& -\frac{3 \mu_3 \sin\theta }{\sqrt{2}}-\lambda_{H\Phi} v \cos\theta +2 \lambda_\Phi v_\phi \sin\theta ~\approx ~ \frac{\sin\theta  \left(M_{h_1}^2+ \frac{4}{3}M_\chi^2\right)}{v_\phi}\,,  \nonumber \\
    h_2 \chi \chi &:& \frac{3 \mu_3 \cos\theta}{\sqrt{2}}-\lambda_{H\Phi} v \sin\theta-2 \lambda_\Phi v_\phi \cos\theta ~\approx ~ -\frac{ \cos\theta \left(M_{h_2}^2+\frac{4}{3}M_\chi^2\right) }{v_\phi} \,, \nonumber \\
    h_1 h_1 \chi \chi &:&  -2 \lambda_\Phi \sin^2\theta-\lambda_{H\Phi} \cos^2\theta ~ \approx ~ -\frac{(M_{h_2}^2-M_{h_1}^2)\sin\theta}{v \, v_\phi} \,, \nonumber  \\
    h_1 h_2 \chi \chi &:& 2 \lambda_\Phi \sin\theta  \cos\theta -\lambda_{H\Phi} \sin\theta  \cos\theta \nonumber  \\
    & \approx & \frac{\sin\theta \left(3 v_\phi \sin\theta  \left( M_{h_1}^2-M_{h_2}^2 \right) + v \left(3 M_{h_2}^2+M_\chi^2 \right)\right)}{3 v \, v_\phi^2} \,, \nonumber \\
    h_2 h_2 \chi \chi &:& \frac{1}{2} \left(-2 \lambda_\Phi \cos^2\theta -\lambda_{H\Phi} \sin^2\theta \right) ~ \approx ~ -\frac{3 M_{h_2}^2+ M_\chi^2}{3 v_\phi^2} \,.
\end{eqnarray}
\subsection{Loop functions}\label{App_loop}
Following are the loop function parameters discussed in Sec.\ref{sec:-Leptogenesis} for the CP-asymmetry diagrams\footnote{Note that before EWSB, the zero temperature SM Higgs mass is considered massless.},
\begin{eqnarray}
F_{i j, R}^{(v)} &=& \sqrt{r_{j i}} \ln \left[\frac{\left(1-r_{j i}\right)-\left(\sigma_i+\sqrt{\delta_{j i}}\right)}{\left(1-r_{j i}\right)-\left(\sigma_i-\sqrt{\delta_{j i}}\right)}\right], \\
F_{i j, L}^{(v)} &=& -\sqrt{\delta_{j i}}+r_{j i} \, \ln \left[\frac{\left(1-r_{j i}\right)-\left(\sigma_i+\sqrt{\delta_{j i}}\right)}{\left(1-r_{j i}\right)-\left(\sigma_i-\sqrt{\delta_{j i}}\right)}\right],\\
 F_{i j k, R R}^{(\mathrm{s})} &=& \frac{\sqrt{r_{j i}} \sqrt{r_{k i}} \sqrt{\delta_{j i}}}{1-r_{j i}}~, ~~~~~
 F_{i j k, R L}^{(\mathrm{s})} = \frac{1}{2} \frac{\sqrt{r_{k i}} \sqrt{\delta_{j i}}\left(1+r_{j i}-\sigma_i\right)}{1-r_{j i}} \, , \\
F_{i j k, L L}^{(\mathrm{s})} &=& \frac{\sqrt{r_{j i}} \sqrt{\delta_{j i}}}{1-r_{j i}} ~,~~~~~~~~~  F_{i j k, L R}^{(\mathrm{s})}=\frac{1}{2} \frac{\sqrt{\delta_{j i}}\left(1+r_{j i}-\sigma_i\right)}{1-r_{j i}} \, .  
\end{eqnarray}

Where $ r_{i j} \equiv M_i^2 / M_j^2, \sigma_i \equiv M_\Phi^2 / M_i^2$ and $\delta_{i j} \equiv \left(1-r_{i j}-\sigma_j\right)^2 - 4 r_{i j} \sigma_j$.
\subsection{Zero Temperature Counter Terms}
\label{cterms}
The following are the expressions of all the counter terms corresponding to each parameter that appears in the tree-level potential:
\begin{align}
  &    \delta\lambda_{H}=\frac{1}{2v^3}(\partial_{h}\Delta V - v \partial^2_{h}\Delta V),\,\,\,
    \delta\lambda_{\Phi}=\frac{1}{2v_{\phi}^3}(\partial_{\phi}\Delta V - v_{\phi}\partial_{\phi}^2 \Delta V),\nonumber\\
  & \qquad \qquad 
    \delta\lambda_{H\Phi}=-\frac{\partial_{h}\partial_{\phi}\Delta V}{v v_{\phi}},\,\,\,  \delta\mu_{3}=0,\nonumber\\
  &  \qquad \delta\mu_{H}^2=\frac{1}{2v }(3\partial_{h}\Delta V  - v_{\phi}\partial_{h}\partial_{\phi}\Delta V - v \partial_{h}^2 \Delta V),\nonumber\\
  & \qquad \delta\mu_{\Phi}^2=\frac{1}{2v_{\phi}}(3\partial_{\phi}\Delta V - v \partial_{\phi}\partial_{h}\Delta V - v_{\phi}\partial_{\phi}^2 \Delta V)\,,
\end{align}

where all the derivatives are evaluated at $h=v$ and $\phi=v_{\phi}$.

\bibliographystyle{JHEP}
\bibliography{ref}

\end{document}